\documentclass[useAMS,usenatbib,fleqn]{mn2e}

\usepackage{aas_macros}

\usepackage{epsfig}
\usepackage{epstopdf}
\usepackage{subfigure}
\usepackage{amsmath}    
\usepackage{longtable}
\usepackage{dcolumn}
\usepackage{amssymb}

\usepackage{upgreek} 

\usepackage{wrapfig}

\DeclareGraphicsExtensions{.jpg,.eps,.EPS,.png,.pdf,.ps}

\usepackage{colordvi}
\usepackage{ulem}

\RequirePackage[colorlinks=true
,urlcolor=blue
,anchorcolor=blue
,citecolor=blue
,filecolor=blue
,linkcolor=blue
,menucolor=blue
,pagecolor=blue
,linktocpage=true
,pdfproducer=medialab
]{hyperref}

\defcitealias{Metzger2012}{M12}
\defcitealias{Fernandez&Metzger2013}{FM13}

\newcommand       \be          {\begin{eqnarray}}
\newcommand       \ee          {\end{eqnarray}}

\usepackage[usenames,dvipsnames]{xcolor}

\title[WD-NS Merger Accretion Flows]{Time dependent models of accretion disks with nuclear burning following the tidal disruption of a white dwarf by a neutron star}

\author[B. Margalit \& B.~D. Metzger]
{Ben Margalit\thanks{E-mail: \href{mailto:btm2134@columbia.edu}{btm2134@columbia.edu}} and Brian D. Metzger \\ 
\normalsize Columbia Astrophysics Laboratory, Columbia University, 538 West 120th St., New York, NY 10027
}

\date{}

\begin{document}
\maketitle

\begin{abstract}
We construct time-dependent one-dimensional (vertically averaged) models of accretion disks produced by the tidal disruption of a white dwarf (WD) by a binary neutron star (NS) companion.  Nuclear reactions in the disk midplane burn the WD matter to increasingly heavier elements at sequentially smaller radii, releasing substantial energy which can impact the disk dynamics.  A model for disk outflows is employed, by which cooling from the outflow balances other sources of heating (viscous, nuclear) in regulating the Bernoulli parameter of the midplane to a fixed value $\lesssim 0$.  We perform a comprehensive parameter study of the compositional yields and velocity distributions of the disk outflows for WDs of different initial compositions.  For C/O WDs, the radial composition profile of the disk evolves self-similarly in a quasi-steady-state manner, and is remarkably robust to model parameters. The nucleosynthesis in helium WD disks does not exhibit this behavior, which instead depends sensitively on factors controlling the disk midplane density (e.g. the strength of the viscosity, $\alpha$).  By the end of the simulation, a substantial fraction of the WD mass is unbound in outflows at characteristic velocities of $\sim 10^9~{\rm cm~s}^{-1}$.  The outflows from WD-NS merger disks contain $10^{-4}-3 \times 10^{-3} M_\odot$ of radioactive $^{56}$Ni, resulting in fast ($\sim$ week long) dim ($\sim10^{40}~{\rm erg~s}^{-1}$) optical transients; shock heating of the ejecta by late time outflows may increase the peak luminosity to $\sim 10^{43}~{\rm erg~s}^{-1}$. The accreted mass onto the neutron star is probably not sufficient to induce gravitational collapse, but may be capable of spinning up the NS to periods of $\sim 10~{\rm ms}$. This is a new possible channel for forming isolated recycled pulsars.
\end{abstract}


\section{Introduction}
The gravitational wave (GW)-driven coalescence of binary compact objects, including white dwarfs (WD), neutron stars (NS), and stellar mass black holes (BH), are widely studied as models for luminous transients.  NS-NS and NS-BH binary mergers are potential central engines of short duration gamma-ray bursts (GRBs; e.g., \citealt{Eichler+89}; \citealt{Berger14} for a review) and other electromagnetic counterparts to the GW signal (e.g.~\citealt{Metzger2010}, \citealt{Margalit&Piran2015}).  The coalescence of WD-WD binaries are likewise believed to be one of the primary channels for producing Type Ia supernovae (SNe; \citealt{Webbink84}).

In this paper we explore the outcome of WD-NS or WD-BH mergers, a class of events which have thus far received far less attention than their NS-NS, NS-BH or WD-WD counterparts.  Roughly twenty WD-NS binaries are known in our Galaxy, of which four are on sufficiently tight orbits that they will merge completely due to GW radiation within a Hubble time.  This population results in an estimated coalescence rate of $\mathcal{R} \sim 10^{-5}$--$10^{-4}$ yr$^{-1}$ per galaxy \citep{OShaughnessy&Kim2010}, comparable within uncertainties to the rate of NS-NS mergers \citep{Kim+15}.

WD-BH mergers were first studied by \cite{Fryer+1999} as a model for long duration GRBs.  They showed that a sufficiently massive WD is tidally disrupted by its BH companion as the binary orbit shrinks due to unstable mass transfer \citep[see also][]{Paschalidis+2009}.  The WD debris is then sheared into an accretion disk with an initial size which is comparable to that of the initial binary at the time of Roche Lobe overflow.  Subsequent accretion of this massive torus was proposed to power a collimated relativistic jet and GRB via $\nu-\bar{\nu}$ annihilation or the Blandford-Znajek process \citep{Fryer+1999}.  \citet{Paschalidis+11} explored the merger of WD-NS mergers using general relativistic hydrodynamical simulations.  They also found that the final state is a NS surrounded by a massive torus, which they argued evolves into a Thorne-Zytkow-like object following the transport of angular momentum outwards.  They described the GW signal that would occur if the central NS collapses to a BH following the cooling and accretion by the NS of the envelope.

The outcome of WD-NS and WD-BH mergers were revisited by \citet[hereafter \citetalias{Metzger2012}]{Metzger2012}, who focused on the steady-state structure of the remnant accretion disk.  \citetalias{Metzger2012} pointed out the importance of nuclear reactions on the structure and dynamics of the accretion flow.  As matter accretes onto the central NS or BH, gravitational energy is converted to internal energy.  This increases the midplane temperature to the point that nuclear fusion converts the inflowing WD matter into increasingly heavier elements at sequentially smaller radii.  Moving inwards through the disk, nucleosynthesis proceeds up to Fe-group elements until, at even higher temperatures, inflowing matter is photodisintegrated into $\alpha$-particles and free nuclei.  \citetalias{Metzger2012} showed that the rate of nuclear energy generated exceeds that of gravity in the outer regions of the disk at hundreds to thousands of gravitational radii, modifying the disk dynamics from those of a standard radiatively inefficient accretion flow.  This novel accretion regime is termed a `nuclear dominated accretion flow', or `NuDAF' (\citetalias{Metzger2012}). 

The high densities and optical depths of the accretion flow following a WD-NS merger prevents matter from efficiently cooling through photon radiation, while the temperatures throughout most of the disk are not enough for neutrino cooling to be dynamically relevant (\citealt{Popham+99}; \citealt{DiMatteo+02} \citealt{Chen&Beloborodov07}).   One dimensional models of such `radiatively inefficient accretion flows' are characterized by positive Bernoulli parameters \citep{Narayan&Yi1995}, indicating the potential importance of unbound outflows on the disk dynamics.  We follow the general framework of \citet{Blandford&Begelman1999}, who postulate that disk winds provide an important cooling mechanism which offsets gravitational (viscous) and nuclear heating (see \citealt{Yuan&Narayan14} for a review).  

Depending on the radial profile of nucleosynthesis within the disk, outflows from the inner regions, where Fe-group elements form, can contain varying amounts of radioactive $^{56}$Ni.  However, the total nickel yield integrated over the lifetime of the torus is generally much less than that produced through shock heating in standard core collapse or Type Ia SNe (\citetalias{Metzger2012}).  This does not exclude WD-NS or WD-BH mergers as progenitors of subluminous, or otherwise exotic, supernova-like transients.

The `Ca-rich gap transients' \citep{Perets+2010,Kasliwal+2012} are a class of recently discovered SNe which are characterized by low luminosities (indicating a small $^{56}$Ni ejecta mass), an ejecta composition rich in calcium (and poor in oxygen), fast temporal evolution (indicating a low ejecta mass of a few tenths of a solar mass), and a puzzling tendency to occur outside the disks of their host galaxies \citep{Perets+2010,Kasliwal+2012}.  Their locations show no evidence for star formation or the presence of an underlying quiescent stellar population, such as a dwarf galaxy or globular cluster (\citealt{Lyman+2014,Lyman+2016}).  Nuclear burning of helium rich matter is a natural explanation for their high Calcium abundances \citep{Perets+2010}, leading \citetalias{Metzger2012} to propose the mergers of a helium WD with a NS as their progenitors.  A large fraction of WD-NS binaries could occur in remote locations if they receive a natal kick from the SN which births the NS (\citetalias{Metzger2012}; \citealt{Lyman+2014}).

\citet[hereafter \citetalias{Fernandez&Metzger2013}]{Fernandez&Metzger2013} followed the 1D steady-state model of \citetalias{Metzger2012} with 2D (axisymmetric) hydrodynamical simulations of radiatively inefficient accretion flows with nuclear burning.  These calculations explored the vertical dynamics of the disk and its interplay with radially-steady burning front, e.g. at which carbon is synthesized to magnesium. \citetalias{Fernandez&Metzger2013} found that if the nuclear energy released at the burning front is large compared to the local thermal energy, then the burning fronts can spontaneously transition into outwards-propagating detonations due to the mixing of hot downstream matter (ash) with cold upstream gas (fuel).  These detonations either falter as the shock propagates into the outer regions of the disk, or completely disrupt the large-scale accretion flow.  Despite this intriguing finding, \citetalias{Fernandez&Metzger2013} note that the detonations they observe could be an artifact of their simplified equation of state, which included only gas pressure and neglected radiation pressure (thus artificially accentuating the temperature discontinuity at the burning front). \citetalias{Fernandez&Metzger2013} also employed only a single nuclear reaction, which prevented them from making detailed predictions for the composition of the disk outflows and their electromagnetic signatures.  

This paper extends the work of \citetalias{Metzger2012} and \citetalias{Fernandez&Metzger2013} by developing a one-dimensional time-dependent $\alpha$-disk model (with outflows) for the remnant accretion disks produced by WD-NS mergers. 
Although we focus primarily on WD-NS mergers, our analysis applies equally to WD mergers with stellar mass BHs.
We use this model to explore the response of the disk to nuclear burning and the resulting time-dependent outflow properties (mass, composition, velocity).  These details bear significantly on the optical light curves and spectra of WD-NS mergers, as well as their radio emission from the interaction of the ejecta with the interstellar medium.  These observational signatures will be investigated in a companion paper.  

The paper is structured as follows.  We begin with a brief discussion of the conditions and processes which lead up to the disruption of the WD by its binary companion, and subsequent formation of an accretion disk (\S\ref{sec:WD_Disruption}). We continue in \S\ref{sec:Disk Model} by describing the disk and outflow model adopted in our work.  We present analytic results in \S\ref{sec:Analytic Results}, the details of which are developed in Appendices \ref{subsec:Appendix_InitialOutflow} and \ref{subsec:Appendix_pExponent}. Results of our numerical simulations are presented in \S\ref{sec:Numerical Results}.  We begin with a detailed analysis of our fiducial C/O WD model (\S\ref{subsec:Fiducial C/O Model}), followed by a parameter study of variations about the fiducial model (\S\ref{subsec:Variation of C/O Model Parameters}).  In \S\ref{sec:He} we explore models for disrupted He WDs, and in \S\ref{sec:hybrid} we explore `hybrid' C/O/He WDs.  We discuss our results in \S\ref{sec:Discussion} and 
conclude 
in \S\ref{sec:Conclusions}.

\section{WD Disruption and Disk Formation} \label{sec:WD_Disruption}

We are interested in binary systems consisting of a WD secondary of mass $M_{\rm WD}$ and a NS primary of mass $M$.  The binary loses orbital energy through GW emission, causing the orbit to shrink and leading to an eventual contact.  The WD experiences Roche lobe overflow (RLOF) once the orbital separation reaches a value \citep{Eggleton1983}
\begin{equation} \label{eq:a_RLOF}
a_\mathrm{RLOF} \approx R_{\rm WD} \frac{0.6 q^{2/3} + \ln \left( 1 + q^{1/3} \right)}{0.49 q^{2/3}},
\end{equation}
where $q = M_{\rm WD} / M$ and $R_{\rm WD}$ is the WD radius.  The latter is well approximated by \citep{Nauenberg1972} 
\begin{equation} \label{eq:R_wd}
R_\mathrm{WD} \approx 10^9 ~ \mathrm{cm} ~ \left( \frac{M_\mathrm{WD}}{0.7 ~ M_\odot} \right)^{-1/3} \left[ 1 - \left( \frac{M_\mathrm{WD}}{M_\mathrm{ch}} \right)^{4/3} \right]^{1/2},
\end{equation}
where $M_{\rm ch} \approx 1.45~M_\odot$ is the Chandrasekhar mass assuming the mean molecular weight per electron of $\mu_e = 2$.

As mass is transferred from the WD to the more massive primary, conservation of angular momentum drives the binary semi-major axis to increase.  On the other hand, as the WD loses mass its radius increases (equation~\ref{eq:R_wd}), which increases the minimal separation for RLOF, $a_{\rm RLOF}$ (equation~\ref{eq:a_RLOF}).  The competition between the two effects is ultimately determined by their timescales --- if $a_{\rm RLOF}$ increases faster than the binary's semi-major axis, the system will progress into runaway mass transfer, effectively disrupting the WD on a dynamical timescale. Otherwise, the binary will slowly drift apart, maintaining stable mass transfer.

\begin{figure} 
\centering
\epsfig{file=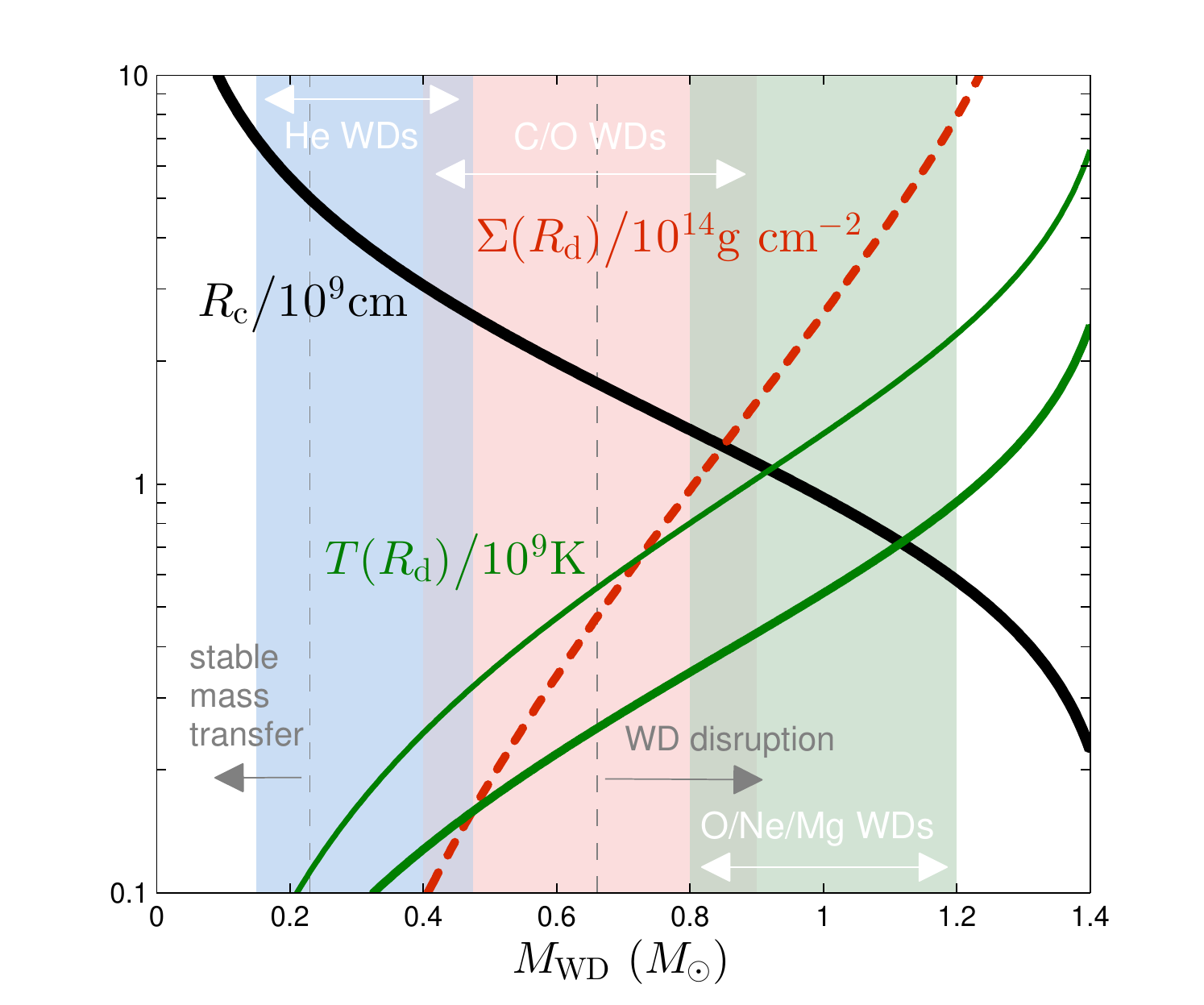,angle=0,width=0.5\textwidth}
\caption{Key parameters of the accretion disk produced by the tidal disruption of a WD by a 1.4$M_{\odot}$ NS binary companion, as a function of the WD mass.  The black curve depicts the circularization radius $R_{\rm c}$ (equation~\ref{eq:R_circ}), which represents the characteristic initial radius of the disk, $R_{\rm d}$. A red dashed curve shows the initial disk surface density $\Sigma_0(R_{\rm d})$ at $r = R_{\rm d}$, calculated  for $m=2$, and $n=7$ (equation~\ref{eq:Density_initial}).  The two green curves bracket the midplane temperature $T(R_{\rm d})$ in the limits that radiation pressure (bottom, thicker curve) and gas pressure (top, lighter curve) dominate, respectively.  The shaded background shows the expected WD composition based on its mass \citep[e.g.][]{Liebert+2005}.  Vertical grey dashed lines mark critical WD masses for unstable mass transfer \citep{Paschalidis+2009}, corresponding to the lower limit set by conservative mass transfer (rightmost line).  The leftmost dashed line provides an estimate of the lower limit on the WD mass allowing unstable mass transfer, in the more realistic case of non-conservative mass transfer.} \label{fig:WDMassDiagram}
\end{figure}

Conservative mass transfer, in which the orbital angular momentum remains constant, is unstable for binaries with mass ratios $q \gtrsim 0.43-0.53$. If, however, orbital angular momentum is deposited into an accretion disk which does not transfer it back into the binary (\citealt{Lubow&Shu1975}), then significantly smaller mass ratios (as small as $q \sim 0.2$) can also lead to unstable mass transfer \citep{Verbunt&Rappaport1988,Paschalidis+2009}.  For a $1.4M_\odot$ NS primary, a conservative lower-limit on the WD mass necessary for disruption is therefore $M_{\rm WD} \gtrsim 0.66M_\odot$. However, in the more realistic case that at least some orbital angular momentum is lost, lower mass WDs can also be disrupted.  The stringent lower-limit of $M_{\rm WD} \gtrsim 0.23M_\odot$ for disruption can in principle extend into the mass range of helium WDs (\citealt{Bobrick+16}; Fig.~\ref{fig:WDMassDiagram}).

If the WD is disrupted by unstable mass transfer at $a_{\rm RLOF}$, its debris will quickly be sheared into an accretion disk of characteristic dimensions proportional to the circularization radius, 
\begin{equation} \label{eq:R_circ}
R_{\rm c} = a_{\rm RLOF} (1 + q)^{-1}.
\end{equation}
This circularization radius is defined as the semi-major axis of a point mass $M_{\rm WD}$ orbiting the central NS/BH, with an angular momentum equal to the that of the binary at the time of disruption.

Detailed hydrodynamical simulations of the WD disruption are required to determine the disk configuration following the disruption (\citealt{Fryer+1999,Paschalidis+11}).  Such a detailed numerical calculation is beyond the scope or purpose of the present work.  We instead adopt a flexible analytic description for the ``initial" disk surface density formed by the disruption, 
\begin{equation} \label{eq:Density_initial}
\Sigma_0 (r) = \mathcal{N}(m,n) \frac{M_\mathrm{d}}{2 \uppi R_{\rm d}^2} {\left(\frac{r}{R_{\rm d}}\right)^m}{\left[1 + \frac{m+2}{n-2} \left(\frac{r}{R_{\rm d}}\right) \right]^{-(m+n)}} ~.
\end{equation}
Here $R_{\rm d} = \mathcal{R}(m,n) R_{\rm c}$ is the characteristic disk radius, at which the local mass $\propto \Sigma_0 r^2$ peaks, $r$ is the cylindrical radial coordinate centered on the NS/BH, and $\mathcal{N}(m,n)$, $\mathcal{R}(m,n)$ are constants $\lesssim 1$ given explicitly in Appendix \ref{subsec:Appendix_InitialConditions}. The latter are calculated assuming that mass and angular momentum are conserved in the disruption process, in which case the total disk mass is $M_{\rm d}=M_{\rm WD}$.

\begin{figure}
\centering
\epsfig{file=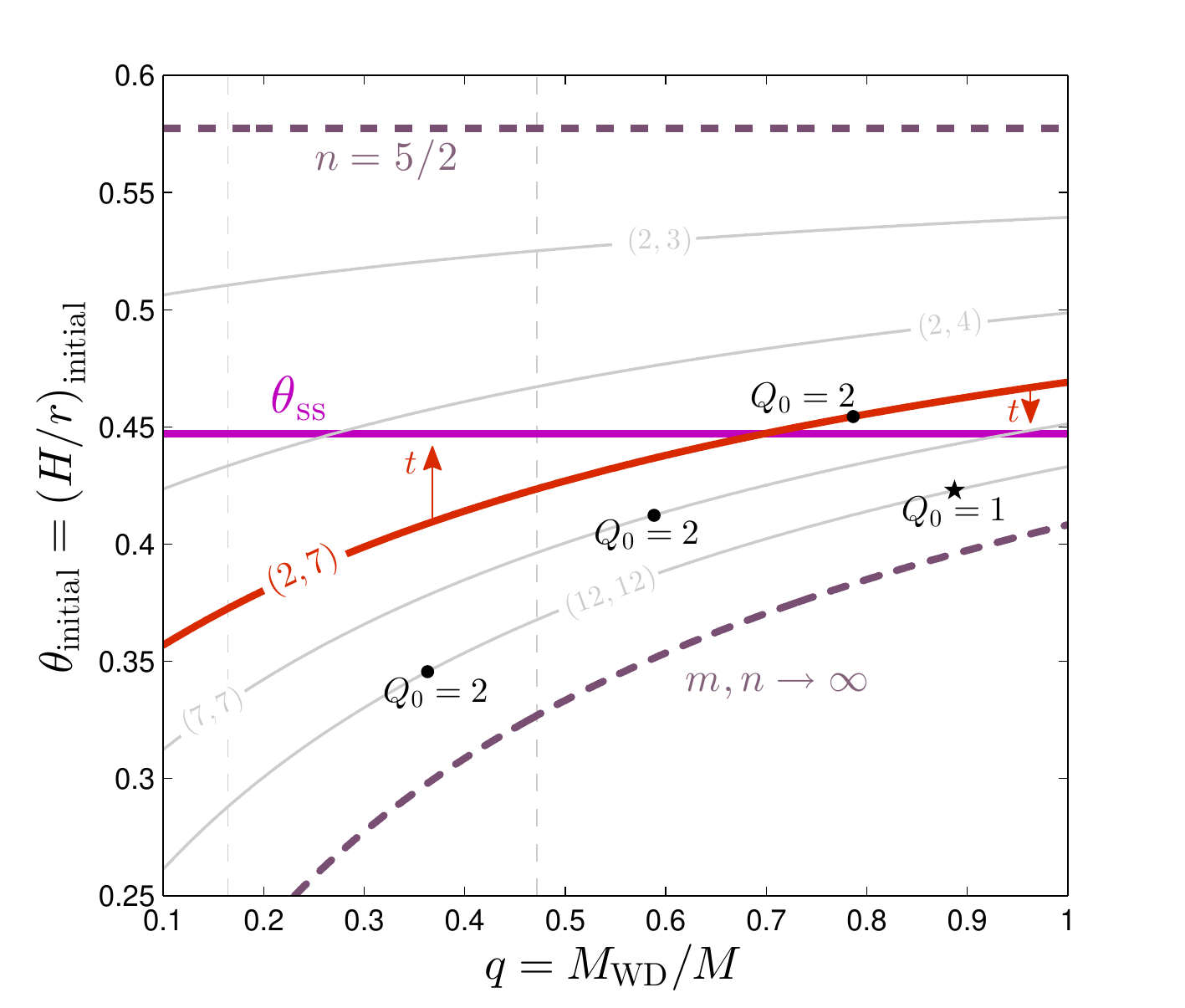,angle=0,width=0.5\textwidth}
\caption{Initial disk aspect ratio, $\theta$, as a function of the binary mass ratio, $q$, for $\gamma = 5/3$ and a range of values of the power-law index parameters $(m,n)$ used to define the initial surface density (labeled along each curve; equation~\ref{eq:Density_initial}). 
The dashed purple curves bracket the permissible range of $\theta_{\rm initial}$.
The horizontal solid purple curve depicts the steady-state value of $\theta = \theta_{\rm ss}$ to which the disk evolves
(taking ${\rm Be}^\prime_{\rm crit} = 0$; equation~\ref{eq:theta_ss}).
If the initial aspect ratio obeys $\theta > \theta_{\rm ss}$, then energy is quickly dissipated by strong outflows until $\theta = \theta_{\rm ss}$.  Alternatively, initial disk configurations with $\theta < \theta_{\rm ss}$ will expand to $\theta_{\rm ss}$ due to viscous and nuclear heating without a significant prompt outflow (red arrows). 
Values of the Toomre parameter $Q_0$ are illustrated by black points and stars. The value of $Q_0$ decreases as one moves along each curve to larger $q$. Only for very large values of $q$ and $(m,n)$ does $Q_0$ drop below unity, indicating that our disk configurations are stable to self-gravitational instabilities.
The right (left) dashed vertical curve approximates the conservative (lower-limit) mass ratio above which the WD is tidally disrupted (see Fig. \ref{fig:WDMassDiagram}).
The vertical axis is readily scaled to different adiabatic indexes; for $\gamma=4/3$, the values of $\theta$ decrease by a factor of $\sqrt{2}$.} \label{fig:theta_initial}
\end{figure}

We further assume that energy is conserved during the process of disk formation because the timescale for energy transport via convection or radiation is orders of magnitude longer than the dynamical timescale over which the disruption occurs.  Equating the orbital energy at disruption to the total initial disk energy (internal+kinetic+gravitational) and using equation (\ref{eq:Density_initial}), we solve for the disk aspect ratio at formation, 
\begin{equation} \label{eq:theta_initial}
\theta_{\rm initial} = \left(\frac{H}{r}\right)_{\rm initial} = \sqrt{\frac{\gamma - 1}{2} \left( 1 - \frac{1}{\left( 1 + q \right) \mathcal{T}(m,n)} \right)} ~,
\end{equation}
which we assume is radially constant.  Here $H$ is the isothermal scaleheight of the disk at radius $r$, $\gamma$ is the adiabatic index (equation \ref{eq:gamma-law_EOS}), $\mathcal{T}(m,n)$ is a constant (Appendix \ref{subsec:Appendix_InitialConditions}), and we have assumed that the disk orbits at the Keplerian rate, $\Omega=\Omega_{\rm k} = (GM/r^{3})^{1/2}$.  Fig.~\ref{fig:theta_initial} shows $\theta_{\rm initial}$ as a function of the binary mass ratio, $q$, for the physically allowed range of the parameters $m$ and $n$.  For comparison a horizontal solid purple line shows the characteristic value of the disk thickness obtained once a steady inflow is achieved ($\S\ref{subsec:MassInflowIndex}$).

The disk is sufficiently massive that we consider the possibility that it becomes susceptible to instabilities arising from self-gravity. The Toomre parameter,
\begin{equation}
Q = \frac{\Omega c_{\rm s}}{\uppi G\Sigma } = \frac{M\theta}{\uppi r^2 \Sigma} \propto \frac{\theta}{q}
\end{equation}
is less than unity for unstable configurations.
The minimal value of this parameter, $Q_0$, is obtained at $t=0$ and $r=R_{\rm d}$. Using equations (\ref{eq:Density_initial}) and (\ref{eq:theta_initial}) for the initial density and disk aspect ratio, we find that $Q_0 > 1$ for most reasonable parameters, indicating that our disks are stable (Fig. \ref{fig:theta_initial}).

The midplane densities and temperatures of WD-NS merger disks span a range of values for which ions, radiation, and (to a lesser extent) degenerate electrons can all contribute significantly to the pressure and energy density of the fluid (\citetalias{Metzger2012}).  At large radii in the disk, and at times soon after disruption, the entropy is relatively low and gas pressure dominates over radiation pressure.  At smaller radii at early times (and for most radii at late times), radiation pressure instead becomes dominant.  In the limits that gas or radiation dominate the midplane pressure, the midplane temperature is given by
\begin{equation} \label{eq:Temp}
T(r) = 
\begin{cases}
(\mu m_p / k_B) \theta^2 \Omega_{\rm k}^2 r^2 ~, ~~& \mathrm{gas} \\
\left[ (3 / 2 a) \theta \Omega_{\rm k}^2 r \Sigma \right]^{1/4} ~, & \mathrm{radiation} 
\end{cases}
\end{equation}
where $\mu$ is the mean molecular weight.
Fig.~\ref{fig:WDMassDiagram} shows the initial disk temperature at $R_{\rm d}$ for various parameters.

\section{Disk and Outflow Model} \label{sec:Disk Model}
This section describes our numerical model for the disk evolution and outflows.  We begin by summarizing the vertically averaged disk equations governing the dynamics, before continuing with details of the mass loss prescription, nuclear burning, and our numerical procedure.

\subsection{Disk Equations}

The vertically integrated continuity equation reads
\begin{equation} \label{eq:continuity}
\partial_t \Sigma + \frac{1}{r} \partial_r \left( r v_r \Sigma \right) + \dot{\Sigma}_\mathrm{w} = 0 ~,
\end{equation}
where $v_r$ is the radial fluid velocity and $\dot{\Sigma}_{\rm w}$ is a sink term which accounts for mass loss from the disk via winds (\S\ \ref{subsec:Wind Prescription}).  Vertical hydrostatic equilibrium is assumed, implying that $H/r \approx c_{\rm s}/v_{\rm k}$, where $c_{\rm s} \equiv \sqrt{P/\rho}$ is the midplane isothermal sound speed, and $v_{\rm k} = r\Omega_k$ is the Keplerian orbital velocity.

The radial momentum equation can be manipulated to obtain the angular velocity,
\be \Omega \approx \Omega_{\rm k} \sqrt{ 1 + \theta^2 \left( {\partial \ln \Sigma}/{\partial \ln r} - 1 \right) }.\ee  
However, because in practice we find that in most cases $\Omega \simeq \Omega_{\rm k}$ to an accuracy of $\lesssim 10\%$, for simplicity we fix $\Omega = \Omega_{\rm k}$ throughout the remainder of this work.

The vertically-averaged azimuthal momentum equation can be rearranged to obtain the radial velocity
\begin{align} \label{eq:v_r}
v_r &\approx -3 \frac{\nu}{r} \frac{\partial \ln \left[ r^2 \nu \Sigma \Omega \right]}{\partial \ln r} \\ \nonumber 
&= -3 \alpha \theta^2 v_{\rm k} \left[ 2 + \frac{\partial \ln \Sigma}{\partial \ln r} + 2 \frac{\partial \ln c_{\rm s}}{\partial \ln r} \right] ~,
\end{align}
where $\nu$ is the kinematic `viscosity', which physically is associated with an anomalous stress. In the second equality we have adopted the standard \citet{Shakura&Sunyaev1973} alpha prescription,
\begin{equation} \label{eq:ShakuraSunyaev_viscosity}
\nu = \alpha c_{\rm s}^2 / \Omega_{\rm k} ~.
\end{equation}
As the magnetorotational instability (MRI) provides one physical mechanism for angular momentum transport \citep{Balbus&Hawley1991}, we adopt values of $\alpha \sim 0.01-0.1$, consistent with those measured by numerical simulations of the MRI \citep[e.g.][]{Davis+2010}.  

The surface density of the disk evolves on the characteristic viscous timescale,
\begin{equation} \label{eq:t_visc}
t_{\rm visc} = \frac{r^2}{\nu} = \alpha^{-1} \theta^{-2} \Omega_{\rm k}^{-1}  ~
\end{equation}
which is longer than the dynamical timescale $\Omega_{\rm k}^{-1}$ by a factor of $\alpha^{-1} \theta^{-2} \gg 1$.

Finally, the specific entropy $s$ and internal energy $u$ evolve according to the first law of thermodynamics,
\begin{equation}
\dot{q}_\mathrm{tot} = \Sigma T (D_t s) = \Sigma (D_t u) - c_{\rm s}^2 (D_t \Sigma),
\label{eq:1st}
\end{equation}
where $D_t \equiv \partial_t + v_r \partial_r$ is the Lagrangian derivative and 
\be
\dot{q}_\mathrm{tot} = \dot{q}_\mathrm{visc} + \dot{q}_\mathrm{nuc} + \dot{q}_\mathrm{wind}
\ee
 is the total disk heating rate per unit area, where
\begin{equation} \label{eq:q_dot_visc}
\dot{q}_\mathrm{visc} = \Sigma \nu \Omega^2 \left( \frac{\partial \ln \Omega}{\partial \ln r} \right)^2 = \frac{9}{4} \alpha \Sigma c_{\rm s}^2 \Omega_{\rm k}
\end{equation}
is the viscous heating rate, $\dot{q}_{\rm nuc}$ is the heating rate due to nuclear burning ($\S\ref{sec:nuclear}$), and $\dot{q}_{\rm wind}$ is the wind cooling rate ($\S\ref{subsec:Wind Prescription}$).

Using continuity (equation~\ref{eq:continuity}), equation (\ref{eq:1st}) can be recast as
\begin{equation} \label{eq:internal_energy}
\partial_t u = \frac{\dot{q}_\mathrm{tot}}{\Sigma} - v_r \partial_r u + c_{\rm s}^2 \left[ \frac{1}{r} \partial_r \left( r v_r \right) + \frac{\dot{\Sigma}_\mathrm{w}}{\Sigma} \right] ~.
\end{equation}

The above equations are closed by imposing an EOS which relates the isothermal sound speed to the internal energy and density $c_s = c_s \left( u, \Sigma \right)$.
For a `gamma-law' EOS, this relation takes the form
\begin{equation} \label{eq:gamma-law_EOS}
u = \frac{c_{\rm s}^2}{\gamma-1} ~.
\end{equation}
Although we incorporate a full EOS in our numerical calculations, equation (\ref{eq:gamma-law_EOS}) is used in analytic estimates.

\subsection{Wind Prescription} \label{subsec:Wind Prescription}
Outflows launched from the disk represent an important sink of mass and energy, as represented by the terms $\propto \dot{\Sigma}_{\rm w}$ in equations (\ref{eq:continuity}) and (\ref{eq:internal_energy}).  We assume that winds do not exert a net torque on the disk and hence neglect their effects on the angular momentum evolution of the disk.

Two parameters are required to prescribe the outflow.  Following \citet{Kohri&Narayan&Piran2005} and \citetalias{Metzger2012}, we define a wind cooling efficiency $\eta_{\rm w}$, which is related to the asymptotic wind velocity by
\be
v_{\rm w} = v_{\rm k} \sqrt{2\eta_{\rm w}}.
\ee
A value $\eta_{\rm w} \sim \mathcal{O}(1)$ corresponds to winds launched at velocities close to the local escape speed.  The corresponding timescale for mass loss is $t_{\rm w} \sim H/v_{\rm w} \sim \theta \Omega_{\rm k}^{-1}$; for $\eta_{\rm w} \sim 1$ this is a factor of $\theta < 1$ times smaller than the local dynamical timescale and a factor of $\alpha \theta^3 \ll 1$ smaller than the accretion timescale. This motivates a prescription for \textit{local} wind cooling, which effectively acts instantaneously.

Another important quantity is the Bernoulli parameter of the disk midplane,
\begin{equation} \label{eq:Bernoulli}
{\rm Be_d} = \frac{1}{2}\Omega^2 r^2 + \frac{1}{2}v_r^2 + u + c_{\rm s}^2 - v_{\rm k}^2 ~,
\end{equation}
and its normalized value ${\rm Be^\prime_d} = {\rm Be_d}/v_{\rm k}^{2}$.  The fact that this quantity is generally positive in one dimensional models of radiatively inefficient accretion flows  \citep{Narayan&Yi1995,Blandford&Begelman1999} shows that matter in principle has sufficient thermal energy to adiabatically expand to infinity.  Using  a $\gamma$-law EOS (equation~\ref{eq:gamma-law_EOS}), the normalized Bernoulli parameter can be written as
 \begin{equation} \label{eq:Bernoulli_theta}
\mathrm{Be^\prime_d} \approx -\frac{1}{2} + \frac{\gamma}{\gamma-1}\theta^2 ~,
\end{equation}
where the radial kinetic energy $\propto \alpha^2 \theta^4 \ll 1$ has been neglected. 

We adopt a wind prescription which cools the disk when the Bernoulli parameter exceeds a fixed value, ${\rm Be^\prime_{crit}} \lesssim 0$.  To conserve energy globally, this is tantamount to assuming that some mechanism (e.g. turbulence or wave damping) heats matter in the corona where the wind is launched at a specific rate exceeding that in the midplane.  In other words, this {\it preferential heating} above the midplane allows some matter to become unbound at the expense of the rest of the disk maintaining Be$'_{\rm d} \lesssim 0$.  Although we do not presume to understand the details of the wind launching process, the properties of the disk/outflow structure that we find by making this assumption show qualitative agreement with global hydrodynamical (e.g.~\citealt{Stone+99}) and MHD (e.g.~\citealt{Hawley&Balbus02}) simulations of radiatively inefficient accretion flows.  These simulations indeed find that the Bernoulli parameter in outflows from the disk at high latitudes are higher than its value in the disk midplane, where ${\rm Be^\prime_d} \lesssim 0$, due to a higher specific heating rate above the midplane.

The picture described above translates into the following functional form for the wind mass loss rate,
\begin{equation} \label{eq:Sigma_dot_w}
\dot{\Sigma}_\mathrm{w} = \Sigma \Omega_{\rm k} \theta^{-1} \sqrt{2 \left( \eta_\mathrm{w} + 1 \right)} \times \Theta\left(\mathrm{Be^\prime_d} - \mathrm{Be^\prime_{crit}}\right) ~,
\end{equation}
where $\Theta\left(x\right)$ is the Heaviside function.  This prescription captures the qualitative expectation that matter is only unbound if the Bernoulli parameter of the disk exceeds a threshold value of ${\rm Be^\prime_{crit}}$.  When outflows are present, it is also consistent with the order of magnitude estimate $\dot{\Sigma}_{\rm w} \sim \Sigma / t_{\rm w}$.

The wind efficiency parameter $\eta_{\rm w}$ essentially equals the specific energy carried away in the wind. The cooling rate of the disk by the wind is therefore given by
\begin{equation} \label{eq:q_dot_w}
\dot{q}_\mathrm{w} = - \dot{\Sigma}_{\rm w} \left( v_{\rm w}^2 /2 - \mathrm{Be_d} \right) = - \dot{\Sigma}_\mathrm{w} v_{\rm k}^2 \left( \eta_\mathrm{w} - \mathrm{Be^\prime_d} \right) ~.
\end{equation}
This general cooling prescription does not depend on the less certain form of $\dot{\Sigma}_{\rm w}$ (equation \ref{eq:Sigma_dot_w}) in the common scenario of a quasi-steady-state disk evolution (see Appendix \ref{subsec:Appendix_pExponent}).
Also note that as long as the mass loss mechanism regulates the disk Bernoulli parameter to $\mathrm{Be^\prime_{crit}}$, then we must require that $\mathrm{Be^\prime_{crit}} < \eta_{\rm w}$, as otherwise the wind cannot cool the disk.  This condition is satisfied if the unbound material has been preferentially heated, as hypothesized above.

\subsection{Nuclear Burning}
\label{sec:nuclear}
The mass fraction of each isotope in the disk, $X_A$, evolves according to an equation of continuity,
\begin{align} \label{eq:X_A}
\partial_t X_A + \frac{1}{r} \partial_r &\left( r v_r X_A \right) + \\ \nonumber 
&\frac{1}{r \Sigma} \partial_r \left[ r \Sigma \nu_{\rm mix} \left(\partial_r X_A \right) \right] + \dot{X}_A^{\rm (nuc)} = 0 ~,
\end{align}
where the second term accounts for the radial advection of the nuclear species with the accretion velocity $v_r$. The third term accounts for mixing of nuclear isotopes with a diffusion coefficient $\nu_{\rm mix}$.  Such mixing is expected due to the same turbulent motions in the disk which drive angular momentum transport, and hence $\nu_{\rm mix}$ is intimately related to the `Shakura-Sunyaev' viscosity $\nu$.  We therefore assume
\begin{equation} \label{eq:alpha_tilde}
\nu_{\rm mix} = \tilde{\alpha} \nu ~.
\end{equation}
Numerical simulations of the MRI which follow the evolution of a passive scalar suggest that $\tilde{\alpha} \approx 0.1$ \citep{Carballido&Stone&Pringle2005}, indicating that `chemical' mixing is less efficient than angular momentum transport. We take this as the fiducial value of $\tilde{\alpha}$ throughout our work, but also vary the value of this parameter, examining its affect on the results.

The last term in equation (\ref{eq:X_A}) represents species-changing nuclear reactions.  For purposes of analytic estimates it is convenient to approximate individual burning rates, $\dot{X}_A^{\rm (nuc)}$, as power-laws near their burning temperature,
\begin{equation} \label{eq:X_A_nuc}
\dot{X}_A^{\rm (nuc)} \propto \rho^\delta X_A^{\delta+1} T^\beta ~.
\end{equation}
For carbon burning, $^{12}$C$(^{12}$C$,\gamma)^{24}$Mg, one can approximate the reaction rate around $\sim 10^9~{\rm K}$ with $\beta=29$, $\delta=1$.

Note that in steady-state, and neglecting the diffusive mixing term, the nuclear reaction rate at the burning front is determined entirely by the accretion velocity, $v_r$, which supplies unburned fuel to the burning front.

In addition to altering the disk composition, nuclear reactions provide a source of heating or cooling, $\dot{q}_{\rm nuc}$, which contributes to the net heating rate $\dot{q}_{\rm tot}$ in equation (\ref{eq:internal_energy}). This term is obtained by summing the energy production rates of all isotopes
\begin{equation}
\frac{\dot{q}_{\rm nuc}}{\Sigma} = \sum_{A,A^\prime} \dot{X}_{A\to A^\prime}^{\rm (nuc)} \frac{Q_{A\to A^\prime}}{m_{A^\prime}} ~,
\end{equation}
where $m_{A^\prime}$ is the mass of isotope $A^\prime$, and $Q_{A\to A^\prime}$ is the Q-value of the reaction turning isotope $A$ into $A^\prime$. The latter neglects energy carried away by neutrinos, which are not trapped for the characteristic densities of the accretion flow.

\subsection{Numerical Procedure} \label{subsec:NumericalProcedure}
We numerically solve equations (\ref{eq:continuity}), (\ref{eq:internal_energy}), and (\ref{eq:X_A}), using expressions for the accretion velocity (\ref{eq:v_r}), mass loss rate (\ref{eq:Sigma_dot_w}), and wind cooling terms (\ref{eq:q_dot_w}). 
We employ the Helmholtz EOS \citep{Timmes&Swesty2000} in relating the thermodynamic variables $c_{\rm s}$, $u$, $\Sigma$, and $T$ (the last of which is necessary to evaluate the nuclear burning rates). This EOS accurately and consistently accounts for an electron-positron gas with arbitrary degree of degeneracy and relativistic motion, an ideal gas of ions, and a Planckian distribution of photons.  

The nuclear reaction rates, $\dot{X}_A$, and nuclear heating term, $\dot{q}_{\rm nuc}$, are numerically evaluated  using the  publicly available\footnote{http://cococubed.asu.edu/code{\_}pages/burn{\_}helium.shtml} 19-isotope $\alpha$-chain reaction network of \cite{Weaver&Zimmerman&Woosley1978}.  This network effectively captures the main burning channels of WD matter ($^{12}$C,$^{16}$O,$^{4}$He,$^{20}$Ne,$^{24}$Mg) up to $^{56}$Ni.  This network takes as input arguments a list of the abundances $\left\{ X_A \right\}_A$, the temperature $T$, and the density $\rho$ at a particular radial and temporal gridpoint, as well as the burning time $dt$, and returns the updated abundances, and energy deposition.

The equations are converted into finite-difference form and solved on a logarithmic radial grid spanning two orders of magnitude above and below the initial peak-density radius $=R_{\rm d} \times \left[ m(n-2) \right]/\left[ n(m+2) \right]$. The initial conditions for $\Sigma$ and $\theta$ are taken according to equations (\ref{eq:Density_initial}) and (\ref{eq:theta_initial}). The variable timestep between each iteration is chosen based on a Courant condition
\begin{equation}
dt = 0.1 \min \left[ \frac{dr^2}{\nu}, \frac{dr}{c_{\rm s}}, \frac{u}{{\rm Eq. ~(\ref{eq:internal_energy}) ~R.H.S.}} \right] ~,
\end{equation}
where the minimum runs also over the entire radial grid on which the three arguments implicitly depend. The third argument of the minimum function ensures that heat deposition in the disk is temporally resolved, which is particularly important considering nuclear heating contributions.

Since nuclear network calls are computationally expensive,
we develop a numerical `steady-state scheme'.
The basic principle is motivated by the fact that
the accretion flow quickly (on a $\lesssim$ viscous timescale) establishes a quasi-steady-state regime, after which, physical quantities vary only secularly with mass loss from the disk.
This means that over short, dynamical, timesteps the temperature, density, and abundance profiles do not change significantly, and consequently neither do the nuclear reaction rates. 

We utilize this property of the accretion flow by logging the nuclear reaction and heating rate at each gridpoint immediately after the nuclear network has been called. At later timesteps, we use the same $\dot{X}_A$ and $\dot{q}_{\rm nuc}$ at this gridpoint in evaluating equations (\ref{eq:X_A}) and (\ref{eq:internal_energy}), instead of calling the nuclear network. We continue using these logged rates until either the temperature, or one of the abundances has fractionally changed by more than $10^{-2}$ since the last network call, at which point we recalculate the rates using the nuclear network. This simple procedure retains nearly perfect fidelity with the full network calculation yet reduces the computational time by factors of several 
(the effective benefit of our method increases with time, as the accretion flow evolves over longer timescales).

\section{Analytic Results} \label{sec:Analytic Results}
We begin by summarizing several key analytic results for the steady-state structure of the disk and outflows, the derivations of which are provided in Appendices \ref{subsec:Appendix_InitialOutflow} and \ref{subsec:Appendix_pExponent}.   

\subsection{Disk Winds} \label{subsec:MassInflowIndex}
Outflows regulate the Bernoulli parameter of the disk midplane (equation \ref{eq:Bernoulli_theta}) to a critical value, ${\rm Be^\prime_d} \simeq {\rm Be^\prime_{crit}}$.  In steady-state, this condition yields a radially constant disk aspect ratio of
\begin{equation} \label{eq:theta_ss}
\theta_\mathrm{ss} = \left(\frac{H}{r}\right)_{\rm ss} \approx \sqrt{\frac{\gamma-1}{2\gamma} \left( 1+2\mathrm{Be^\prime_{crit}} \right)} ~.
\end{equation}
Any disk structure will achieve this universal aspect-ratio on a short timescale set either by the outflow or thermal time (depending on whether initially $\theta > \theta_{\rm ss}$ or $\theta < \theta_{\rm ss}$, respectively). This result is independent of the specific implementation of our wind prescription as long as ${\rm Be^\prime_d} = {\rm Be^\prime_{crit}}$.

The aspect ratio provides a measure of the thermal energy of the disk (equation \ref{eq:Bernoulli_theta}). If the disk is initially too hot, such that the initial aspect ratio (equation \ref{eq:theta_initial}) exceeds its steady-state value, $\theta_{\rm ss}$, then strong winds will act to quickly cool the flow, until ${\rm Be^\prime_d} \simeq {\rm Be^\prime_{crit}}$. The total mass lost from the disk during this brief `precursor' phase is approximately (Appendix \ref{subsec:Appendix_InitialOutflow})
\begin{equation} \label{eq:M_precursor_text}
M_{\rm w}^{\rm (precursor)} \approx \frac{1+2{\rm Be^\prime_{crit}}}{\gamma \left(\eta_{\rm w} - {\rm Be^\prime_{crit}} \right)} \left(\frac{\Delta \theta}{\theta_{\rm ss}}\right) \times M_{\rm d} ~,
\end{equation}
where $\Delta \theta = \theta_{\rm initial}-\theta_{\rm ss}$ is the difference between the initial and steady-state value of the disk aspect ratio.

Only for large binary mass ratio $q$ does $\theta_{\rm initial}$ (solid red line in Fig.~\ref{fig:theta_initial}) exceed the steady-state value $\theta_{\rm ss}$ (horizontal solid purple curve).  Even in this case, however, $\Delta \theta / \theta_{\rm ss} \lesssim 5\times10^{-2}$ is sufficiently small that $M^{\rm (precursor)}_{\rm w} \lesssim 3\times10^{-2} M_{\rm d}$ for fiducial values of the relevant parameters. 

Following standard notation \citep[e.g.,][]{Blandford&Begelman1999}, we define the mass inflow exponent 
\begin{equation} \label{eq:p_index_definition}
p \equiv \frac{\partial \ln \dot{M}_{\rm in}}{\partial \ln r}, 
\end{equation}
where $\dot{M}_{\rm in} = 2 \uppi r v_r \Sigma$ is the local mass inflow rate.  The value of $p$ is constrained by energy and mass conservation to be in the range $0 \leq p < 1$ for normal accretion disks without nuclear burning as an additional source of energy.

\begin{figure}
\centering
\epsfig{file=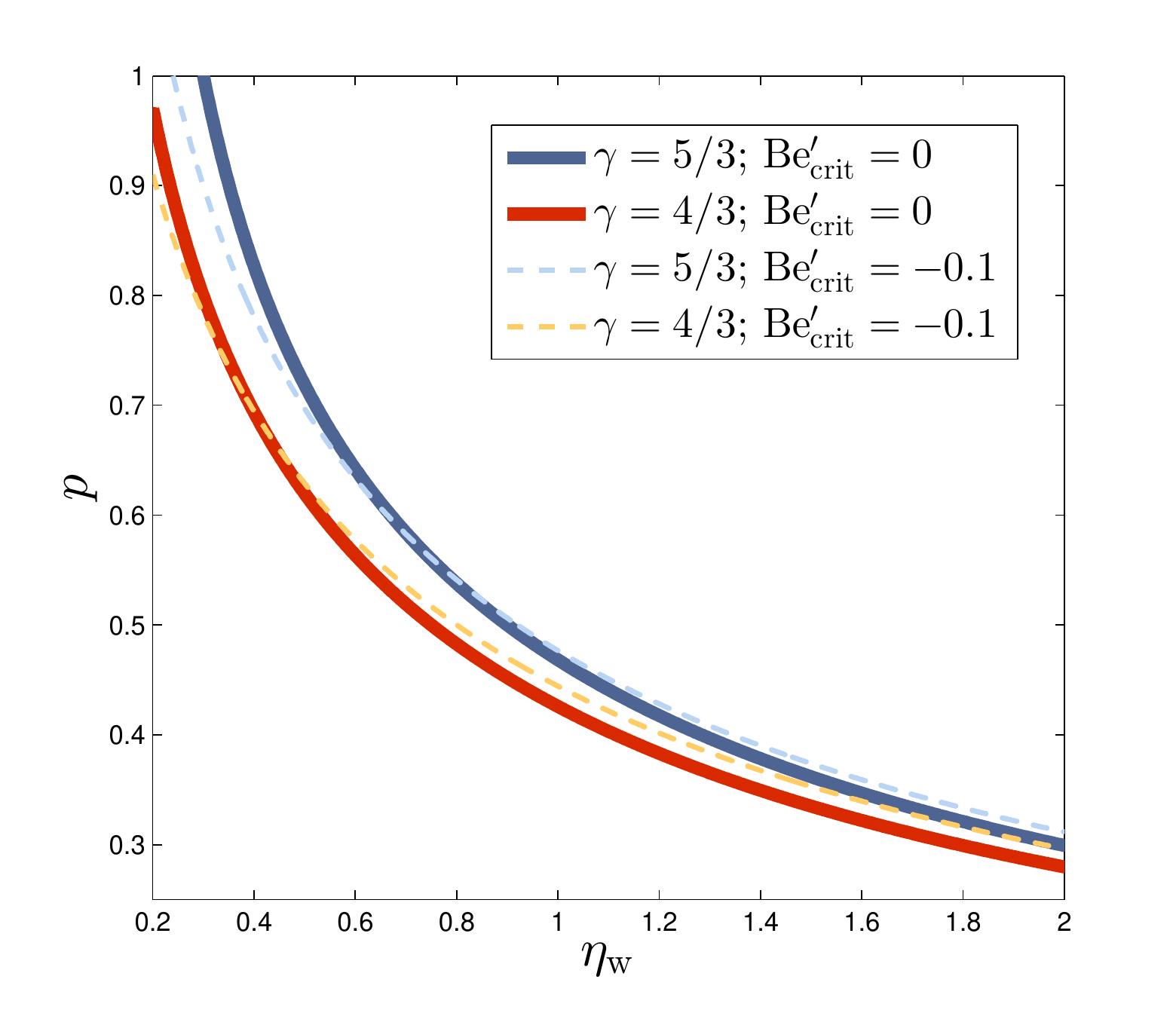,angle=0,width=0.5\textwidth}
\caption{Mass inflow exponent $p$ (equation~\ref{eq:p_index_definition}) as a function of the wind efficiency parameter $\eta_{\rm w}$ based on the analytic expression derived in Appendix \ref{subsec:Appendix_pExponent}.  Blue and red tinted curves are calculated for an adiabatic index of $\gamma=5/3$ and $\gamma=4/3$, respectively, and hence bracket the allowed range.  The dependence of $p$ on the disk Bernoulli regulation parameter ${\rm Be^\prime_{crit}}$ is much weaker.  Thick solid curves are calculated assuming ${\rm Be^\prime_{crit}}=0$, whereas dashed light curves are for ${\rm Be^\prime_{crit}}=-0.1$.  Global hydrodynamical and MHD simulations of radiatively inefficient accretion flows typically find values of $p \gtrsim 0.5$, suggesting preferred values of $\eta_{\rm w} \lesssim 1$.} \label{fig:p_index}
\end{figure}

As shown in Appendix \ref{subsec:Appendix_pExponent}, combining the wind cooling prescription (equation~\ref{eq:q_dot_w}) with mass and energy conservation (equations~\ref{eq:continuity},\ref{eq:internal_energy}) under steady-state conditions ($\partial_t = 0$) fully determines the value of $p=p(\eta_{\rm w}, {\rm Be^\prime_{crit}}, \gamma)$.  Fig. \ref{fig:p_index} shows that for physically reasonable choices of $\eta_{\rm w} \approx 1$ and ${\rm Be^\prime_{crit}} \approx 0$, one obtains values of $p \gtrsim 0.5$ which are in broad agreement with the results of hydrodynamical and MHD simulations of radiatively inefficient accretion flows \citep{Stone+99,Igumenshchev+00,Hawley+01,Narayan+12,McKinney+12,Yuan+12}.

Our analytic solution does not account for nuclear heating, $\dot{q}_{\rm nuc}$, which breaks the self-similarity of the problem by introducing additional energy and time scales.  Nuclear heating competes with viscous heating in locally balancing wind cooling (advective cooling is approximately a fixed fraction of $\dot{q}_{\rm visc}$ in steady-state; Appendix \ref{subsec:Appendix_pExponent}). Since more mass must be lost to winds to offset additional nuclear burning at fixed $\eta_{\rm w}$, nuclear heating increases the value of $p$ locally near the burning front, thus decreasing the mass-inflow rate in this region accordingly.

\subsection{Late-time Disk Evolution}
\label{sec:latetime}

Most of the disk mass accretes over a characteristic timescale equal to the viscous time $t_{\rm visc}$ (equation~\ref{eq:t_visc}) evaluated at the initial characteristic disk radius $\sim R_{\rm d}$.  At times $t \gg t_{\rm visc}$ the disk evolution approaches a self-similar state.  Following known solutions for accretion disks with outflows \citep[e.g.][and references therein]{Metzger+2008}, the characteristic disk radius expands as 
$R_{\rm d} \propto t^{2/3}$, and the mass inflow rate scales as 
\begin{equation} \label{eq:Mdot_in_scaling}
\dot{M}_{\rm in} \propto r^p t^{-4(p+1)/3}.
\end{equation}
Note that the radial scaling applies only in the steady-state part of the disk ($r < R_{\rm d}$) and that terms of order $(r_* / R_{\rm d})^p \ll 1$ have been neglected, where $r_*$ is the inner boundary of the disk.  

Combining \ref{eq:Mdot_in_scaling} with equations (\ref{eq:p_index_definition}), (\ref{eq:v_r}), and (\ref{eq:Temp}), the disk surface density evolves as
\begin{equation}
\Sigma \propto r^{p-1/2} t^{-4(p+1)/3} ~,
\end{equation}
and the midplane temperature (equation~\ref{eq:Temp}) as
\begin{equation} \label{eq:Temp_scaling}
T \propto 
\begin{cases}
r^{-1} t^0 ~, ~~& \mathrm{gas} \\
r^{(p-5/2)/4} t^{-(p+1)/3} ~, & \mathrm{radiation} 
\end{cases}
\end{equation}
where the latter has been separated into gas and radiation pressure-dominated regimes.

Most nuclear reaction rates depend more sensitively on temperature than density (an important exception sometimes being the triple-$\alpha$ reaction).  Burning fronts therefore typically track the evolution of constant temperature surfaces. For the radiation dominated case of most relevance at late times and small radii, equation (\ref{eq:Temp_scaling}) is inverted to find
\begin{equation} \label{eq:Temp_scaling_r_propto_t}
r \left( T = T_{\rm rad} = {\rm const.} \right) \propto t^{-4(p+1) / 3(5/2-p)} ~.
\end{equation}
Under the assumptions that (1) the burning front of an isotope $A$ peaks around $r(T_{\rm burn})$, and (2) the radial shape of the abundance profile $X_A$ and its peak value are constant in time, then the mass ejection rate in this isotope is approximately given by
\begin{equation}
\left. \dot{M}_{\rm w} \left( X_A \right) \approx X_A \dot{\Sigma}_{\rm w} r^2 \right\vert_{r(T_{\rm burn})} \propto \dot{M}_{\rm in} \left[ r\left(T_{\rm burn}\right) \right] ~.
\end{equation}
The total outflow rate (integrated across all radii) then evolves as
\begin{equation} \label{eq:Mdot_w_scaling}
\dot{M}_{\rm w} \propto t^{-(2p+4)/3} ~,
\end{equation}
intimating that the factional mass loss rate in isotope $X_A$ decreases at times $t \gg t_{\rm visc}$ as
\begin{equation} \label{eq:Mdot_w_XA_scaling}
\frac{\dot{M}_{\rm w}\left(X_A\right)}{\dot{M}_{\rm w}} \propto t^{-2p/3 - 4p(p+1)/3(5/2-p)} ~.
\end{equation}
Physically, this temporal decrease of $\dot{M}_{\rm w}(X_A)$ is driven by the inward migration of the burning fronts as the disk temperature decreases with time.  A lower temperature reduces the radius at which a particular isotope is first formed, thus reducing its contribution to the disk outflows.  This result will prove useful later in extrapolating our numerical results to times later than the end of the simulation.

\section{Numerical Results} \label{sec:Numerical Results}
Following the procedure described in \S \ref{subsec:NumericalProcedure}, we have performed a suite of accretion disk/outflow simulations, as summarized in Table \ref{tab:ModelParameters}, corresponding to different model parameters and compositions of the disrupted WD.

\begin{table*}
\caption{Model parameters of simulations performed in this paper.
(a) Initial mass fractions $X_A$ of the WD or disk. 
(b) Initial mass of WD or disk in solar masses. 
(c) Mass of WD binary companion (NS or BH) in solar masses. 
(d) Shakura-Sunyaev alpha viscosity parameter (equation~\ref{eq:ShakuraSunyaev_viscosity}). 
(e), (f) Wind efficiency parameter and critical Bernoulli parameter respectively (\S \ref{subsec:Wind Prescription}). (g) Initial disk density power-law parameters (equation~\ref{eq:Density_initial}). (h) Normalized mixing efficiency parameter (equation~\ref{eq:alpha_tilde}).
}
\begin{tabular}{c  c  c  c  c  c  c  c  c  c}
\hline
Model & Initial $X_A$ $^{\rm (a)}$ & $M_{\rm WD}$ $^{\rm (b)}$ & $M$ $^{\rm (c)}$ & $\alpha$ $^{\rm (d)}$ & $\eta_{\rm w}$ $^{\rm (e)}$ & ${\rm Be^\prime_{crit}}$ $^{\rm (f)}$ & $(m,n)$ $^{\rm (g)}$ & $\tilde{\alpha}$ $^{\rm (h)}$ & Comments \\ \hline

$\mathtt{CO\_Fid}$ & $X_{\rm {}^{12}C} = X_{\rm {}^{16}O} = 0.5$ & $0.6$ & $1.4$ & $0.1$ & $1$ & $0$ & $(2,7)$ & $0.1$ & fiducial model \\ 

$\mathtt{CO\_Nuc}$ & $-$ & $-$ & $-$ & $-$ & $-$ & $-$ & $-$ & $-$ & $\dot{q}_{\rm nuc} = 0$ \\

$\mathtt{CO\_Mix1}$ & $-$ & $-$ & $-$ & $-$ & $-$ & $-$ & $-$ & $0$ & no mixing \\

$\mathtt{CO\_Mix2}$ & $-$ & $-$ & $-$ & $-$ & $-$ & $-$ & $-$ & $1$ & strong mixing \\

$\mathtt{CO\_Alpha}$ & $-$ & $-$ & $-$ & $0.01$ & $-$ & $-$ & $-$ & $0.1$ & weak viscosity \\

$\mathtt{CO\_Wnd1}$ & $-$ & $-$ & $-$ & $0.1$ & $-$ & $-0.1$ & $-$ & $-$ & wind param. \\

$\mathtt{CO\_Wnd2}$ & $-$ & $-$ & $-$ & $-$ & $-$ & $+0.1$ & $-$ & $-$ & $-$ \\

$\mathtt{CO\_Wnd3}$ & $-$ & $-$ & $-$ & $-$ & $0.5$ & $0$ & $-$ & $-$ & $-$ \\

$\mathtt{CO\_Wnd4}$ & $-$ & $-$ & $-$ & $-$ & $2$ & $-$ & $-$ & $-$ & $-$ \\

$\mathtt{CO\_Den}$ & $-$ & $-$ & $-$ & $-$ & $1$ & $-$ & $(4,7)$ & $-$ & initial density \\

$\mathtt{CO\_Comp1}$ & $X_{\rm {}^{12}C} = 0.4, X_{\rm {}^{16}O} = 0.5$ & $-$ & $-$ & $-$ & $-$ & $-$ & $(2,7)$ & $-$ & initial $X_A$ \\

$\mathtt{CO\_Comp2}$ & $X_{\rm {}^{12}C} = 0.6, X_{\rm {}^{16}O} = 0.4$ & $-$ & $-$ & $-$ & $-$ & $-$ & $-$ & $-$ & $-$ \\

\hline
$\mathtt{He\_Fid}$ & $X_{\rm {}^{4}He} = 1$ & $0.3$ & $1.2$ & $0.1$ & $1$ & $0$ & $(2,7)$ & $0.1$ & He fiducial model \\ 

$\mathtt{He\_Nuc}$ & $-$ & $-$ & $-$ & $-$ & $-$ & $-$ & $-$ & $-$ & no nuclear heating \\

$\mathtt{He\_Mix1}$ & $-$ & $-$ & $-$ & $-$ & $-$ & $-$ & $-$ & $0$ & no mixing \\

$\mathtt{He\_Mix2}$ & $-$ & $-$ & $-$ & $-$ & $-$ & $-$ & $-$ & $1$ & strong mixing \\

$\mathtt{He\_Alpha}$ & $-$ & $-$ & $-$ & $0.01$ & $-$ & $-$ & $-$ & $0.1$ & weak viscosity \\

$\mathtt{He\_Wnd3}$ & $-$ & $-$ & $-$ & $0.1$ & $0.5$ & $-$ & $-$ & $-$ & wind param. \\

$\mathtt{He\_Wnd4}$ & $-$ & $-$ & $-$ & $-$ & $2$ & $-$ & $-$ & $-$ & $-$ \\

$\mathtt{He\_Den}$ & $-$ & $-$ & $-$ & $-$ & $1$ & $-$ & $(4,7)$ & $-$ & initial density \\

$\mathtt{He\_Mass}$ & $-$ & $0.4$ & $-$ & $-$ & $-$ & $-$ & $(2,7)$ & $-$ & WD mass \\

\hline
$\mathtt{CO\_He1}$ & \begin{tabular}[c]{@{}c@{}} $X_{\rm {}^{12}C} = X_{\rm {}^{16}O} = 0.4$, \\ $X_{\rm {}^{4}He} = 0.2$ \end{tabular} & $0.6$ & $0.4$ & $0.1$ & $1$ & $0$ & $(2,7)$ & $0.1$ & `hybrid' WD \\

$\mathtt{CO\_He2}$ & \begin{tabular}[c]{@{}c@{}} $X_{\rm {}^{12}C} = X_{\rm {}^{16}O} = 0.475$, \\ $X_{\rm {}^{4}He} = 0.05$ \end{tabular} & $-$ & $-$ & $-$ & $-$ & $-$ & $-$ & $-$ & $-$ \\

\hline
\end{tabular}
 \label{tab:ModelParameters}
\end{table*}

\subsection{Fiducial Model} \label{subsec:Fiducial C/O Model}

Our fiducial model, $\mathtt{CO\_Fid}$, corresponds to the merger of a $0.6 M_\odot$ C/O WD with a $1.4 M_\odot$ NS.  The initial composition of the WD, and hence of the disk, is half (by mass) carbon and half oxygen, $X_{\rm {}^{12}C} = X_{\rm {}^{16}O} = 0.5$.  We employ a Shakura-Sunyaev alpha viscosity parameter of $\alpha = 0.1$ and a composition mixing parameter (equation \ref{eq:alpha_tilde}) of $\tilde{\alpha} = 0.1$ \citep{Carballido&Stone&Pringle2005}. The fiducial wind efficiency parameter and critical (normalized) Bernoulli parameter are taken to be $\eta_{\rm w}=1$ and ${\rm Be^\prime_{crit}}=0$, respectively.  In steady-state, these parameters describe a marginally bound disk with a mass inflow index of $p \approx 0.43$--$0.47$ for adiabatic indexes $\gamma=1.33$--$1.67$ (Fig.~\ref{fig:p_index}).  The power-law parameters of the initial disk density profile (see equation~\ref{eq:Density_initial}) are taken to be $m=2$ and $n=7$.

The characteristic initial radius of the disk is $R_{\rm d} \simeq 1.8 \times 10^9 ~{\rm cm}$, corresponding to an initial viscous timescale of $t_{\rm visc,0} \simeq 68~{\rm s}$ measured at the radius where the initial density distribution peaks.  We terminate our simulations at the time $t_{\rm end}=2 t_{{\rm visc,0}}$, at which point roughly half the initial mass of the disk has either been lost to outflows or has been accreted through the inner boundary of the grid. By $t = t_{\rm end}$ the burning fronts creating Fe-group elements begin crossing through the inner boundary of our grid.

\subsubsection{Accretion/Outflow Rates}
\label{sec:accretionrates}

\begin{figure*}
\centering
\begin{subfigure}[]{\label{fig:Mdot_a} \epsfig{file=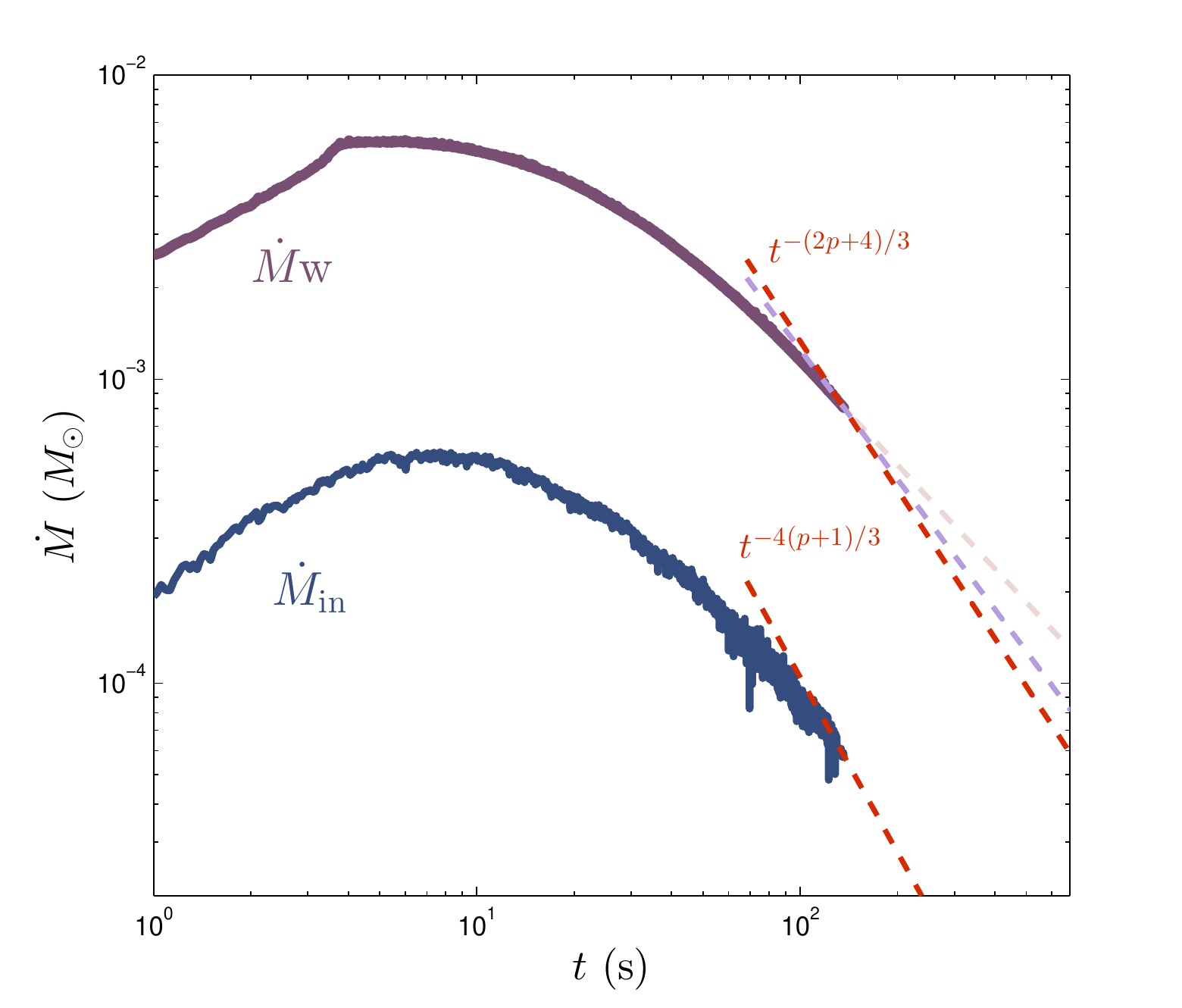,angle=0,width=0.45\textwidth} }\end{subfigure}
~
\begin{subfigure}[]{\label{fig:Mdot_b} \epsfig{file=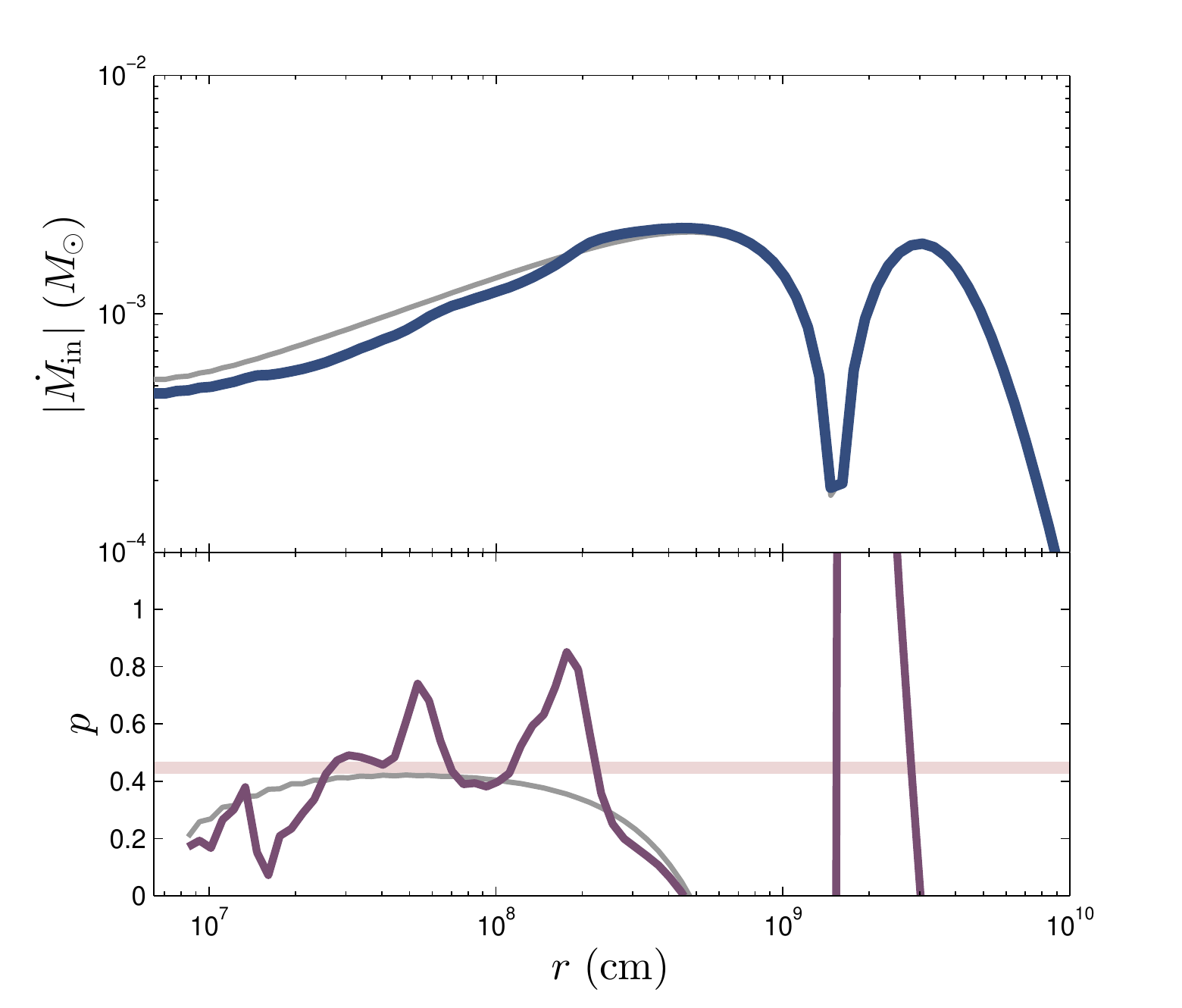,angle=0,width=0.45\textwidth} }\end{subfigure}
\caption{{\bf (a)} Mass inflow rate, $\dot{M}_{\rm in}(r_*)$, through the inner radial boundary at $r_* \simeq 7\times 10^{6}$ cm (blue) and total wind outflow rate, $\dot{M}_{\rm w}$ (purple), as a function of time following disk formation.  
The accretion rate peaks on a timescale of $t_{\rm visc} \approx 8~{\rm s}$.
The late-time power-law evolution of the inflow rate predicted by a self-similar model (equations~\ref{eq:Mdot_in_scaling}, \ref{eq:Mdot_w_scaling}) are shown as dashed red curves.  The light pink curve is a direct power-law extrapolation of $\dot{M}_{\rm w}$ from the simulation end time, while the dashed purple curve shows an intermediate power-law extrapolation based on mass conservation (equation~\ref{eq:Mdot_w_extrap}).
{\bf (b)} Snapshot of the radial profile of the mass inflow rate, $\dot{M}_{\rm in}$, at $t = 16 ~{\rm s} \gtrsim t_{\rm visc}$, for our fiducial model $\mathtt{CO\_Fid}$.  The bottom panel shows the mass inflow exponent, $p \equiv \partial \ln \dot{M}_{\rm in}/\partial \ln r$.  The pink shaded region shows the range of $p$ predicted for a steady-state disk (see Appendix \ref{subsec:Appendix_pExponent}).  Grey curves in both panels show a model $\mathtt{CO\_Nuc}$ in which nuclear heating is manually turned off.  Local peaks in $p(r)$, relative to the $\mathtt{CO\_Nuc}$ model, are caused by strong localized nuclear heating from, e.g., $^{12}$C and $^{16}$O burning fronts.  The local minimum in $p(r)$ at $r \lesssim 2\times10^7~{\rm cm}$ is the result of cooling from endothermic photodisintegration. 
} \label{fig:Mdot}
\end{figure*}

Fig.~\ref{fig:Mdot} shows the time evolution (top panel) and radial profile (bottom panel) of the total mass inflow and outflow rates.  The maximum inflow rate at $R_{\rm d}$ can be estimated by $\dot{M}_{\rm in}\left(R_{\rm d}\right) \sim M_{\rm d} / t_{\rm visc,0} \sim 9 \times 10^{-3} ~M_\odot~{\rm s}^{-1}$.  However, most of this inflow is ultimately lost to outflows, with only a fraction $\sim \left( r_* / R_{\rm d} \right)^p \ll 1$ reaching the inner boundary at $r = r_*$.   This is illustrated explicitly in Fig. \ref{fig:Mdot_a}, which shows that $\dot{M}_{\rm in}\left(r_*\right) \ll \dot{M}_{\rm w}$.  Physically, $r_*$ represents the NS surface, but in our case it represents the inner boundary of our radial grid at $r_* \simeq 7 \times 10^{6}$ cm.  The wind outflow rate $\dot{M}_{\rm w} \sim \dot{M}_{\rm in}\left( R_{\rm d} \right)$ peaks at roughly the same value as the accretion rate, although it rises to a maximum on a timescale $t_{\rm visc} \simeq 8~{\rm s}$ which is shorter than $t_{\rm visc,0}$. 

Dashed lines show a range of power-law extrapolations of the mass inflow and outflow rates. A light pink line shows an extrapolation of $\dot{M}_{\rm w}\left( t_{\rm end} \right)$ based on the best-fit logarithmic slope  measured near the end of the simulation run.  Red curves show the late-time self-similar evolution predicted by equations (\ref{eq:Mdot_in_scaling}) and (\ref{eq:Mdot_w_scaling}), which are generally steeper because they represent the asymptotic power-law towards which the solution is evolving.  An intermediate extrapolation shown with a purple line is derived by requiring that the integrated mass loss rate obey mass conservation, viz.~
\begin{equation}
M_{\rm acc}(t_{\rm end}) + M_{\rm w}(t_{\rm end}) + \int_{t_{\rm end}}^{\infty} \dot{M} \left( t > t_{\rm end} \right) \, dt = M_{\rm d},
\end{equation}
where $M_{\rm acc}(t_{\rm end})$ and $M_{\rm w}(t_{\rm end})$ are the total mass accreted through the inner grid boundary and lost to wind outflows by the simulation end time, respectively. Solving for the appropriate wind mass loss exponent $\zeta$, defined by
\begin{equation} \label{eq:Mdot_w_extrap}
\dot{M}_{\rm w}\left(t>t_{\rm end}\right) = \dot{M}_{\rm w}\left(t_{\rm end}\right) \times \left( \frac{t}{t_{\rm end}} \right)^{-\zeta} ~, 
\end{equation}
we obtain
\begin{equation} \label{eq:Mdot_w_extrap_powerlawindex}
\zeta = 1 + \frac{\dot{M}_{\rm w}(t_{\rm end}) \times t_{\rm end}}{M_{\rm d} - M_{\rm acc}(t_{\rm end}) - M_{\rm w}(t_{\rm end})} ~.
\end{equation}
We employ this power-law scaling when we extrapolate the properties of outflows from the final timestep of our numerical simulations $t_{\rm end}$ to late times, $t = \infty$. 

Fig. \ref{fig:Mdot_b} shows the radial profile of the inflow rate $\dot{M}_{\rm in}$ at a fixed time, $t = 16~{\rm s} \sim 2 t_{\rm visc}$. As expected, a steady-state power-law scaling $\dot{M}_{\rm in} \sim r^p$ is obtained for radii $r \lesssim R_{\rm d}$ (equation~\ref{eq:Mdot_in_scaling}).  The local dip in $\dot{M}_{\rm in}$ and the apparent discontinuity in its derivative near $r = 2\times 10^9 ~{\rm cm}$ is an artifact of the absolute value and logarithmic scale of the vertical axis. This location corresponds to a turnover point, where the radial velocity $v_r$ passes through zero.  Outside of this radius, where the radial velocity is positive, a small amount of mass carries angular momentum to large radii.  

The bottom panel of Fig. \ref{fig:Mdot_b} shows the radial profile of the mass loss index $p$ (equation~\ref{eq:p_index_definition}).  If no nuclear burning were present, then in the steady-state portion of the disk at $r \lesssim R_{\rm d}$ we would expect $p$ to vary about the theoretically expected range, as depicted by the shaded pink region for adiabatic index in the range $\gamma = 1.33-1.67$ (equation~\ref{eq:Appendix_p_index}).  Indeed, this range is reasonably well matched by the grey curves, which show an otherwise identical model, $\mathtt{CO\_Nuc}$, but with the effects of nuclear burning artificially turned off.  
Localized spikes in $p(r)$, such as those located at $r \approx 2 \times 10^8~{\rm cm}$ and $r \approx 5 \times 10^7~{\rm cm}$ which break from the smooth trend exhibited by the grey $\mathtt{CO\_Nuc}$ solution, occur at the $^{12}$C and $^{16}$O burning fronts.  The significant amounts of energy released by nuclear burning at these locations (Fig. \ref{fig:qdot}) must be offset by greater cooling of the disk (stronger outflows) than in disks heated purely by viscosity.  These local maxima in the mass outflow rate 
are accompanied by a decrease in $\dot{M}_{\rm in}$ (as required by mass conservation), which reflect as local peaks in the mass inflow exponent $p$.

\begin{figure}
\centering
\epsfig{file=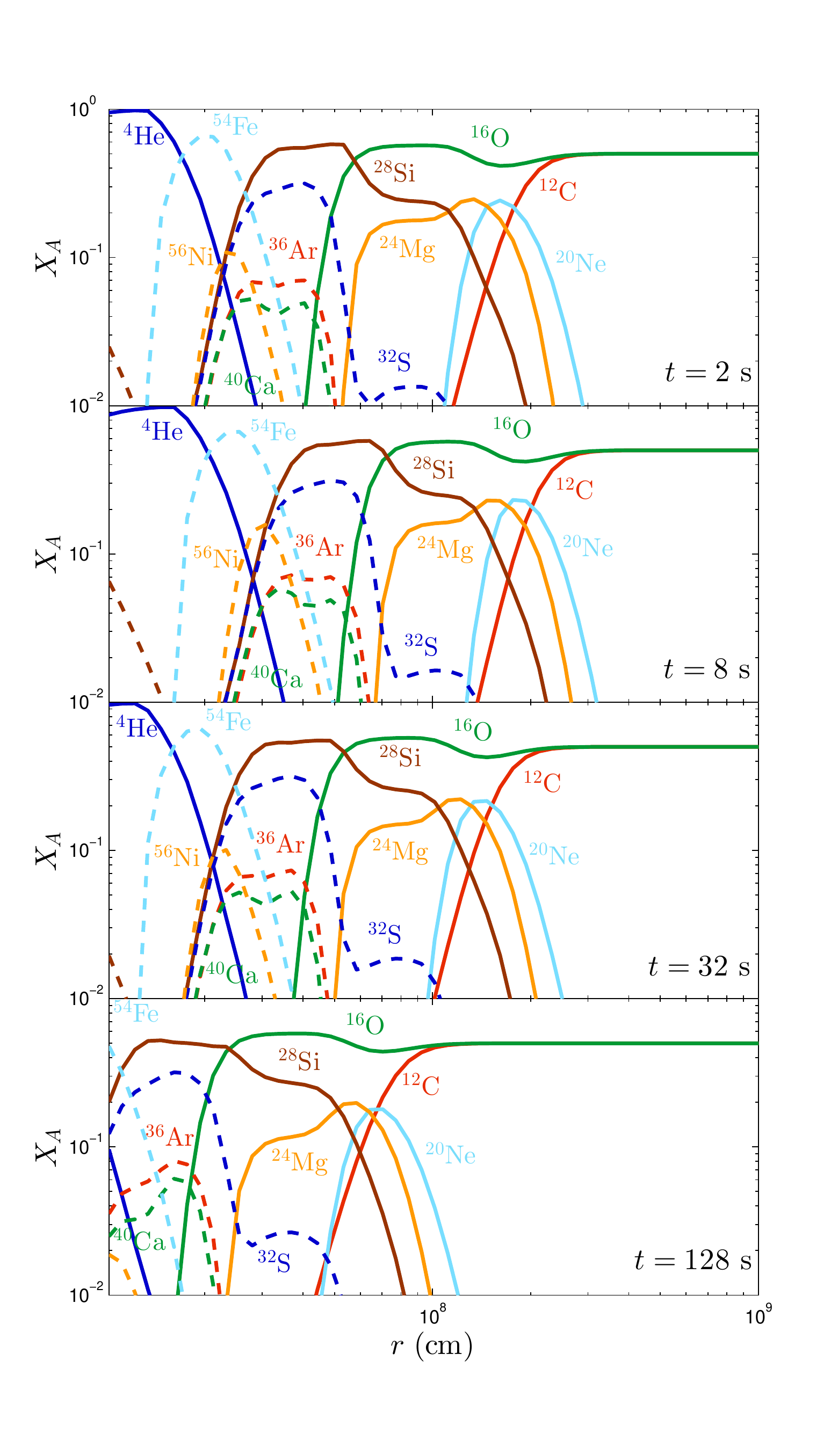,angle=0,width=0.45\textwidth}
\caption{Snapshots in the evolution of the radial profile of the mass fraction $X_A$ of key isotopes.  The second panel ($t=8~{\rm s}$) roughly corresponds to the time of peak accretion.  
The final panel ($t=128~{\rm s}$) approximately corresponds to the simulation end time.  The composition profiles exhibit a self-similar evolution, with the overall abundance pattern shifting as a whole to larger (smaller) radii before (after) the peak accretion timescale, respectively.  This `steady state' self-similar behaviour characterizes the disk composition already from very early times $\ll t_{\rm visc}$.} \label{fig:X_A_Evolution}
\end{figure}

\subsubsection{Disk Composition}

Fig. \ref{fig:X_A_Evolution} shows snapshots of the radial profile of the mass fraction $X_A(r)$ of key isotopes.  The disk composition assumes an onion-skin structure, reminiscent of that of evolved massive stars, in which successively heavier elements burn at sequentially smaller radii.  At radii $r\lesssim 2\times 10^8~{\rm cm}$, the temperature of the disk midplane becomes sufficiently high, $T \gtrsim 10^9~{\rm K}$, to initiate burning of the initial carbon/oxygen composition, generating $^{20}$Ne and $^{24}$Mg.  At smaller radii, the temperature increases further, fusing these isotopes into $^{28}$Si.  At $r\sim 6\times 10^7~{\rm cm}$, $^{32}$S is created, which quickly burns to $^{36}$Ar, $^{40}$Ca, $^{44}$Ti, $^{48}$Cr and $^{52}$Fe, and finally up to $^{54}$Fe and $^{56}$Ni.  Near the innermost radii, $r \lesssim 3\times10^7 ~{\rm cm}$, photo-disintegrations breaks these heavy elements apart into $^{4}$He ($\alpha$-particles) and free nucleons.  

The same qualitative picture holds at each snapshot in time because the key nuclear reactions are temperature limited.  The composition profiles at different times therefore remain nearly identical to one another, modulo rescaling of the radial axis.  This apparent self-similarity in $X_A(r,t)$ is a direct consequence of the self-similarity in the temperature profile (equation~\ref{eq:Temp_scaling}), insofar as the burning fronts reside in regions of the disk dominated by radiation pressure and at radii $\lesssim R_{\rm d}$ characterized by a steady inward accretion rate.  The composition profile in 
Fig. \ref{fig:X_A_Evolution} is similar to that obtained by the steady-state model of \citetalias{Metzger2012}.  At any time the composition is well described by a steady-state model, with the mass feeding rate $\dot{M}_{\rm in}(R_{\rm d})$ varying secularly in time.

At early times $t < t_{\rm visc}$, the density and temperature at a fixed radius $r < R_{\rm d}$ are small, with a correspondingly small burning front radius (first panel of Fig.~\ref{fig:X_A_Evolution}).  As gas fills the inner disk and accretes onto the NS, the temperature rises and the burning fronts move outwards, reaching their peak values on a timescale $t \sim t_{\rm visc}$ (second panel).  Finally, at times $t > t_{\rm visc}$, as the disk mass and density decrease, the constant temperature regions again move inwards to smaller radii, and the burning fronts and composition profiles shift steadily in the same fashion (third and fourth panels).

\begin{figure}
\centering
\epsfig{file=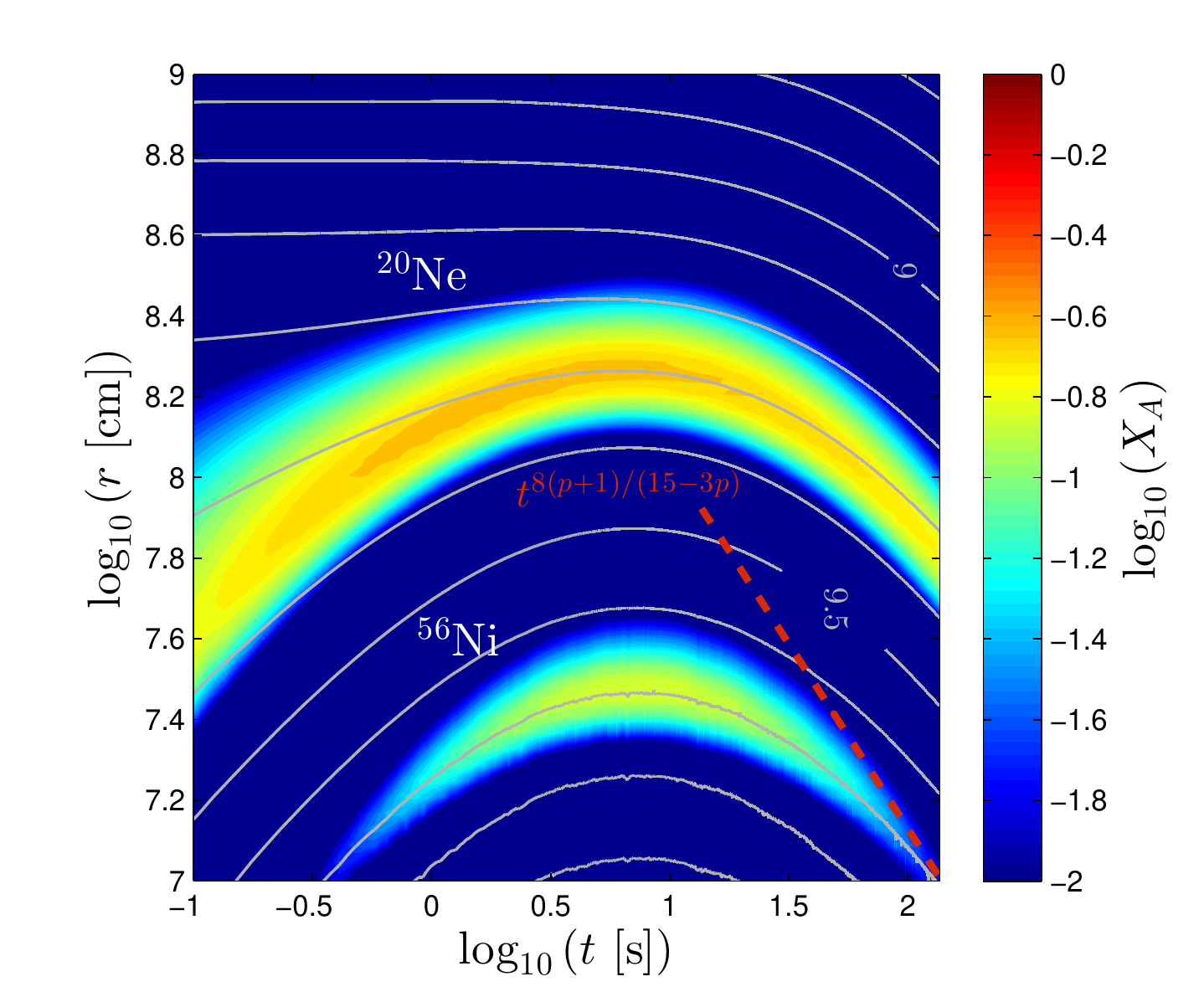,angle=0,width=0.45\textwidth}
\caption{Contours of the mass fractions of $^{56}$Ni and $^{20}$Ne as a function of radius ${\rm log}_{10} (r)$ and time ${\rm log}_{10} (t)$. As shown in Fig. \ref{fig:X_A_Evolution}, the two burning fronts track one another in a self-similar manner, first increasing to larger radii at initial times $\lesssim t_{\rm visc}$, and then decreasing at later times. Curves of constant temperature are overplot with grey lines, spaced equally in intervals of $\Delta {\rm log}_{10}(T~[{\rm K}]) = 0.1$, with $T = 10^9~{\rm K}$ and $10^{9.5}~{\rm K}$ labeled for reference.  The nuclear burning fronts, which are traced by the isotope abundances, track the temperature evolution closely.  A dashed red curve shows the self-similar power-law scaling of $r(T_{\rm rad}={\rm const.})$ which is achieved at late times (equation~\ref{eq:Temp_scaling_r_propto_t}).}
 \label{fig:X_A_Contour}
\end{figure}

Fig. \ref{fig:X_A_Contour} further illustrates this evolution by showing contours of the mass fraction of two sample elements, $^{56}$Ni and $^{24}$Mg, in the space of radius and time.  The peak mass fractions of each element rise to larger radii at $t \lesssim t_{\rm visc}$, and decrease after $t \gtrsim t_{\rm visc}$.  Contours of constant temperature are overplot with grey curves.  The fact that the composition and temperature contours track one another again illustrates that the relevant nuclear reactions are temperature limited.  The constant temperature curves at $r < R_{\rm d}$ and $t \gg t_{\rm visc}$ also agree well with the predicted late-time self-similar evolution in the radiation-dominated regime (equation~\ref{eq:Temp_scaling_r_propto_t}), which we have overplot with a dashed red line.

\begin{figure*}
\centering
\begin{subfigure}[]{\label{fig:qdot_a} \epsfig{file=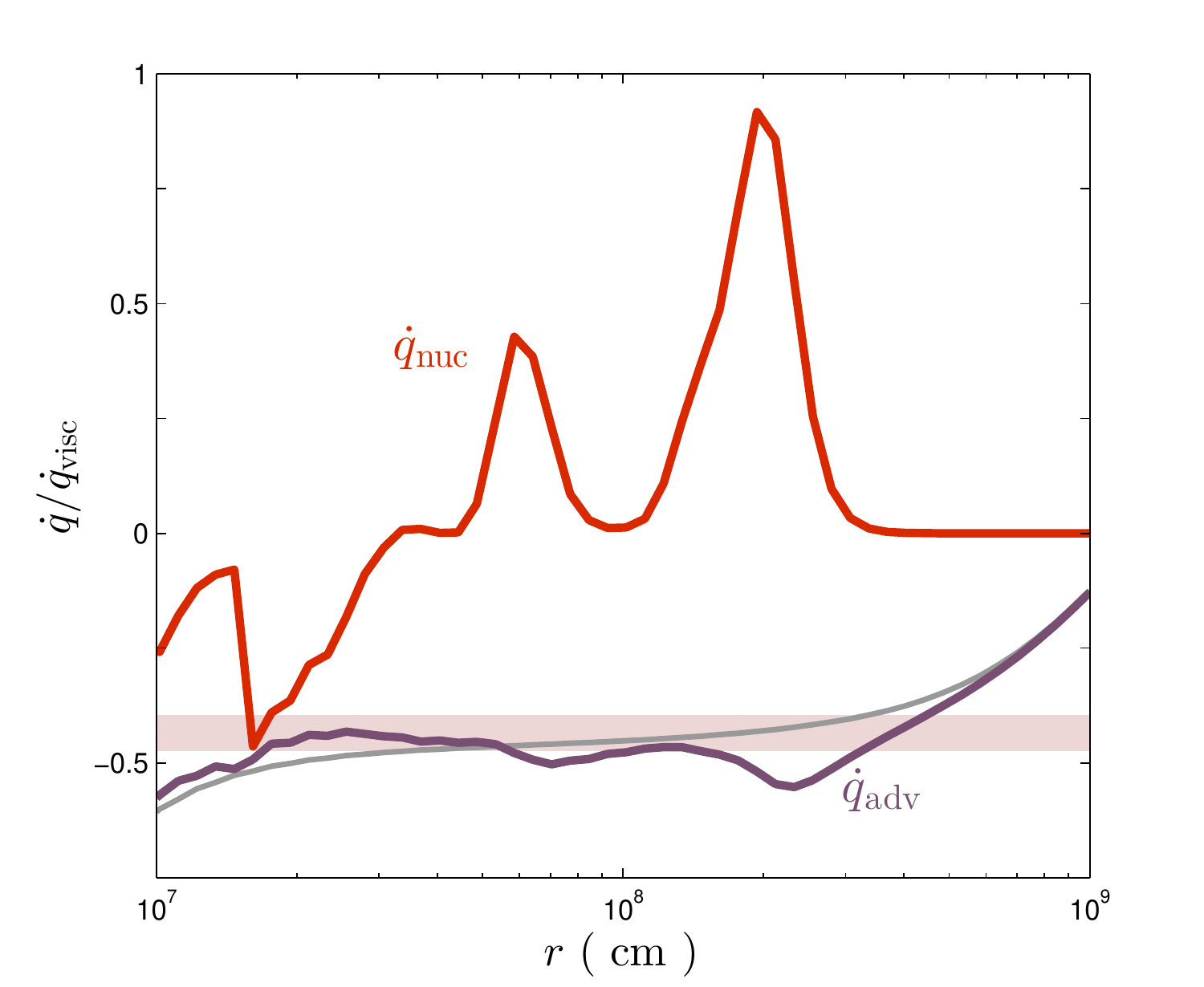,angle=0,width=0.45\textwidth} }\end{subfigure}
~
\begin{subfigure}[]{\label{fig:qdot_b} \epsfig{file=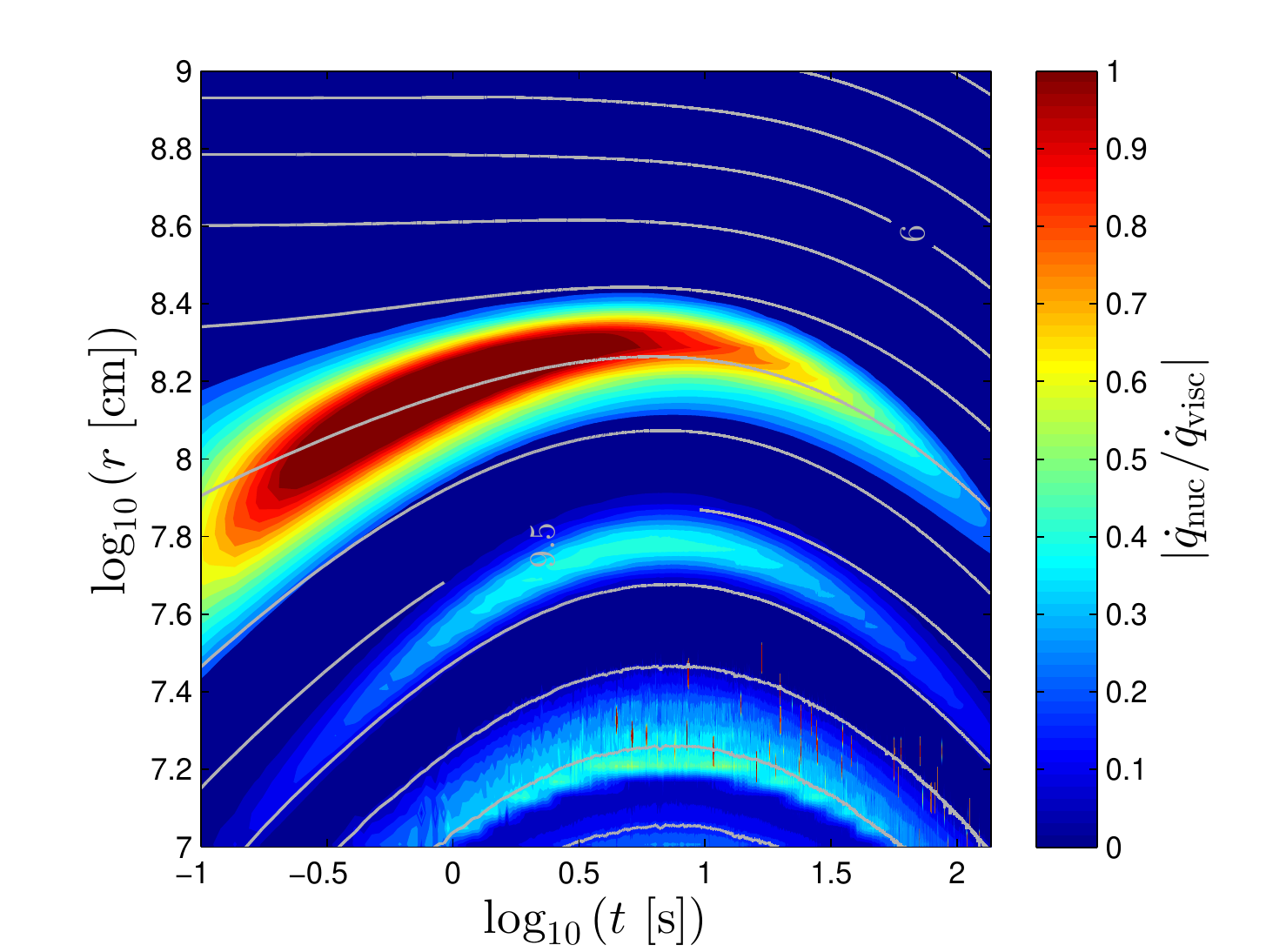,angle=0,width=0.45\textwidth} }\end{subfigure}
\caption{{\bf (a)} Snapshot of the radial profile of heating and cooling rates in the disk midplane (equation~\ref{eq:internal_energy}) at $t=8~{\rm s}$.  Red and purple curves show, respectively, the nuclear heating rate and advective cooling rate, normalized to the viscous heating rate (equation~\ref{eq:q_dot_visc}).  Nuclear burning has an order unity impact on the disk and outflow dynamics at locations where $\dot{q}_{\rm nuc} \sim \dot{q}_{\rm visc}$, specifically near the $^{12}$C and $^{16}$O burning fronts at $r \approx 2\times 10^8~{\rm cm}$ and $r \approx 6\times 10^7~{\rm cm}$ (Fig.~\ref{fig:X_A_Evolution}).  At small radii $r \lesssim 2\times 10^7~{\rm cm}$, endothermic photodisintegrations provide a source of nuclear {\it cooling}.  A grey curve shows the advective cooling rate for an otherwise identical model, $\mathtt{CO\_Nuc}$, with nuclear heating artificially turned off.  The pink shaded 
region shows the theoretically expected range of $\dot{q}_{\rm adv}$ (equation~\ref{eq:Appendix_q_dot_adv}) for a steady-state disk with $\gamma=1.33-1.67$.  Wind cooling, which is not illustrated here, provides additional cooling of the disk, such that the net heating rate $\Sigma_i \dot{q}_i \approx 0$.
{\bf (b)} Contours of the nuclear heating rate normalized to the viscous heating rate in the space of disk radius ${\rm log}_{10}(r)$ and time ${\rm log}_{10}(t)$.  Peaks in the nuclear heating rate again closely follow the carbon and oxygen burning fronts (cf.~Fig.~\ref{fig:X_A_Contour}).  Nuclear heating is most significant at early times $t \lesssim t_{\rm visc}$ at the outermost $^{12}$C burning front, where the gravitational potential well is shallow.
} \label{fig:qdot}
\end{figure*}

Beyond generating a rich radial abundance distribution, nuclear burning can have dynamically important influence on the disk and its outflows. Fig. \ref{fig:qdot_a} compares contributions to the net heating $\dot{q}_{\rm tot}$ in equation (\ref{eq:internal_energy}) at a snapshot around the time $t \sim t_{\rm visc}$. The nuclear heating rate, $\dot{q}_{\rm nuc}$, as a function of radius is shown with a solid red curve, in units of the viscous heating rate $\dot{q}_{\rm visc}$ (equation~\ref{eq:q_dot_visc}). The two clear peaks, at around the radii $r \approx 2\times 10^8~{\rm cm}$ and $r \approx 6\times 10^7~{\rm cm}$ correspond to the carbon and oxygen burning fronts, respectively.  In the first case nuclear heating rate is locally as important as viscous heating, i.e. $\dot{q}_{\rm nuc} \sim \dot{q}_{\rm visc}$ (\citetalias{Metzger2012}; \citetalias{Fernandez&Metzger2013}).  For a steady-state disk, the advective cooling rate $\dot{q}_{\rm adv}$ (purple line) is a constant fraction of $\dot{q}_{\rm visc}$ (equation~\ref{eq:Appendix_q_dot_adv}), as depicted by the horizontal lightly shaded pink region. As in Fig. \ref{fig:Mdot_a}, our numerical results roughly agree with this expectation for $r \lesssim R_{\rm d}$, especially in the comparison model, $\mathtt{CO\_Nuc}$, for which nuclear burning has been artificially turned off (grey curve).

Fig. \ref{fig:qdot_b} shows contours of $\dot{q}_{\rm nuc} / \dot{q}_{\rm visc}$ in the space of disk radius and time.  Comparison with Fig. \ref{fig:X_A_Evolution} shows that $\dot{q}_{\rm nuc}$ follows the $^{12}$C and $^{16}$O burning fronts, and is most important relative to viscous heating at early times $t \lesssim t_{\rm visc}$ when the burning fronts occur at larger radii in the disk. Despite the importance of nuclear burning near the burning fronts prior to peak accretion, it is subdominant to viscous heating across most radii (away from the burning fronts) and at late times $t \gg t_{\rm visc}$.

\subsubsection{Outflow Properties}

\begin{figure*}
\centering
\begin{subfigure}[]{\label{fig:Wind_a} \epsfig{file=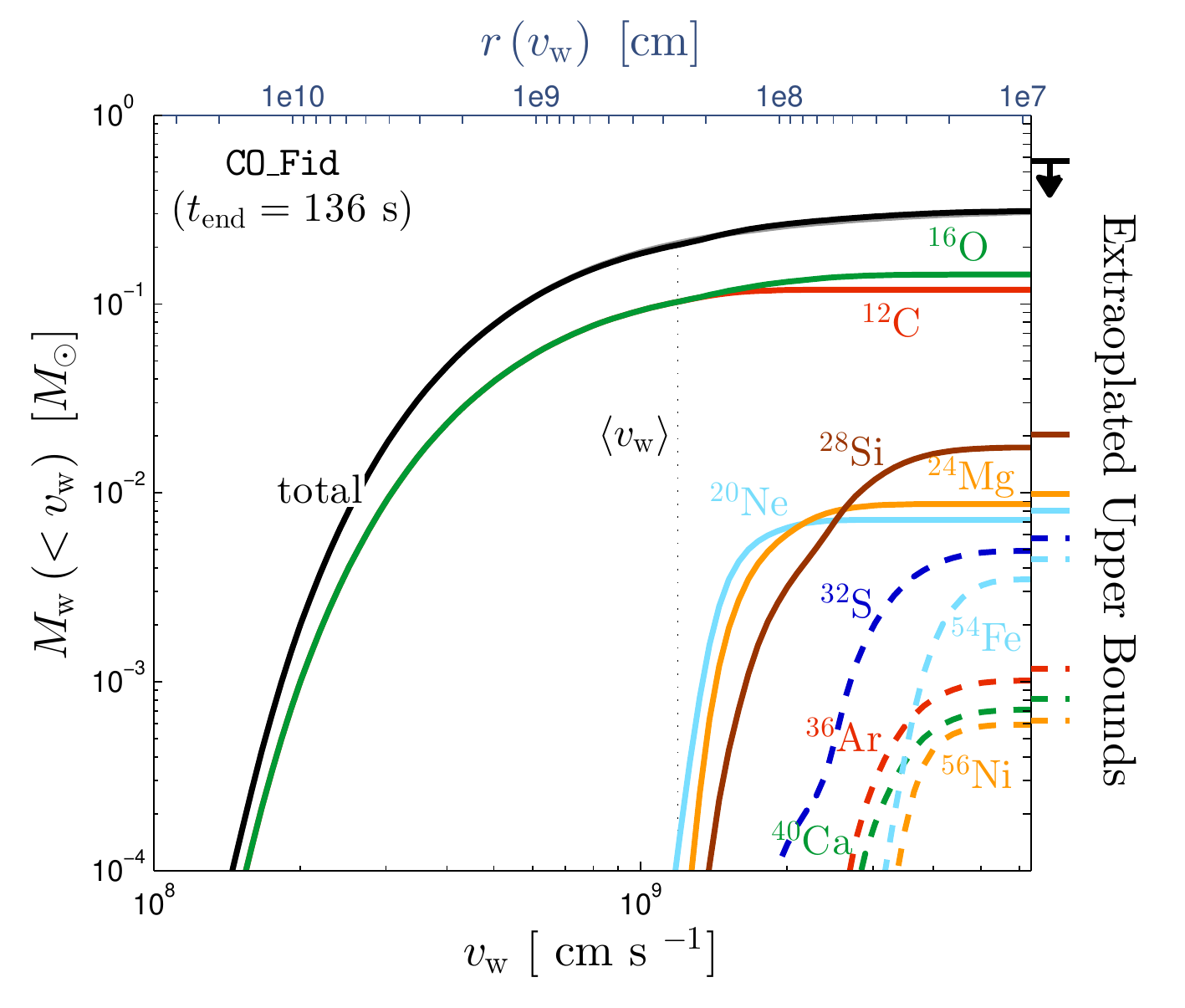,angle=0,width=0.45\textwidth} }\end{subfigure}
~
\begin{subfigure}[]{\label{fig:Wind_b} \epsfig{file=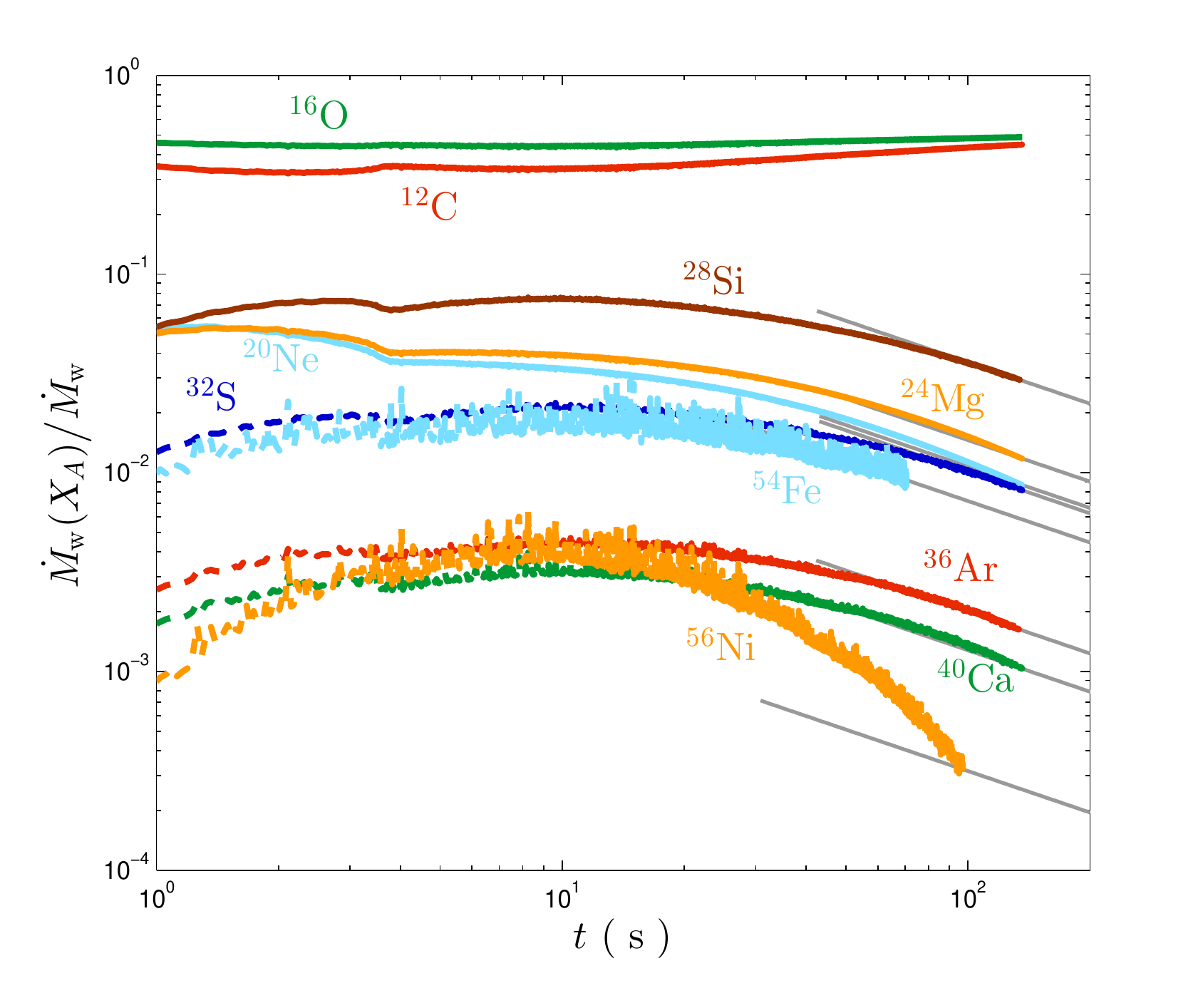,angle=0,width=0.45\textwidth} }\end{subfigure}
\caption{{\bf (a)} Cumulative mass distribution $M_{\rm w}(<v_{\rm w})$ of the disk outflows below a given outflow velocity $v_{\rm w}$ in the C/O fiducial model, evaluated at the final snapshot and shown separately for each isotope.  Color and style conventions are the same as in Fig. \ref{fig:X_A_Evolution}, apart for the additional black curve illustrating the total (i.e. summed over all elements) wind distribution for the fiducial model. The horizontal blue axis across the top of the plot equivalently shows the distribution in the disk radii from which the outflow was ejected. 
Short horizontal curves outside the right axis show the total outflow mass in various elements, extrapolated from the end of the simulation to $t \to \infty$.
{\bf (b)} Fractional mass outflow rates of various elements $\dot{M}_{\rm w} \left(X_A\right) / \dot{M}_{\rm w}$ as a function of time.  Intermediate mass isotopes are well approximated by the theoretically motivated power-law extrapolation given by equation (\ref{eq:Mdot_w_XA_scaling}), as illustrated by solid grey curves beginning near the simulation end time.
} \label{fig:Wind}
\end{figure*}

Fig. \ref{fig:Wind_a} shows the cumulative mass distribution $M_{\rm w}(<v_{\rm w})$ of the disk outflows below a given outflow velocity $v_{\rm w}$, separately for each isotope.  The horizontal blue axis along the top shows the corresponding radius $r = 2 \eta_{\rm w} G M/ v_{\rm w}^2$ from which matter leaves the disk.  A solid black curve shows the total mass (all isotopes). Short horizontal curves extending beyond the axis depict extrapolated upper bounds on the total mass ejected in various isotopes at $t \to \infty$. For most isotopes, these extrapolations are only very small corrections to the ejected mass at $t_{\rm end}$, apart for unburned carbon and oxygen (not shown) which increase by a factor of $\sim$ two (see total ejecta extrapolation; black curve).

Of the total mass $M_{\rm w} = 0.31 M_\odot$ unbound by the end of the simulation, approximately $0.12 M_\odot$ is unburned carbon and $0.14 M_\odot$ is unburned oxygen.  Heavier isotopes are ejected with smaller abundances and at higher velocities, which is understood by the fact that they originate from smaller radii in the disk, where $v_{\rm w}$ is larger.
The average mass weighted outflow velocity of the ejecta is $\langle v_{\rm w} \rangle \simeq 1.2 \times 10^9~{\rm cm}$ (vertical dotted line in Fig. \ref{fig:Wind_a}).

Fig. \ref{fig:Wind_b} shows the fraction of the total mass outflow rate in different isotopes, 
$\dot{M}_{\rm w} \left( X_A \right) / \dot{M}_{\rm w}$, as a function of time.  
In $\S\ref{sec:latetime}$ we described an analytic method for extrapolating the mass outflow rates from the disk to times later than the endpoint of the simulation.  Solid grey lines show this power-law extrapolation of the mass loss rates for different isotopes from equation (\ref{eq:Mdot_w_XA_scaling}).  Although this provides a reasonable description for intermediate mass elements such as $^{40}$Ca and $^{36}$Ar, other isotopes do not fare as well.  The abundances of the unburned isotopes carbon and oxygen obviously to not peak around a particular burning front, but rather extend to the outer edge of the disk.  The lowest mass isotopes, $^{4}$He and free nucleons (not illustrated in Fig.~\ref{fig:Wind_b}), which are only present at small radii, are plagued by a similar problem; their radial domain is broad and extends inside the range captured by our numerical grid.  

Fig. \ref{fig:Wind_b} also shows that the mass fraction of $^{56}$Ni decreases more rapidly with time near the end of our simulation than predicted by equation (\ref{eq:Mdot_w_XA_scaling}; see also Figs.~\ref{fig:X_A_Evolution} and \ref{fig:X_A_Contour}).  This disagreement stems from the assumption that the peak value of $X_A$ is constant in time, while for $^{56}$Ni it decreases.  The same issue affects the intermediate isotopes discussed previously, albeit to a lesser extent. For these reasons, our extrapolated values for the total ejecta are best taken as upper limits.

\subsection{Variations about the Fiducial Model} \label{subsec:Variation of C/O Model Parameters}

\subsubsection{Nuclear Heating}

Fig. \ref{fig:Mdot_a} shows clear differences between the accretion inflow rate in our fiducial model $\mathtt{CO\_Fid}$ (dark blue curve) and that with heating from nuclear burning turned off, $\mathtt{CO\_Nuc}$ (light grey curve).  In the fiducial case $\dot{M}_{\rm in}$ increases faster with radius than the smooth power-law decline of $\mathtt{CO\_Nuc}$, predominantly in two `steps' at the carbon and oxygen burning fronts. As was already discussed, these differences are the result of nuclear burning increasing the wind outflow rate near the burning fronts.

\begin{figure*}
\centering
\begin{subfigure}[]{\label{fig:XA_Nuc} \epsfig{file=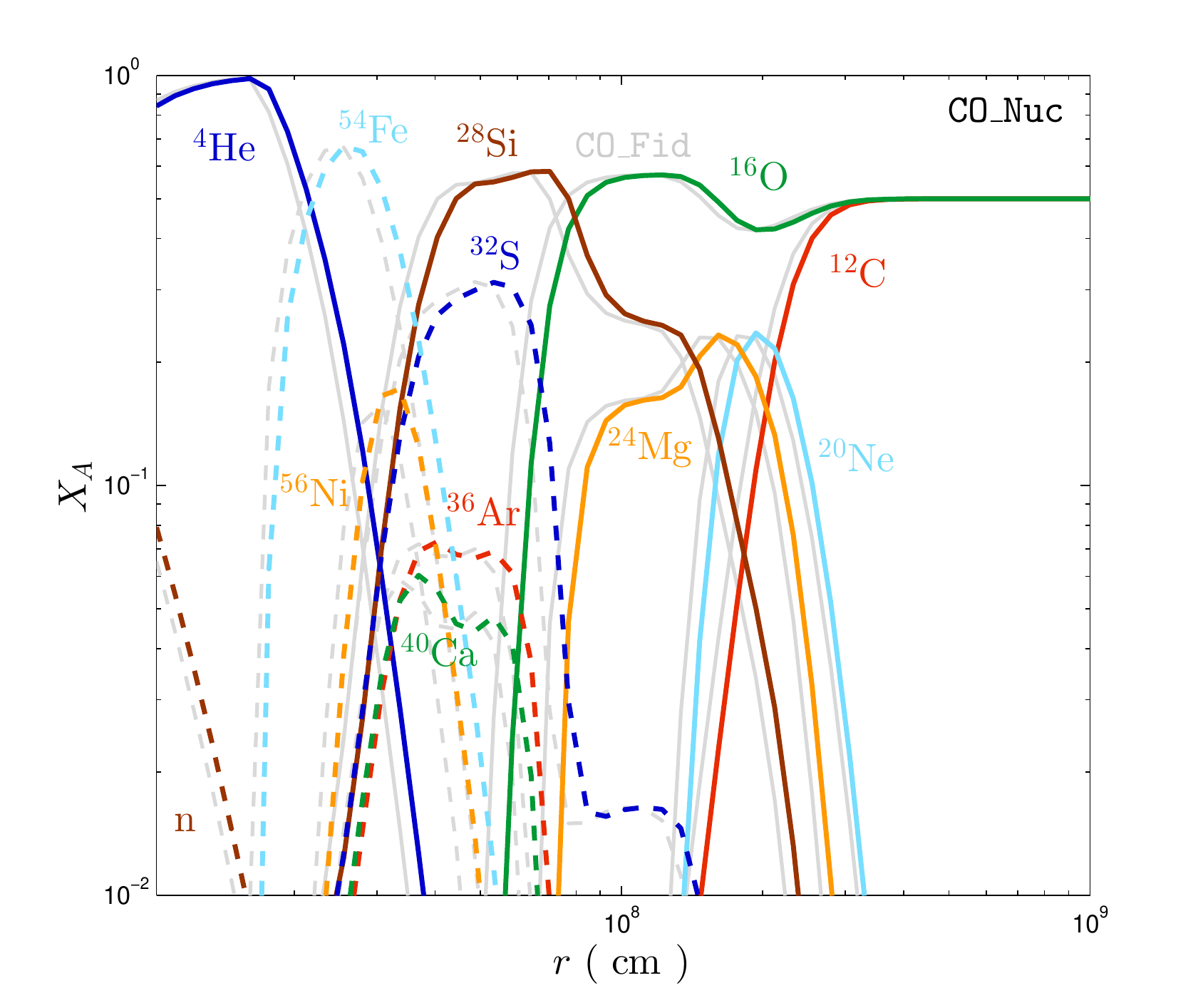,angle=0,width=0.45\textwidth} }\end{subfigure}
~
\begin{subfigure}[]{\label{fig:XA_Mix2} \epsfig{file=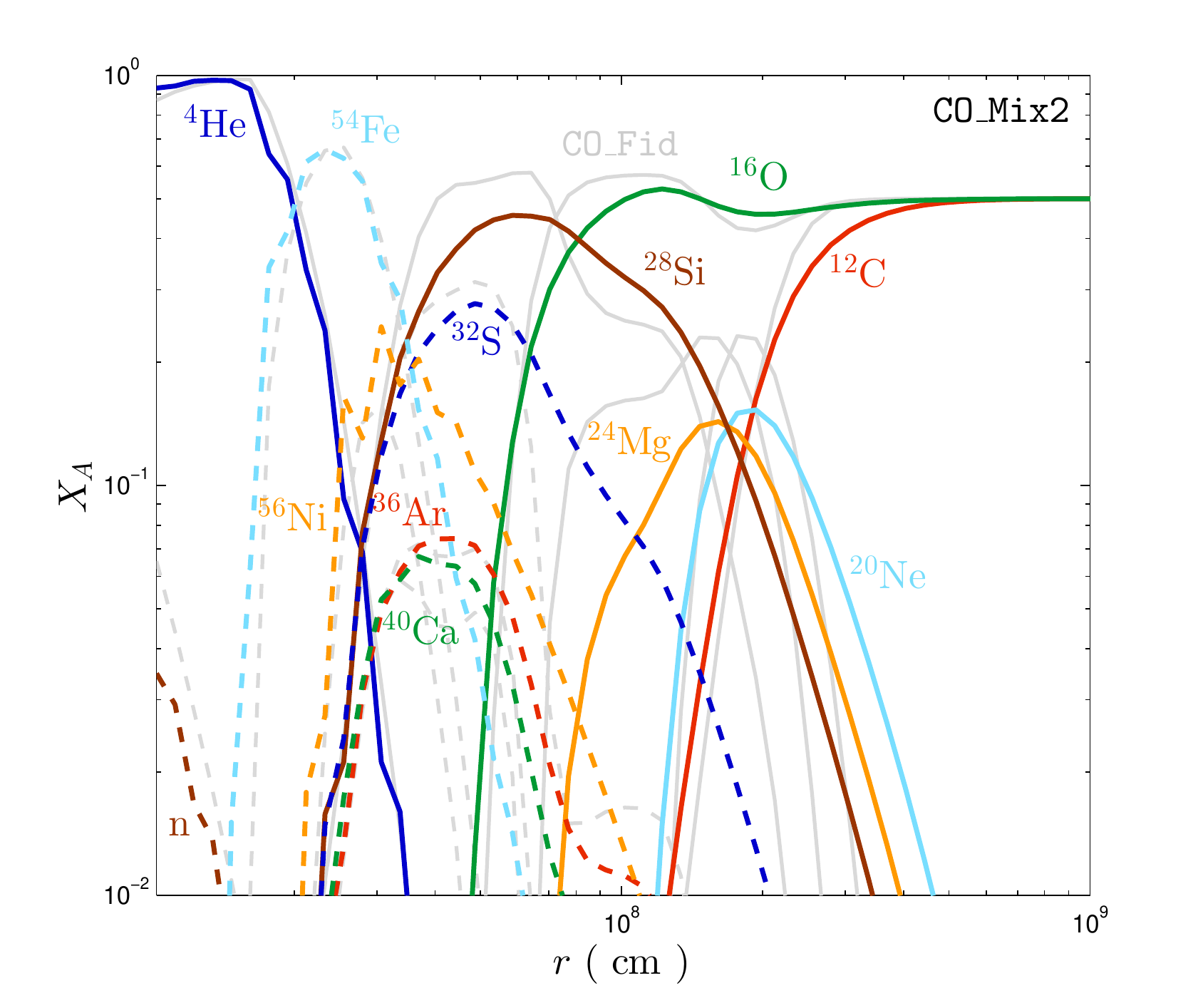,angle=0,width=0.45\textwidth} }\end{subfigure}
\caption{{\bf (a)} Disk composition profiles for model $\mathtt{CO\_Nuc}$ at time $t=8~{\rm s}$ (colored curves), in comparison with the fiducial model profiles (grey curves). The profiles are nearly identical in shape, yet systematically shifted outwards in radius due to the increased temperature at fixed $r$ when nuclear heating is turned off.
{\bf (b)} Same as panel (a), but for model $\mathtt{CO\_Mix2}$ with a mixing parameter which is 10 times its value in the fiducial case. Strong mixing changes the composition profiles significantly, and generates extended tails of burned `ashes' which diffuse upstream (to larger radii).
}
\end{figure*}

Fig. \ref{fig:XA_Nuc} compares the disk composition in the $\mathtt{CO\_Nuc}$ model (colored curves) to the fiducial case (light grey curves).  The composition profiles are nearly identical in shape, yet systematically shifted to slightly larger radii, as compared to the fiducial case.  Because less mass is lost to outflows, the correspondingly larger inflow rate increases the disk temperature, which in turn moves the burning fronts outwards.  Despite the outflow rate  being locally enhanced near the burning fronts, the total (radial- and time-integrated) mass loss rate is not affected significantly.  
This generic result is a consequence of the fact that if $\left( r_* / R_{\rm d} \right)^p \ll 1$, then the total outflow rate is controlled by the outer feeding rate $\dot{M}_{\rm in}(R_{\rm d})$, which is unaffected by nuclear burning.
The total ejecta mass and its velocity distribution are therefore nearly identical to the fiducial case.

\subsubsection{Chemical Mixing Efficiency}
Models $\mathtt{CO\_Mix1}$ and $\mathtt{CO\_Mix2}$ explore the effect of changing the dimensionless mixing parameter to values of $\tilde{\alpha}=0$ and $\tilde{\alpha} =1$, respectively, as compared to the fiducial model with $\tilde{\alpha} = 0.1$.  Mixing should have its greatest impact on the radial composition profile, as diffusive mixing smooths out strong gradients and discontinuities in $X_A(r)$.

With mixing turned off ($\mathtt{CO\_Mix1}$), the results are nearly indistinguishable from those of the fiducial case.  From this we can conclude that if turbulence is indeed less efficient at mixing passive scalars (such as $X_A$) as compared to transporting angular momentum, i.e. $\tilde{\alpha} \ll 1$, then the effects of mixing can to high accuracy be neglected altogether.  Although this is a trivial result for a truly passive scalar, in our case the composition $X_A$ enters the nuclear reaction rates, which feedback on the dynamical structure of the disk.

In the opposite case of strong mixing, the results change more significantly.  Fig. \ref{fig:XA_Mix2} compares the composition profile for model $\mathtt{CO\_Mix2}$ to the fiducial case. As expected, mixing smooths out sharp features in the composition and generally distributes the burning products across a wider range of radii.  Matter is seen to diffuse upstream to larger radii, as shown most clearly in the case of $^{32}$S and $^{56}$Ni.  Diffusion downstream also occurs, but it is not readily observed in the composition profiles because the inner profile of the mass fraction is truncated by nuclear burning, which occur sharply inside a fixed radius in the disk, largely irrespective of $X_A$. 

Although the total mass of the ejecta is also found to be insensitive to $\tilde{\alpha}$, the abundance of particular isotopes can be altered significantly, generally increasing in comparison with the fiducial model (except for $^{24}$Mg). Most significantly, the ejected $^{56}$Ni mass increases by a factor of $\sim$four (see Table \ref{tab:OutflowProperties} for numerical values for representative isotopes).

\subsubsection{Strength of Turbulent Viscosity}
Modifying the value of the viscosity parameter affects the evolution timescale of the disk, which is determined by the viscous timescale $t_{\rm visc} \propto \alpha^{-1}$ (equation~\ref{eq:t_visc}).  Model $\mathtt{CO\_Alpha}$ is calculated for $\alpha=0.01$, as compared to our fiducial model with $\alpha = 0.1$.  

\begin{figure}
\centering
\epsfig{file=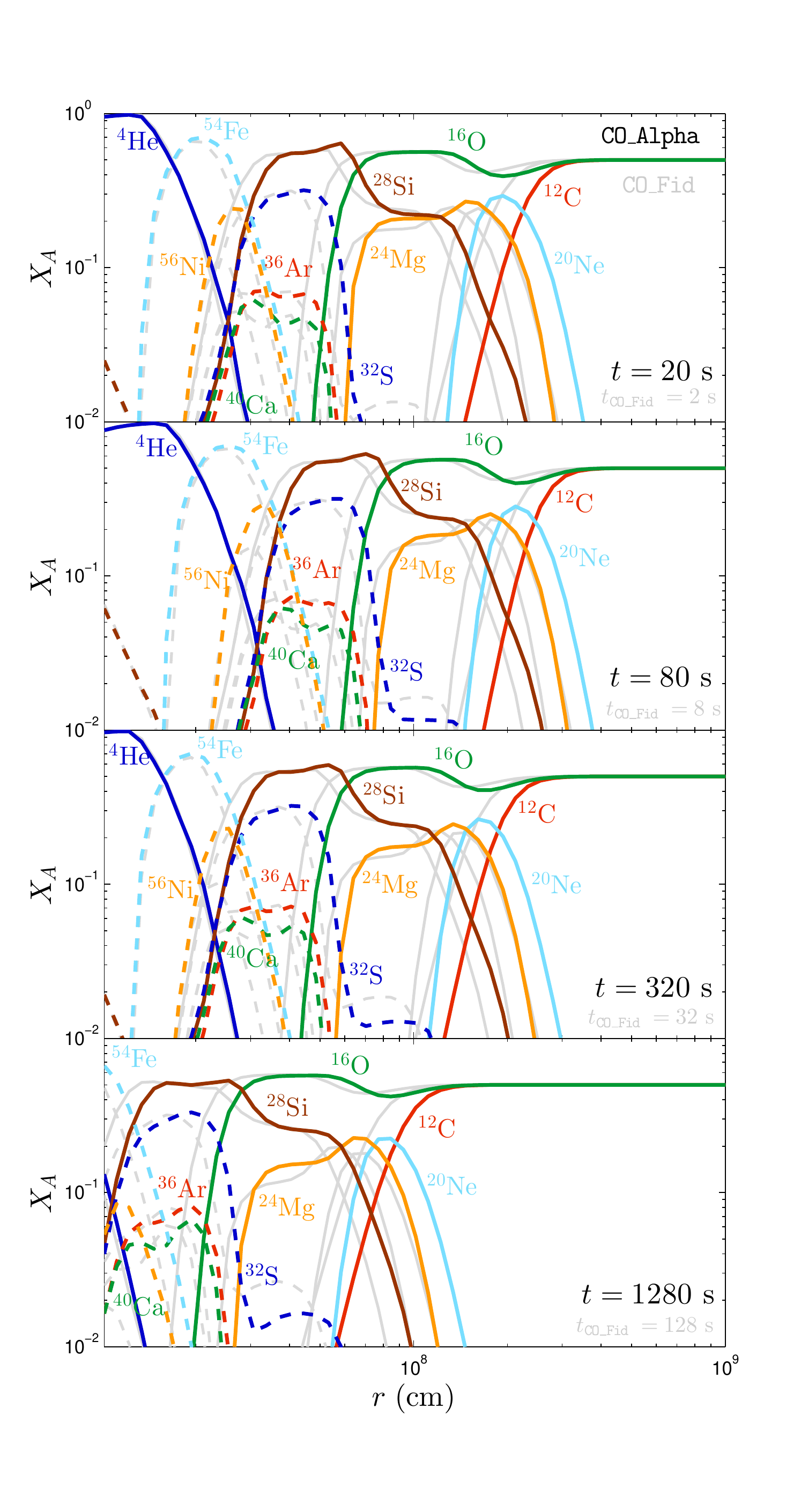,angle=0,width=0.45\textwidth}
\caption{Evolution of the radial composition for the model $\mathtt{CO\_Alpha}$ (colored curves), as in Fig. \ref{fig:X_A_Evolution}, compared to the fiducial model, $\mathtt{CO\_Fid}$ (grey curves). The different panels are plotted at snapshots such that $t_{\mathtt{CO\_Fid}} = (\alpha_{\mathtt{CO\_Alpha}} / \alpha_{\mathtt{CO\_Fid}}) \times t_{\mathtt{CO\_Alpha}}$, where $(\alpha_{\mathtt{CO\_Alpha}} / \alpha_{\mathtt{CO\_Fid}})=0.1$, and are equivalent to the panels in Fig.~\ref{fig:X_A_Evolution} for the fiducial model. Notably, the composition profiles and evolution of both models scaled this way are identical, apart from a slight shift in radii, which is well explained by equation (\ref{eq:r_burn_alpha_scaling}).
} \label{fig:X_A_Alpha}
\end{figure}

Fig.~\ref{fig:X_A_Alpha} compares the composition profile for the $\mathtt{CO\_Alpha}$ run to that of the fiducial model at four snapshots, taken at times normalized to the same fraction of the viscous time, i.e. $t_{\mathtt{CO\_Fid}} = 0.1 \times t_{\mathtt{CO\_Alpha}}$.  Applying this mapping, the overall composition at a given time remains nearly identical, which is a non-trivial result because nuclear reactions break the self-similarity of the $\alpha$-disk.  The smaller value of $\alpha$ does cause a small shift in the composition profiles of the $\mathtt{CO\_Alpha}$ model to larger radii, with the $^{12}$C burning front increasing by $\approx 20\%$.\footnote{
This result is well explained by 
the steady-state version of
equation (\ref{eq:X_A}). Using the analytic power-law approximation to the burning rates, equation (\ref{eq:X_A_nuc}), with $\delta=1$, 
neglecting mixing, 
and defining the burning front as the location of some fixed logarithmic derivative $\partial \ln X_A / \partial \ln r$, one finds that
\begin{equation} \label{eq:r_burn_alpha_scaling}
r_{\rm burn} \propto \alpha^{1/\left[ p-(5/2-p)\beta/4 \right]} ~.
\end{equation}
Using $\beta=29$ as appropriate for the $^{12}$C$(^{12}$C$,\gamma)^{24}$Mg reaction, we obtain $r_{\rm burn}(\alpha=0.01) / r_{\rm burn}(\alpha=0.1) \approx 1.17$, in perfect agreement with the numerical results.}  This shift causes the ratio of nuclear to viscous heating rates, $\dot{q}_{\rm nuc} / \dot{q}_{\rm visc}$, to increase by a modest factor of $\sim r_{\rm burn}(\alpha=0.01) / r_{\rm burn}(\alpha=0.1)$.  However, this difference is much smaller than the factor of $10$ difference one would expect {\it if} the burning fronts occurred at the same radius independent of $\alpha$ ($\dot{q}_{\rm visc} \propto \alpha$, while $\dot{q}_{\rm nuc}$ does not depend on $\alpha$).

Finally, the total mass and composition of the disk outflows are also nearly independent of $\alpha$, with the important exception of the $^{56}$Ni mass, which increases by a factor of $\sim$three for $\alpha = 0.01$ as compared to the fiducial model.

\subsubsection{Initial Density Profile}

We also explore the sensitivity of our results to the initial density profile of the disk, which is uncertain because it depends on the details of how the WD is disrupted.  Model $\mathtt{CO\_Den}$ explores the impact of increasing the radial power-law index of the inner initial density profile (equation~\ref{eq:Density_initial}) to $m=4$ from its fiducial value of $m=2$.  This slightly increases the initial radius of the disk, $R_{\rm d}$, and, more importantly, decreases the initial density at $r < R_{\rm d}$.  Although the composition profile of model $\mathtt{CO\_Den}$ are nearly identical to those of the fiducial models at late times $\gtrsim 12~{\rm s} \sim t_{\rm visc}$, the differences at early times are more pronounced.  In $\mathtt{CO\_Den}$ the burning fronts occur at smaller radii than the fiducial case because of the lower normalization of the temperature profile resulting from the lower initial density.

Although the initial density distribution of the disk impacts its evolution only at early times, $t \lesssim t_{\rm visc}$, the final (integrated) outflow distribution does exhibit some significant differences, most notably in that the mass distributions of some isotopes extend to higher velocities.  This is because, at early times when the burning fronts are located at smaller radii than in the fiducial model, nucleosynthesis occurs deeper in the potential well where the outflow velocity $v_{\rm w} \propto r^{-1/2}$ is larger.

\subsubsection{Initial Composition}
Exploring the sensitivity of the disk composition and associated nuclear burning to variation of the initial C/O mixture (models $\mathtt{CO\_Comp1}$, $\mathtt{CO\_Comp2}$) revealed only very weak dependence on this parameter. Model $\mathtt{CO\_Comp2}$, which has slightly larger carbon abundances ($X_{^{12}{\rm C}}=0.6$) produced somewhat larger peak $^{20}$Ne and $^{24}$Mg abundances, although the composition profile morphology is otherwise identical to Fig. \ref{fig:X_A_Evolution}. Similarly, model $\mathtt{CO\_Comp1}$, which is slightly carbon deficient (and appropriately oxygen rich), $X_{^{12}{\rm C}}=0.4$, led to weaker carbon burning and subsequent $^{20}$Ne, $^{24}$Mg abundances. These traits, in addition to the zeroth order effect of larger or smaller initial carbon/oxygen abundances in each model, were the only observable differences between the ejecta distribution of these models and the fiducial model.

\subsubsection{Wind Prescription}
Finally, we explore the sensitivity of our results to the parameters of the wind outflow model.  Models $\mathtt{CO\_Wnd1}$, $\mathtt{CO\_Wnd2}$ vary the fiducial critical Bernoulli parameter from zero, to ${\rm Be^\prime_{crit}=\pm0.1}$, and models $\mathtt{CO\_Wnd3}$, $\mathtt{CO\_Wnd4}$ alter the wind efficiency parameter from its nominal value of one, to $\eta_{\rm w} = 0.5$, $2$ respectively.

In model $\mathtt{CO\_Wnd1}$ the initial aspect ratio of the disk (equation~\ref{eq:theta_initial}) is larger than its steady-state value, $\theta_{\rm ss}$ (equation~\ref{eq:theta_ss}).  As discussed in \S \ref{subsec:MassInflowIndex} and Appendix \ref{subsec:Appendix_InitialOutflow}, this initial configuration results in a strong transient wind phase lasting a short time $\sim t_{\rm w}$ that cools the disk to its steady-state Bernoulli parameter, ${\rm Be^\prime_{crit}}$. In this specific case, the initial aspect ratio of the disk is $\theta_{\rm initial}=0.418$ (same as for the fiducial model) is 10$\%$ larger than its steady-state value at $R_{\rm d,0}$ of $\theta_{\rm ss} \approx 0.38$ (evaluated numerically).  From equation~(\ref{eq:M_precursor_text}), we predict a prompt ejection of $\sim 3 \times10^{-2} M_\odot$, in excellent agreement with the $2.8 \times 10^{-2} M_\odot$ outflow mass measured from the model at early times.
Besides this precursor outflow, the evolution and final outflow composition for model $\mathtt{CO\_Wnd1}$ is nearly identical to the fiducial model.   Model $\mathtt{CO\_Wnd2}$ also shows no significant deviations from the fiducial model at times $t \gtrsim t_{\rm visc}$.  This is not surprising because Fig. \ref{fig:p_index} shows that the mass-inflow index $p$ is not sensitive to the value of the critical Bernoulli parameter.  

The results are more sensitive to the wind efficiency parameter, $\eta_{\rm w}$. Increasing $\eta_{\rm w}$ by a factor of two, as in model $\mathtt{CO\_Wnd4}$, decreases the mass-inflow exponent $p$ noticeably (as also predicted by Fig.~\ref{fig:p_index}). This means that more material accretes inwards, at the expense of weaker outflows. This causes the disk density and thereby temperature at any given radius to increase in comparison with the fiducial model, shifting the burning fronts to larger radii but preserving the shape and evolution of the composition profiles.  The outflow distribution, on the other hand, changes qualitatively.

\begin{figure*}
\centering
\begin{subfigure}[]{\label{fig:Wind_Wnd3} \epsfig{file=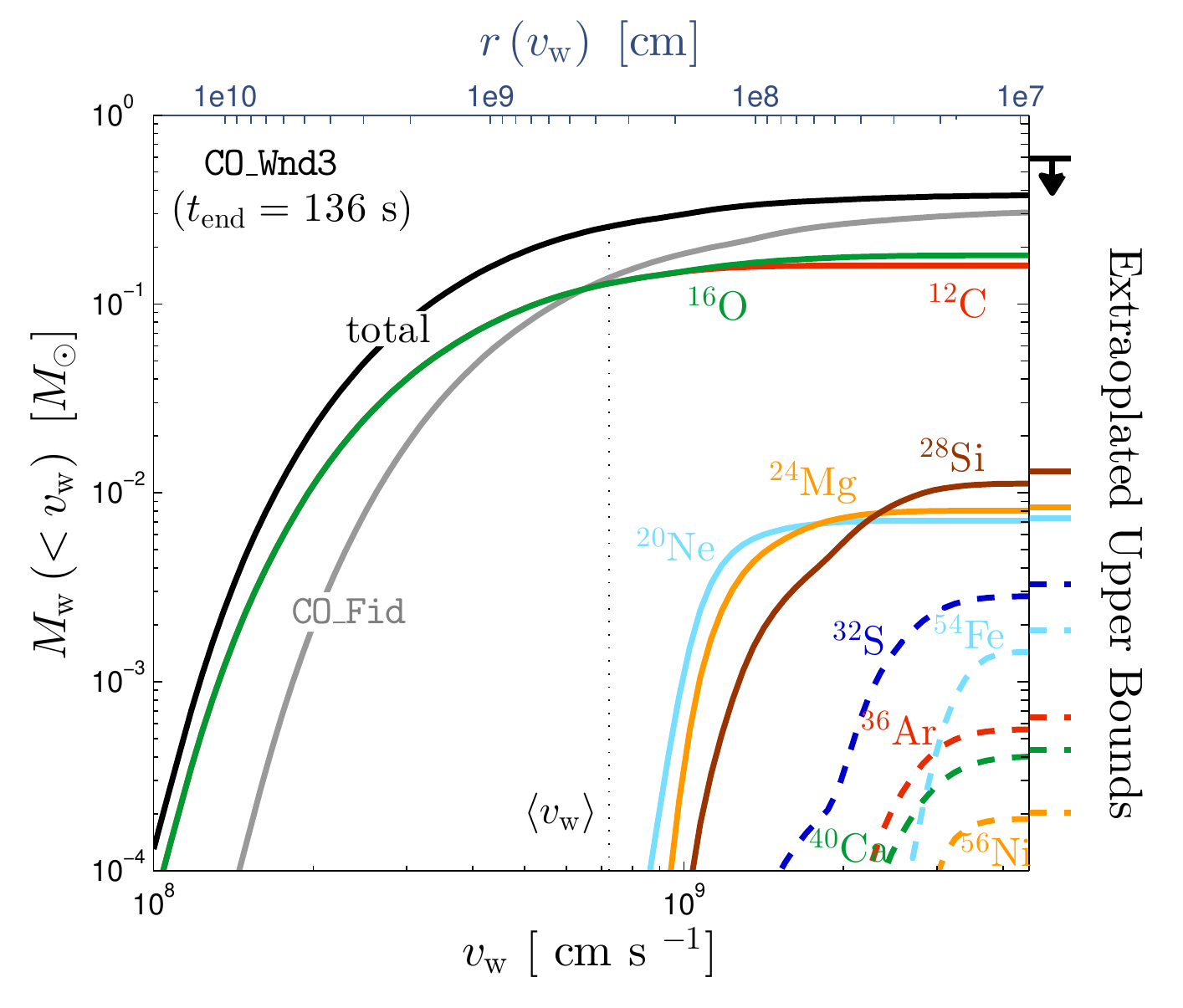,angle=0,width=0.45\textwidth} }\end{subfigure}
~
\begin{subfigure}[]{\label{fig:Wind_Wnd4} \epsfig{file=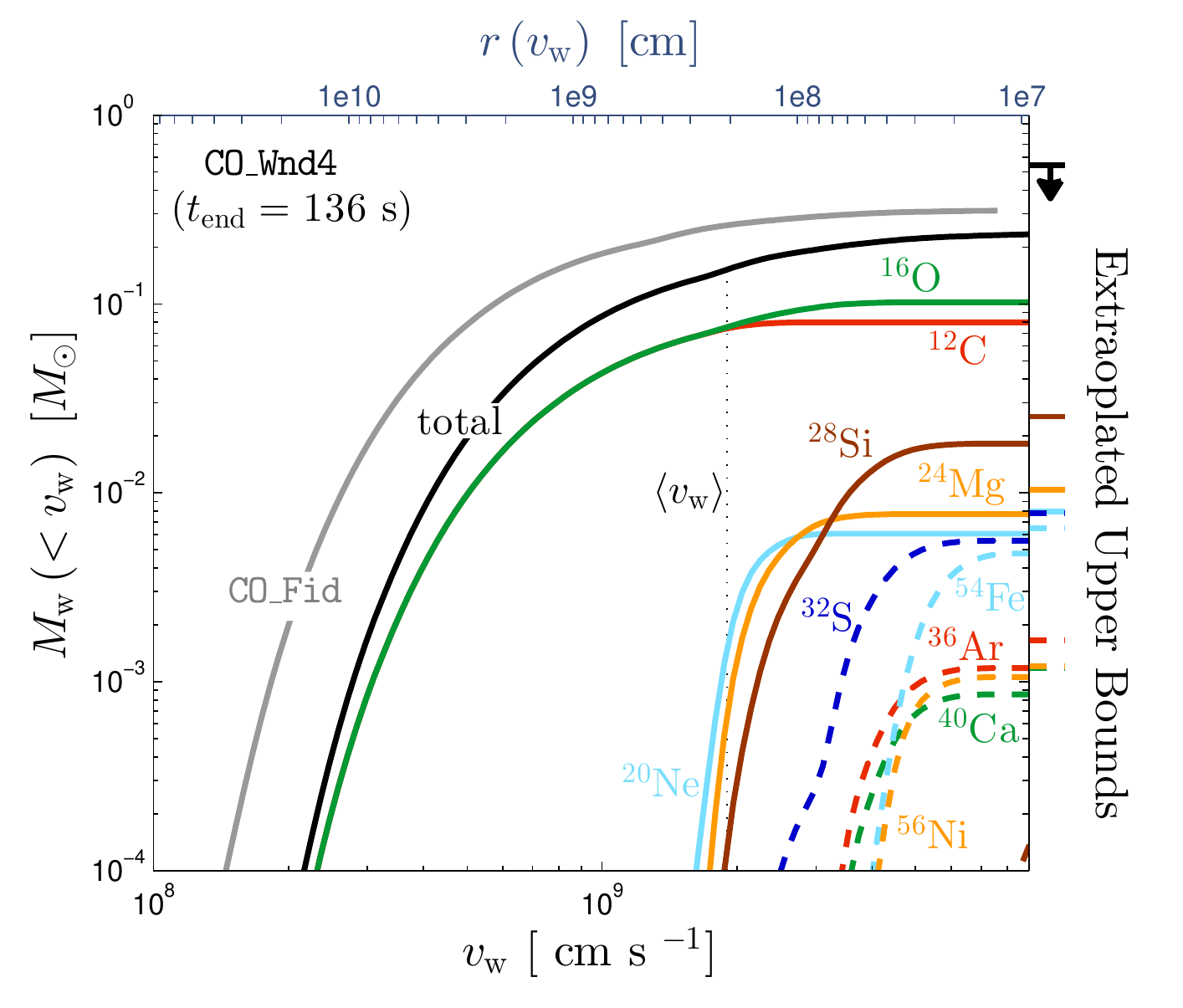,angle=0,width=0.45\textwidth} }\end{subfigure}
\caption{Velocity distribution of wind ejecta for models $\mathtt{CO\_Wnd3}$, {\bf (a)}, and $\mathtt{CO\_Wnd4}$, {\bf (b)}. Different style/color curves represent the distribution in various isotopes (same as in Fig. \ref{fig:Wind_a}). Black curves plot the total outflow mass distribution, which can be compared with the solid grey curves, illustrating the same quantity for model $\mathtt{CO\_Fid}$ (note though that the top x-axis does not apply to model $\mathtt{CO\_Fid}$).
}
\end{figure*}

Fig. \ref{fig:Wind_Wnd4} shows the final wind distribution for model $\mathtt{CO\_Wnd4}$ in comparison to the fiducial model $\mathtt{CO\_Fid}$.  The most noticeable change is that the wind distribution extends to larger velocities, and the total ejecta mass decreases (although particular isotope yields do increase). The first of these trends is straightforward to understand because the wind efficiency parameter directly determines the ejecta velocity.  However, even by rescaling the velocity axis by $\eta_{\rm w}^{1/2}$, the total $\mathtt{CO\_Wnd4}$ ejecta distribution curves shows an excess of mass at high velocities, due to the fact that more mass flows to smaller radii (the small value of $p$).  A similar argument explains why most high mass isotopes, such as $^{56}$Ni are overproduced. Quantitatively, the total ejecta mass for this model decreases by $\sim 25\%$ to $0.23 M_\odot$, while the $^{56}$Ni yield increases by a factor of two to $1.3 \times 10^{-3} M_\odot$. 
Model $\mathtt{CO\_Wnd3}$, in which the wind efficiency parameter is decreased, can be explained by similar arguments (Fig. \ref{fig:Wind_Wnd3}).

\subsection{He WD Models}
\label{sec:He}
We additionally consider models for the accretion of a disrupted helium WD, the properties of which differ qualitatively from the C/O models discussed above. Our fiducial model, $\mathtt{He\_Fid}$, describes a typical $0.3 M_\odot$ He WD which merges with a $1.2 M_\odot$ NS companion. The model parameters, $\alpha=0.1$, $\eta_{\rm w}=1$, ${\rm Be^\prime_{crit}}=0$, $\tilde{\alpha}=0.1$, and $(m,n)=(2,7)$, are the same as for the fiducial C/O model (see Table \ref{tab:ModelParameters}).

\begin{figure}
\centering
\epsfig{file=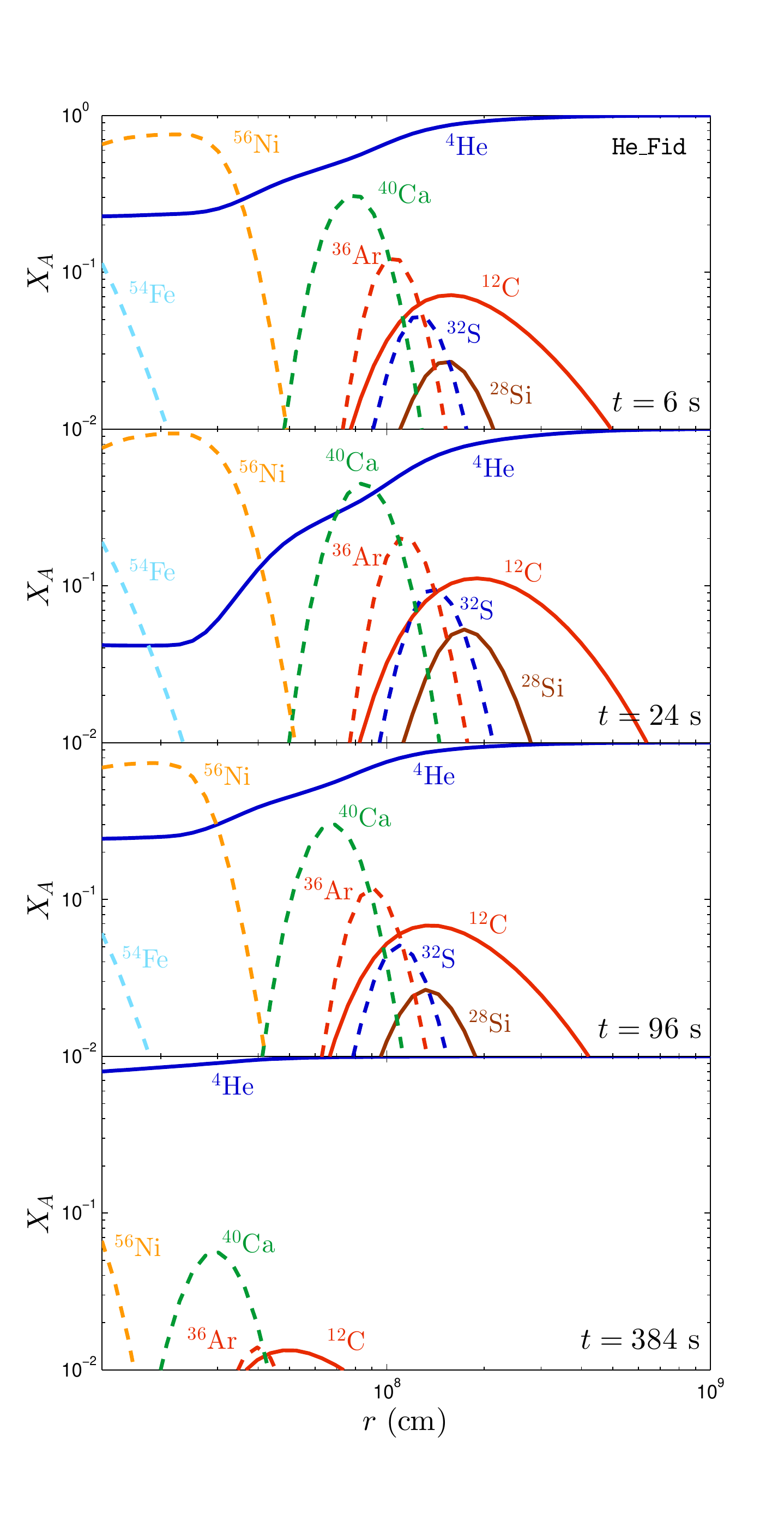,angle=0,width=0.45\textwidth}
\caption{Evolution of the disk composition, similar to Fig. \ref{fig:X_A_Evolution}, but for the fiducial helium WD model, $\mathtt{He\_Fid}$. }\label{fig:X_A_Evolution_He}
\end{figure}

The initial disk radius for this model, $R_{\rm d} \simeq 3.4 \times 10^9~{\rm cm}$, is larger than in the case of our C/O models due to the smaller mass of He WDs (Fig.~\ref{fig:WDMassDiagram}). The peak accretion (or outflow) rate is $\simeq 8 \times 10^{-4}~M_\odot~{\rm s}^{-1}$ and is achieved on a timescale of $t_{\rm visc} \approx 25~{\rm s}$.  

Fig. \ref{fig:X_A_Evolution_He} shows snapshots of the disk composition at four representative timesteps, similar to Fig. \ref{fig:X_A_Evolution} for $\mathtt{CO\_Fid}$. At early times (top panel), the density limited triple-$\alpha$ reaction begins fusing $^{12}$C at $r \sim 4 \times 10^8~{\rm cm}$. Due to the high $^{4}$He abundance, rapid $\alpha$-captures onto the seed $^{12}$C nuclei immediately fuse into higher mass elements, $^{28}$Si, $^{32}$S, $^{36}$Ar, $^{40}$Ca, and peaking at $^{56}$Ni. The intermediate elements $^{16}$O, $^{20}$Ne and $^{24}$Mg, which have extremely high $\alpha$-capture rates serve as `stepping stones' in this process, but are severely underproduced themselves, reaching peak abundances of only $10^{-4}$, $3 \times 10^{-4}$, and $10^{-3}$, respectively. As the density increases towards the peak accretion time (second panel), the triple-$\alpha$ reaction becomes more effective, increasing the seed carbon abundance and thereby the higher mass elements' mass fractions as well ($^{56}$Ni reaches peak abundances of $\sim 1$). At later times (third panel from top) the disk density decreases, inhibiting the triple-$\alpha$ reaction and increasing the helium abundance while the high mass isotopes decrease, until at late times the disk reverts to a nearly pure helium composition (fourth panel).

Qualitatively, this nucleosynthesis is dramatically different than that of our previous C/O WD models, producing large $^{56}$Ni and $^{40}$Ca abundances (along of course with a large unburned $^{4}$He abundance) despite extremely low $^{16}$O mass fractions. Additionally, the evolution of the composition profiles differs qualitatively from the C/O WDs --- the composition is set almost entirely by the triple-$\alpha$ reaction which, while effective at $\sim$peak-accretion time when the density is highest, becomes very inefficient at late times in the disk evolution. This causes the mass fraction $X_A$ profiles to steadily decrease after $t \gtrsim t_{\rm visc}$, and essentially disappear at late times. In comparison, the C/O model composition profiles preserved their morphology and normalization in a self-similar manner, merely shifting inwards to smaller radii at late times.

The inefficiency of the triple-$\alpha$ reaction at early/late times is a direct consequence of its strong density dependence, and on the fact that the reverse reaction, $^{12}{\rm C} \to 3\alpha$, is in contrast a temperature sensitive reaction.
Both the forward and reverse reactions scale in the same way with density, as $\dot{X}^{\rm (nuc)}_{3\alpha} \propto \rho^2$, so that the two reactions balance each other at a fixed temperature, $T_{\rm lim}$. Using analytic reaction rates from \cite{Caughlan&Fowler1988}, we find
\begin{equation} \label{eq:triple_alpha_Tlim}
T_{\rm lim} \simeq 1.685 \times 10^9~{\rm K} ~.
\end{equation}
At temperatures $\gtrsim T_{\rm lim}$, the triple-$\alpha$ reaction cannot effectively fuse $^{12}$C, since any carbon would immediately be disintegrated back into $^{4}$He by the dominant reverse triple-$\alpha$ process.
Although triple-$\alpha$ may successfully occur around $\sim t_{\rm visc}$, at early (late) times the disk density rises (drops) at a faster rate than the disk temperature, so that at some point 
$T \left( \rho_{3\alpha} \right) > T_{\rm lim}$,
and carbon fusion effectively ceases. Here $\rho_{3\alpha}$ is the `burning density' at which the triple-$\alpha$ process occurs. Since the seed carbon nuclei are key to forming successively heavier elements through rapid $\alpha$-captures, this affects the entire disk composition for elements above $^{4}$He.

\begin{figure*}
\centering
\begin{subfigure}[]{\label{fig:X_A_contour_He} \epsfig{file=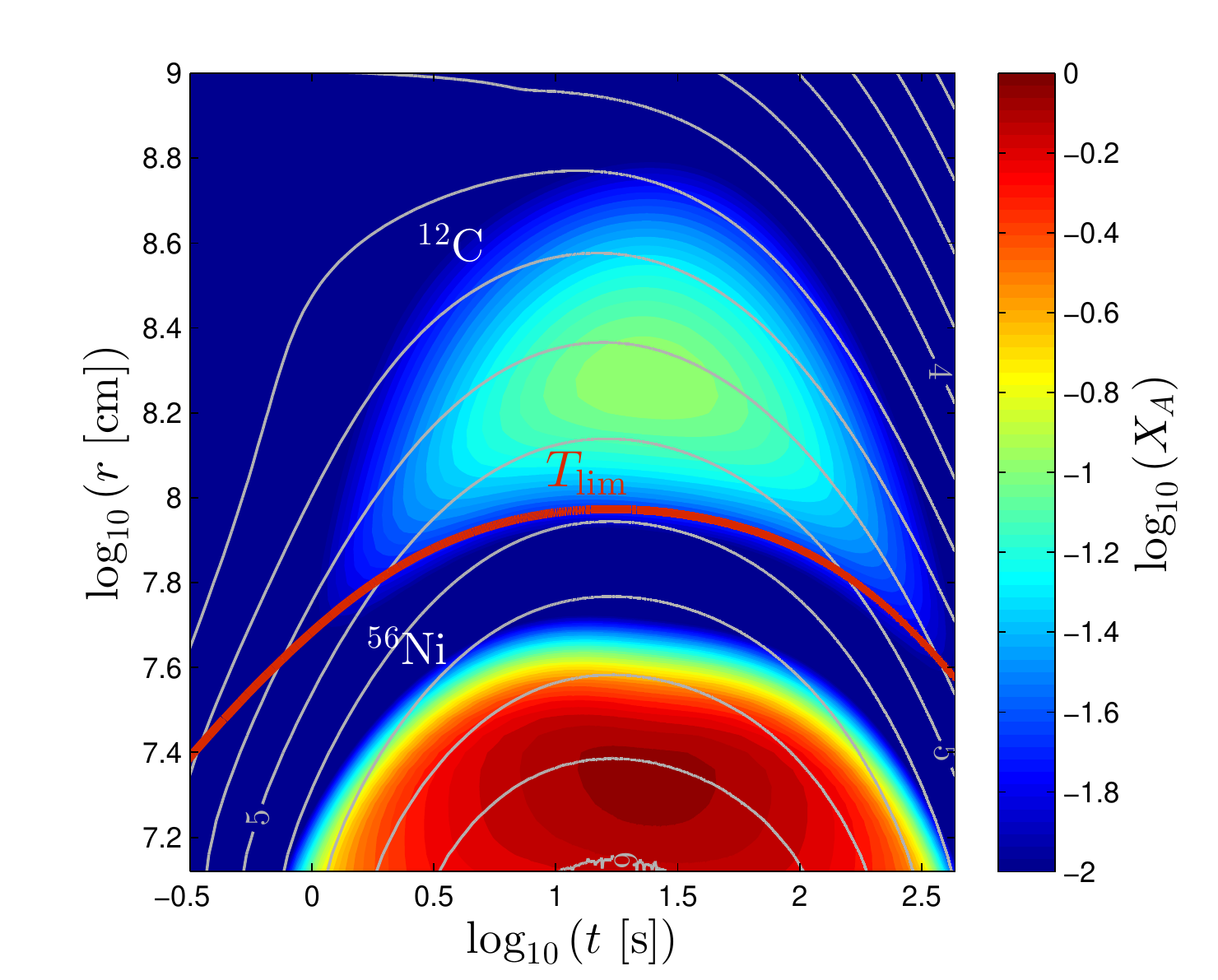,angle=0,width=0.45\textwidth} }\end{subfigure}
~
\begin{subfigure}[]{\label{fig:qdot_He} \epsfig{file=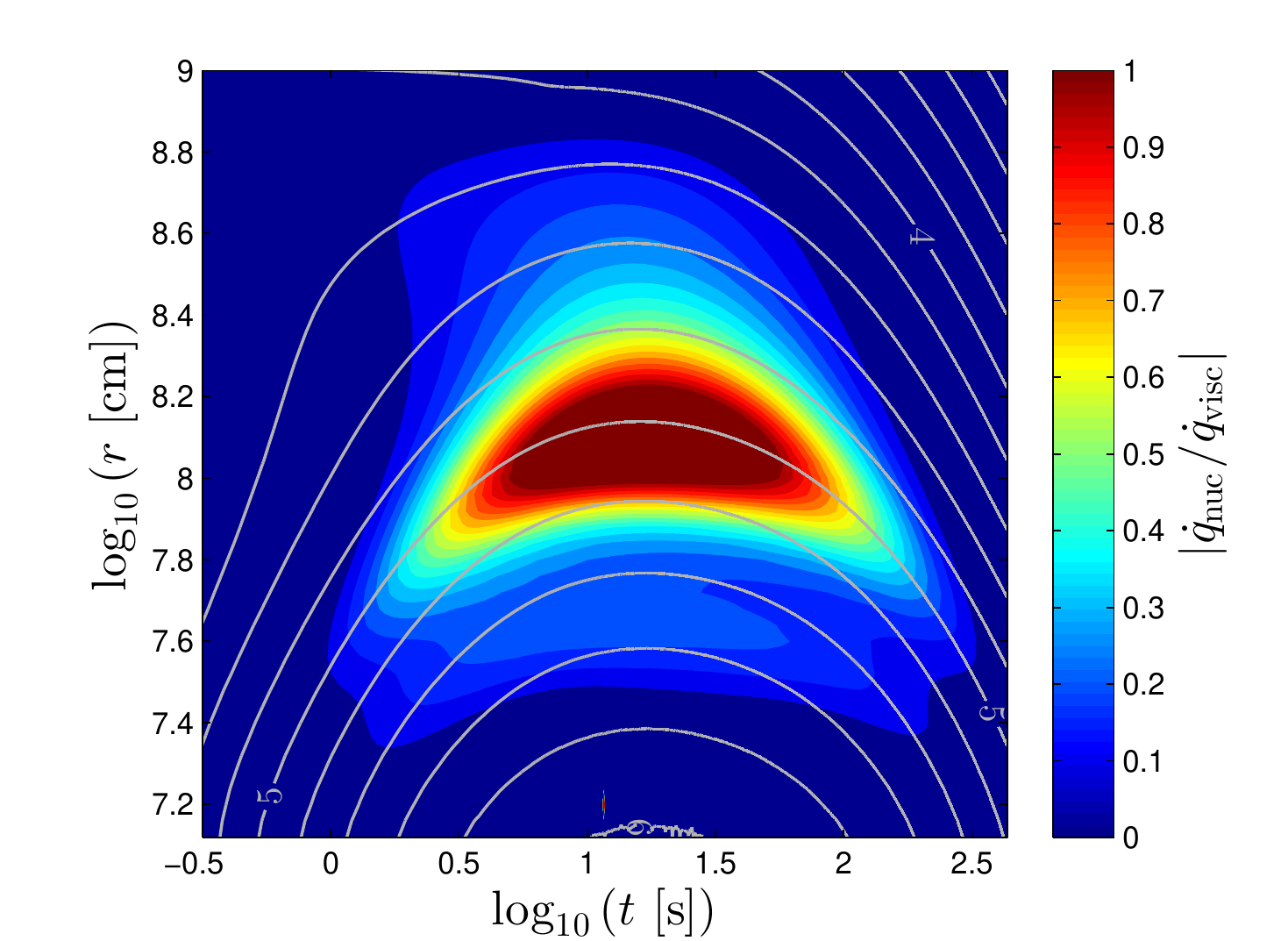,angle=0,width=0.45\textwidth} }\end{subfigure}
\caption{{\bf (a)} Contour plot of two representative isotope abundances, $^{12}$C and $^{56}$Ni, in the fiducial helium disk model as a function of $r$ and $t$. The carbon abundance traces the triple-$\alpha$ limiting reaction in the flow. The inset grey contours depict curves of constant density, $\rho$, logarithmically equally spaced by $\Delta {\rm log}_{10}(\rho~[{\rm g~cm}^{-3}]) = 0.2$, and labeled at $\rho=10^4, 10^5, 10^6 ~{\rm g~cm}^{-3}$. The strongly density dependent triple-$\alpha$ reaction is seen to roughly track these curves, but effectively shuts off after $t\sim 200~{\rm s}$, when the triple-$\alpha$ `burning density', $\rho_{3\alpha} \sim 10^5~{\rm g~cm}^{-3}$, approaches the $T=T_{\rm lim}$ constant temperature curve (thick red).  Above this temperature (equation \ref{eq:triple_alpha_Tlim}) the reverse triple-$\alpha$ reaction rate, $^{12}{\rm C} \to 3\alpha$, exceeds the forward $3\alpha \to {}^{12}{\rm C}$ rate, and carbon cannot effectively be fused.
{\bf (b)} Nuclear heating rate in the ${\rm log}_{10}(r)$, ${\rm log}_{10}(t)$ plane. Nuclear reactions deposit a significant amount of energy in the disk in the outer radii at which the triple-alpha reaction commences, and are dynamically more important than in the C/O burning case (see Fig.~\ref{fig:qdot_b}).
}
\end{figure*}

Fig. \ref{fig:X_A_contour_He} illustrates this point by showing a contour plot of the evolution of two representative isotopes --- $^{56}$Ni and $^{12}$C, the second of which is a direct tracer of the triple-$\alpha$ burning. Additionally, curves of constant density are plotted in logarithmic spacings of $\Delta {\rm log}_{10}(\rho~[{\rm g~cm}^{-3}]) = 0.2$. To first order, the triple-$\alpha$ burning front tracks the density evolution and peaks around $\rho_{3\alpha} \sim 10^5~{\rm g~cm}^{-3}$. The thick red curve plots  
a constant temperature contour at $T=T_{\rm lim}$.
It is clear that at late times, $\gtrsim 200~{\rm s}$, the condition $T \left( \rho_{3\alpha} \right) > T_{\rm lim}$ is satisfied and the carbon abundance drops significantly. The $^{56}$Ni abundance also drops starting at this time, illustrating how the triple-$\alpha$ reaction effectively limits the entire disk composition.

As in our previous discussion of C/O WDs, nuclear burning deposits significant energy in the disk. Fig.~\ref{fig:qdot_He} plots the nuclear heating rate relative to the viscous disk heating, similar to Fig.~\ref{fig:qdot_b} for the C/O fiducial model. Nuclear heating is an important energy source as long as the triple-$\alpha$ reaction is effective, and dominates the total disk heating around $\sim 10^8~{\rm cm}$ for a significant portion of the disk evolution.

\begin{figure}
\centering
\epsfig{file=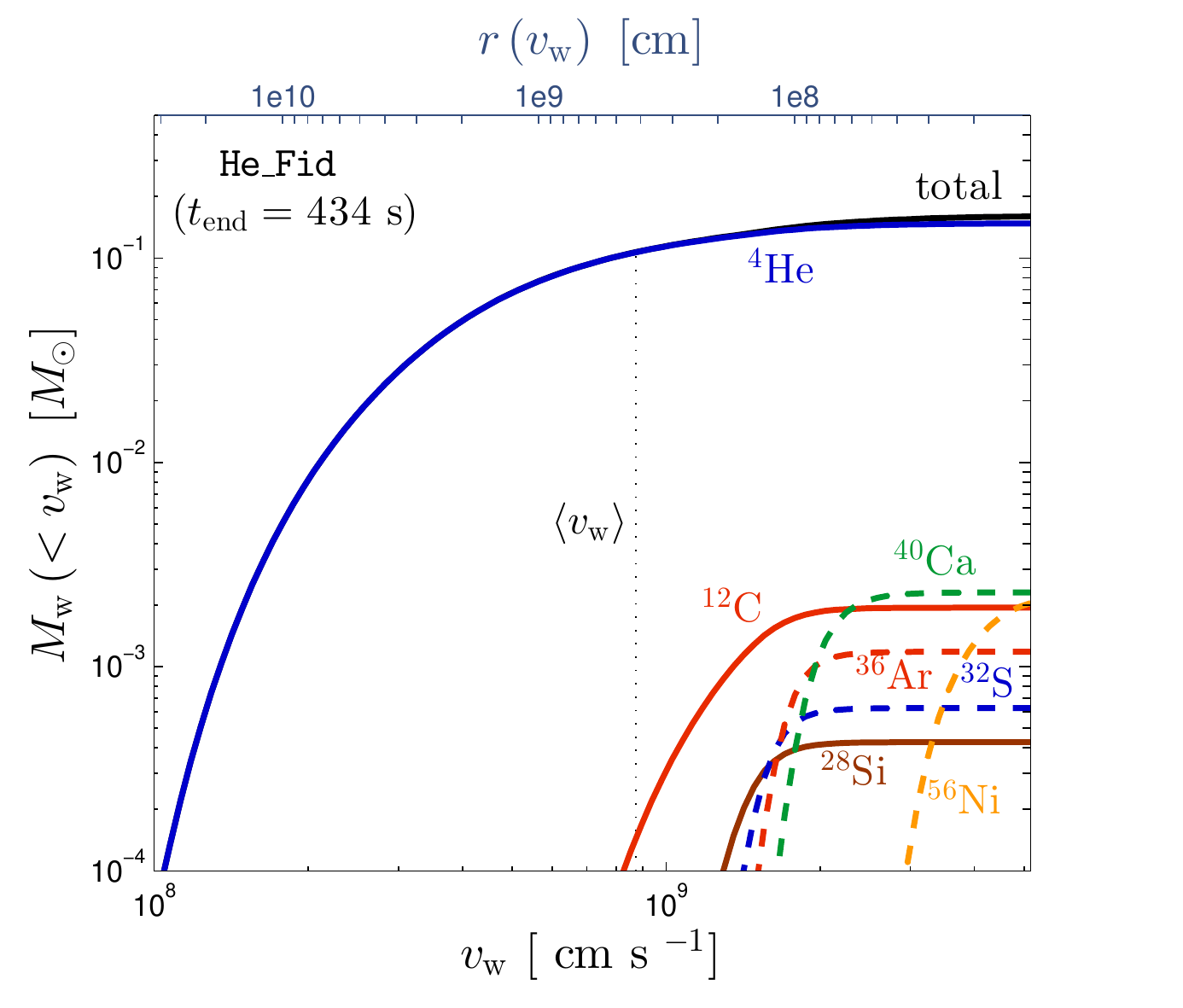,angle=0,width=0.45\textwidth}
\caption{Velocity distribution of the wind ejecta, from the fiducial helium WD model, $\mathtt{He\_Fid}$, evaluated at the simulation end time. The distribution of different isotopes are colored as in Fig.~\ref{fig:Wind_a}. A solid black curve shows the total mass distribution (summed over all elements). 
The $^{56}$Ni distribution extends up to the largest velocities (smallest radius) captured by our numerical grid, indicating that we do not resolve the entire $^{56}$Ni outflow, and thus that our model provides only a lower limit on the nickel mass in the ejecta.
} \label{fig:Wind_He_Fid}
\end{figure}

Finally, in Fig.~\ref{fig:Wind_He_Fid} we plot the outflow velocity distribution at the simulation termination time, $t_{\rm end}=434~{\rm s}$. At this time, $0.16M_\odot$ or roughly half of the initial WD mass has been ejected (black curve), predominantly as unburnt $^{4}$He (solid blue curve), at characteristic velocities of $\langle v_{\rm w} \rangle \simeq 8.7 \times 10^8~{\rm cm~s}^{-1}$.

We note that the numerically obtained values of the $^{56}$Ni and $^{54}$Fe ejecta mass are only lower limits on their true values because these isotopes' composition profile extends interior to our numerical inner boundary, and therefore their contributions to the ejecta are not entirely captured (see Fig.~\ref{fig:X_A_Evolution_He}).
Additionally, we do not extrapolate the outflow composition to $t \to \infty$ as we did for the C/O models, since the triple-$\alpha$ reaction which sets the disk composition does not obey the analytic scaling of equation (\ref{eq:Mdot_w_XA_scaling}) which was developed for temperature limited nuclear reactions. Despite this fact, the ejecta mass in various isotopes at time $t_{\rm end}$ is likely a reliable estimator of the ejected mass at time $t \to \infty$ (except for the case of $^{56}$Ni and $^{54}$Fe discussed above, and for helium which tracks the total ejecta mass and is expected to reach values of $\sim M_{\rm d}=0.3 M_\odot$). This is because by the simulation termination time, the peak mass fractions of isotopes heavier than $^{4}$He decrease below $\sim 10^{-2}$ (see Fig.~\ref{fig:X_A_Evolution_He}), so that at subsequent times, the disk composition and accompanying outflow is essentially purely helium.

The total outflow distribution of model $\tt He\_Nuc$ is overall very similar to the fiducial model, except that $\mathtt{He\_Fid}$ exhibits a slight excess of ejected matter around $\sim 10^9~{\rm cm~s}^{-1}$. This occurs due to the significant (in fact dominant) contribution of nuclear burning to the disk heating rate (Fig.~\ref{fig:qdot_He}), which is locally balanced by stronger wind cooling, i.e., larger outflows.

The lack of nuclear feedback in model $\mathtt{He\_Nuc}$ causes an increase in the disk density (compared with the fiducial model) at radii $\lesssim 2\times10^8~{\rm cm}$ (near the triple-$\alpha$ burning front), which in turn increases the efficiency of the density-limited triple-$\alpha$ burning. This changes the composition profiles somewhat more substantially than by merely shifting the burning fronts to larger radii (as was the case for the temperature limited reactions of the C/O disk, see Fig.~\ref{fig:XA_Nuc}), and in particular, more $^{56}$Ni is synthesized.

As in the C/O WD scenario, the mixing parameter $\tilde{\alpha}$ has little effect on the results. With mixing effectively turned off (model $\mathtt{He\_Mix1}$), the results are essentially identical in every respect to the fiducial model, indicating once again that small mixing parameters can be well approximated by neglecting mixing altogether. For model $\mathtt{He\_Mix2}$, in which the mixing parameter is increased to $\tilde{\alpha}=1$, the composition profiles show prominent tails towards larger radii due to burned ash diffusing upstream, in parallel with the results of C/O WD mixing illustrated in Fig. \ref{fig:XA_Mix2}. This does not have a significant effect on the outflow composition, except on the $^{56}$Ni yield, which increases by a modest factor of $\sim 1.5$.

Similarly, varying the initial density distribution of the disk, as in model $\mathtt{He\_Den}$, has little effect on the outcome. Just as for the C/O models, the composition evolution changes slightly at early times $\lesssim t_{\rm visc}$, but is identical to the fiducial model at later times. 

\begin{figure}
\centering
\epsfig{file=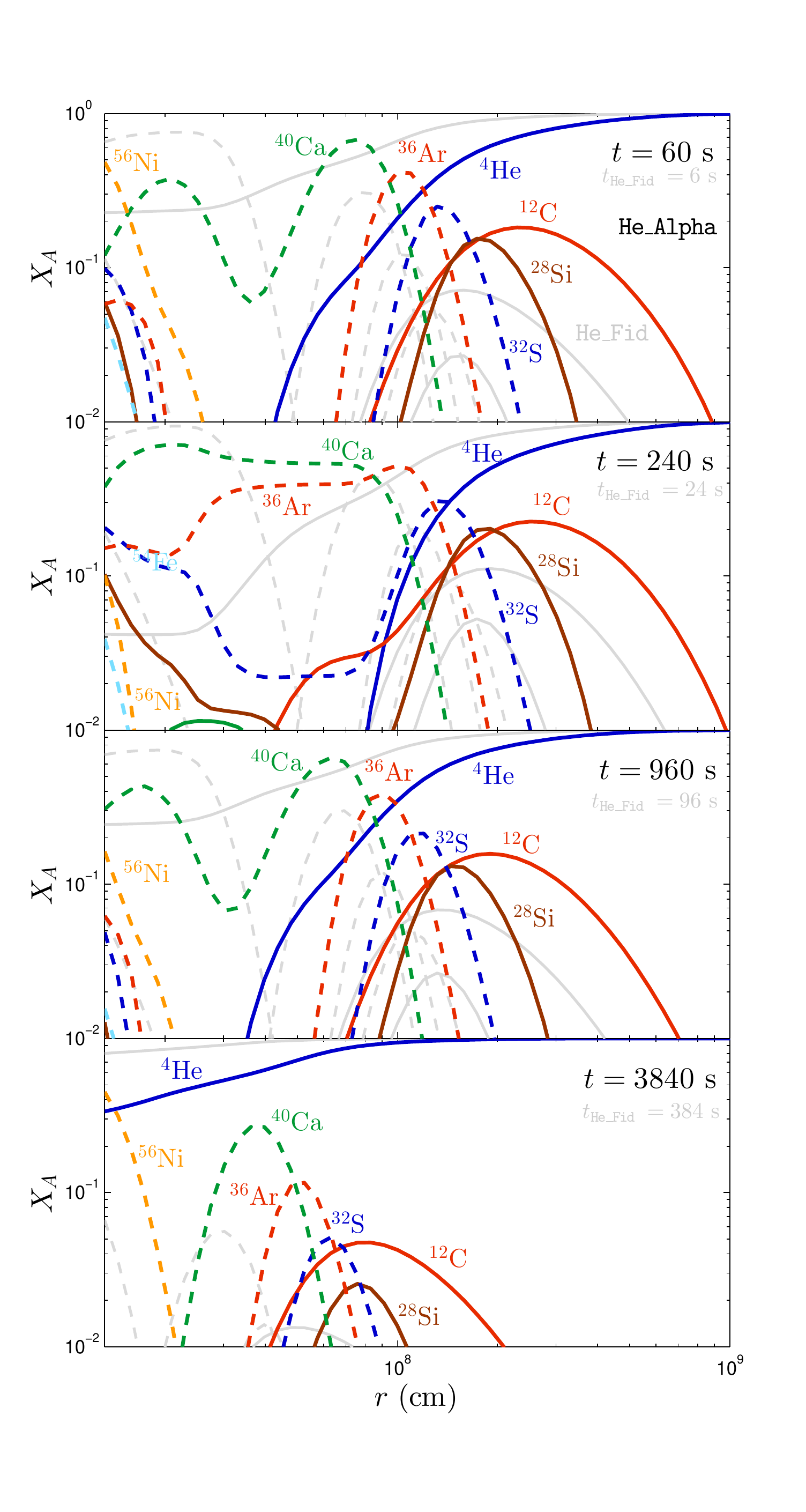,angle=0,width=0.45\textwidth}
\caption{Evolution of the nuclear composition for the model $\mathtt{He\_Alpha}$ (colored curves), as in Fig.~\ref{fig:X_A_Evolution_He}, compared to the fiducial model, $\mathtt{He\_Fid}$ (grey curves).  Different panels correspond to snapshots defined by $t_{\mathtt{He\_Fid}} = (\alpha_{\mathtt{He\_Alpha}} / \alpha_{\mathtt{He\_Fid}}) \times t_{\mathtt{He\_Alpha}}$, where $(\alpha_{\mathtt{He\_Alpha}} / \alpha_{\mathtt{He\_Fid}})=0.1$, and are equivalent to the panels in Fig.~\ref{fig:X_A_Evolution_He} for the fiducial model.  Note that the composition profiles are qualitatively different in model $\mathtt{He\_Alpha}$ due to the density sensitivity of the limiting triple-$\alpha$ reaction.  This is in contrast to C/O models, in which scaling the viscosity parameter, $\alpha$, changes (to first order) only the overall timescale of disk evolution (Fig.~\ref{fig:X_A_Alpha}).
} \label{fig:X_A_He_Alpha}
\end{figure}

On the other hand, varying the alpha-viscosity parameter (model $\mathtt{He\_Alpha}$) impacts the results much more significantly than for the C/O models. Fig. \ref{fig:X_A_He_Alpha} shows the evolution of the composition profile for this model at four representative timesteps. These are usefully compared with the fiducial helium model composition (Fig.~\ref{fig:X_A_Evolution_He}), which are overplot with light grey curves. In the case of C/O WDs, the composition profile preserved its radial shape in time.  However, for helium accretion this is clearly not the case.  At the time of peak accretion (second panel), the $^{4}$He abundance decreases below $X_A \lesssim 10^{-2}$ at radii $r \lesssim 8 \times 10^7~{\rm cm}$, resulting in the significantly larger amounts of intermediate elements such as $^{40}$Ca and $^{36}$Ar being synthesized further in. The isotopes $^{56}$Ni and $^{54}$Fe are produced almost entirely interior to our inner grid boundary, precluding a reliable prediction of their ejecta abundances.  

Changing the helium WD mass also changes the outcome significantly. Model $\mathtt{He\_Mass}$ corresponds to a $0.4M_\odot$ He WD with the same, nominal $1.2M_\odot$ binary companion.  The results differ from the fiducial model and to some extent continue the trend apparent in $\mathtt{He\_Alpha}$ of strong triple-$\alpha$ burning. 

Finally, in model $\mathtt{He\_Wnd4}$ we increase the wind efficiency parameter from its fiducial value of $\eta_{\rm w}=1$.  This also has a substantial affect on the results, mainly by decreasing the mass inflow exponent $p$ (Fig.~\ref{fig:p_index}).  The resulting higher mass inflow rate increases the disk density at each radius, which as previously discussed is intimately related to the efficiency of nuclear burning. For larger values of $\eta_{\rm w}$, the triple-$\alpha$ process remains effective for a longer period of time, increasing the nucleosynthesis of heavy elements.  The dynamical significance of helium burning is also increased accordingly, with $\vert \dot{q}_{\rm nuc}/\dot{q}_{\rm visc} \vert$ reaching peak values of $\sim 20$.  The same reasoning explains why nucleosynthesis is less effective for model $\mathtt{He\_Wnd3}$, for which $\eta_{\rm w}$ is smaller than its fiducial value.

\subsection{Hybrid WDs}
\label{sec:hybrid}
We conclude by discussing results for WDs composed of both C/O and He, so-called `hybrid' WDs (\citealt{Han+00}).  Given the rather speculative nature of this type of WD, we run only a couple models and do not perform a full parameter space survey as was done for C/O and He WDs.  The model parameters are identical to the fiducial C/O case (\S \ref{subsec:Fiducial C/O Model}), except for the initial composition of $X_{\rm {}^{12}C} = X_{\rm {}^{16}O} = 0.4$, $X_{\rm {}^{4}He} = 0.2$ and $X_{\rm {}^{12}C} = X_{\rm {}^{16}O} = 0.475$, $X_{\rm {}^{4}He} = 0.05$ for models $\mathtt{CO\_He1}$ and $\mathtt{CO\_He2}$ respectively.

\begin{figure}
\epsfig{file=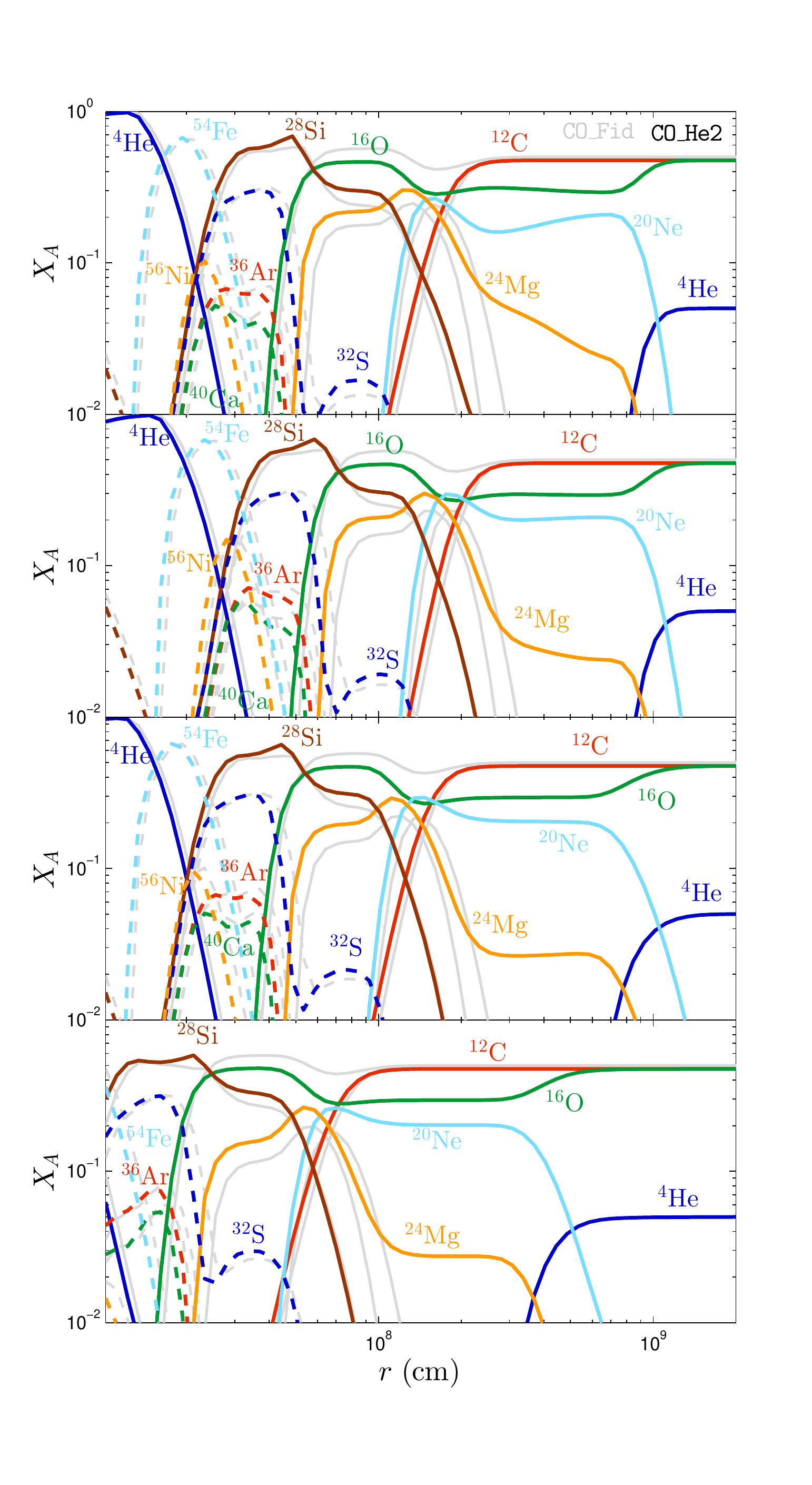,angle=0,width=0.45\textwidth} 
\caption{Composition profiles of hybrid WD model 
$\mathtt{CO\_He2}$. The background grey curves 
plot the composition of the fiducial C/O WD model ($\mathtt{CO\_Fid}$).
This model, which contains only $5\%$ initial $^{4}$He abundances is comparatively similar to the fiducial C/O composition profiles except for $^{16}$O $\alpha$-captures which fuse $^{20}$Ne and $^{24}$Mg at large radii $\sim 10^9~{\rm cm}$.
} \label{fig:X_A_Hybrid}
\end{figure}

The composition profile of model $\mathtt{CO\_He2}$ and its evolution are illustrated in Fig.~\ref{fig:X_A_Hybrid}.  Grey curves show for comparison the results of the fiducial C/O WD model, $\mathtt{CO\_Fid}$. The composition profiles are generally similar to the C/O model. The primary difference is at large radii, where $\alpha$-captures onto $^{16}$O fuse $^{20}$Ne and subsequently $^{24}$Mg already at $\sim 10^9~{\rm cm}$. This increases these isotopes' abundances in the wind significantly, but does not alter the profiles at small radii appreciably.  The composition profile of model $\mathtt{CO\_He1}$, which has a large initial helium abundance, schematically extends the same trend apparent in $\mathtt{CO\_He2}$. $\alpha$-captures efficiently burn the initial oxygen content into ${^{20}}$Ne, ${^{24}}$Mg and even ${^{28}}$Si at large radii.  Carbon burning at $r \sim 10^8~{\rm cm}$ replenishes the depleted $^{16}$O abundance, and at higher temperatures high mass isotopes are synthesized up to $^{56}$Ni.

\begin{figure*}
\centering
\begin{subfigure}[]{ \epsfig{file=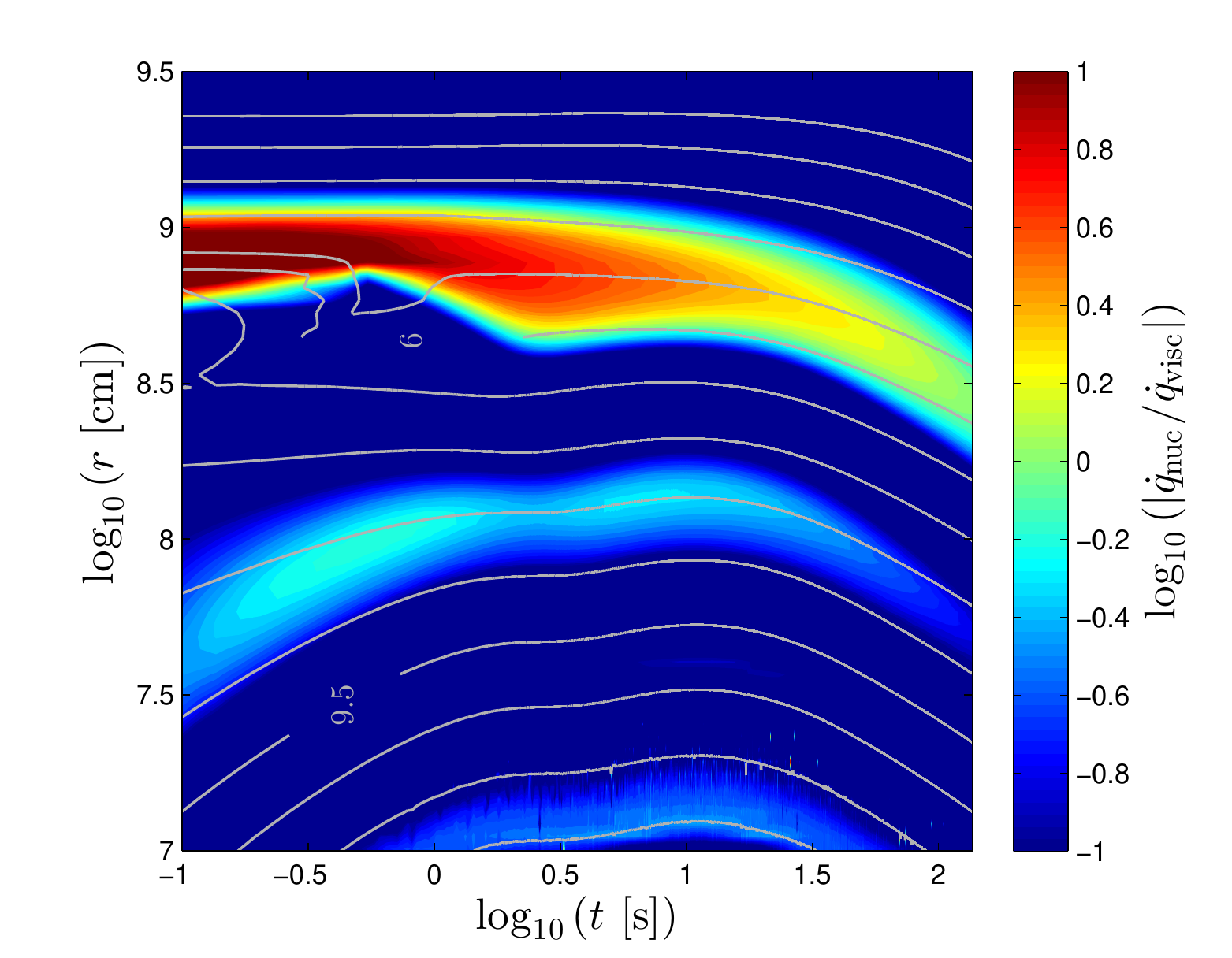,angle=0,width=0.45\textwidth} }\end{subfigure}
~
\begin{subfigure}[]{ \epsfig{file=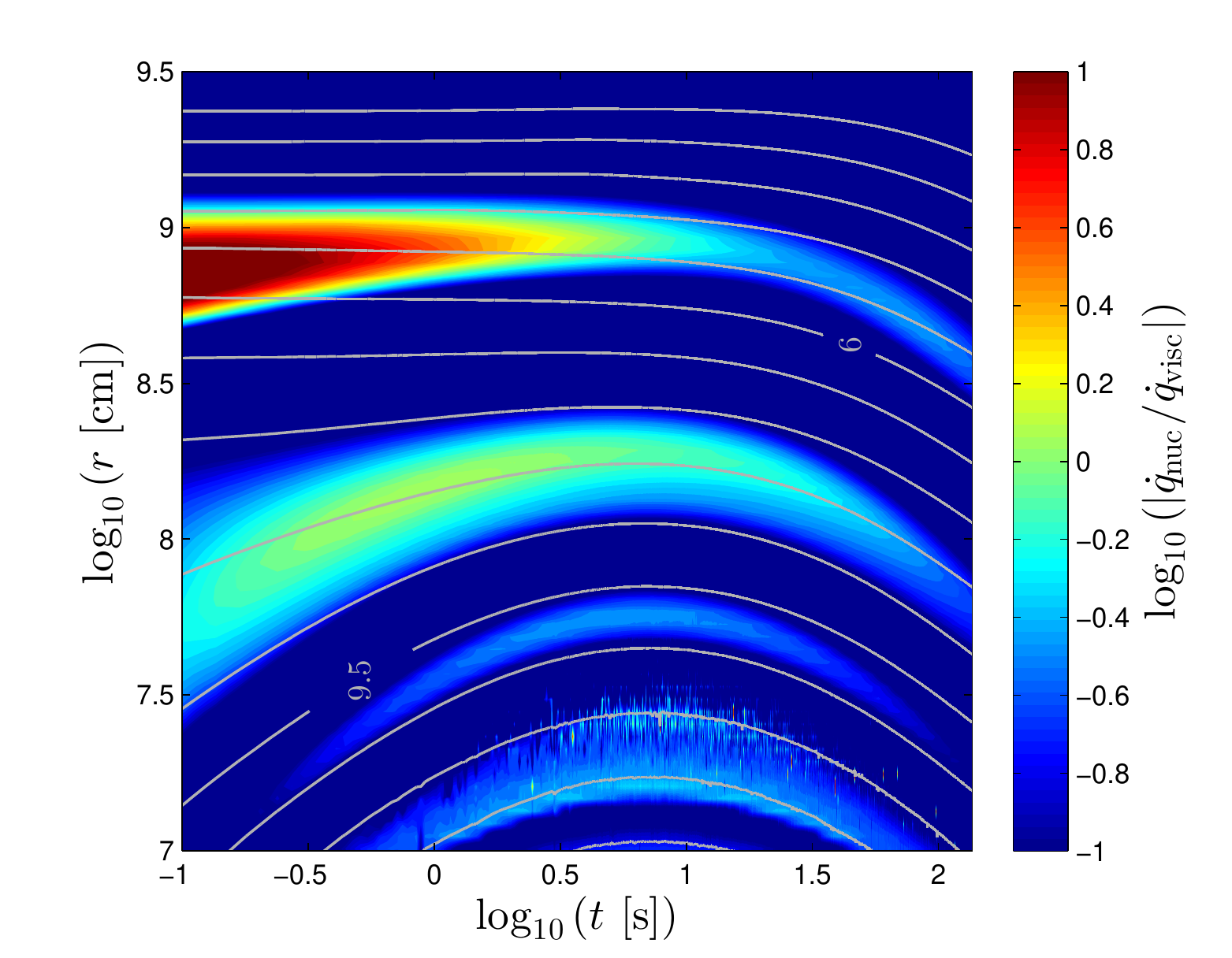,angle=0,width=0.45\textwidth} }\end{subfigure}
\caption{Nuclear heating rate relative to the viscous heating rate, $\vert \dot{q}_{\rm nuc} / \dot{q}_{\rm visc} \vert$, for the hybrid WD models $\mathtt{CO\_He1}$ {\bf (a)}, and $\mathtt{CO\_He2}$ {\bf (b)}. In contrast to previous plots of this quantity for other models (see Figs.~\ref{fig:qdot_b}, \ref{fig:qdot_He}), in this case the colorbar is logarithmically spaced. The nuclear heating rate exceeds the viscous heating rate by more than an order of magnitude at early times and around the $^{16}{\rm O}(\alpha,\gamma)^{20}{\rm Ne}$ burning front ($\sim 10^9~{\rm cm}$).
} \label{fig:qdot_Hybrid}
\end{figure*}

One small but noticeable difference of the hybrid models is that the burning fronts of heavy isotopes shift slightly to smaller radii, indicating that the density, and hence temperature at a given radius are smaller than for the fiducial C/O model. The reason is the familiar argument --- nuclear reactions, in this case at the $^{16}{\rm O}(\alpha,\gamma)^{20}{\rm Ne}$ burning front, deposit large amounts of energy at large radii, which launches substantial outflows and decreases the density at smaller radii. This is illustrated in Fig.~\ref{fig:qdot_Hybrid}, which shows contours of $\vert \dot{q}_{\rm nuc} / \dot{q}_{\rm visc} \vert$ (spaced logarithmically this time). The nuclear heating rate exceeds the viscous heating rate by over an order of magnitude at early times around $r \lesssim 10^9~{\rm cm}$.

The short timescales and large energy release associated with the $^{16}$O $\alpha$-captures suggest that this burning may realistically produce a detonation instead of a steady inflow. Such a detonation would not be captured by our numerical scheme, and we therefore cannot resolve this in our present work. Estimating the ratio of the burning to dynamical timescales, we find for the hybrid WD models
\begin{equation}
\frac{t_{\rm nuc}}{t_{\rm dyn}} \sim \left. \frac{u \Omega_{\rm k}}{\dot{q}_{\rm nuc}} \right\vert_{r_{\rm burn}} < 1 ~,
\end{equation}
indicating that nuclear burning proceeds dynamically.
Importantly, none of the other C/O or He WD models constructed in our work satisfy this criterion, illustrating that this is a direct feature of composite He/O matter burning.

\section{Discussion} \label{sec:Discussion}

Disk outflows from WD-NS mergers are capable of powering short lived supernova-like optical transients (\citetalias{Metzger2012}). These fast transients peak on a characteristic timescale of (\citealt{Arnett82})
\begin{align}
t_{\rm pk} &= \left( \frac{3 \kappa M_{\rm w}}{4 \uppi c \langle v_{\rm w} \rangle} \right)^{1/2} \\ \nonumber
&\approx 6.5~{\rm days}~\left( \frac{M_{\rm w}}{0.4 M_\odot} \right)^{1/2} \left( \frac{\langle v_{\rm w} \rangle}{10^9~{\rm cm~s}^{-1}} \right)^{-1/2}, 
\end{align}
where $\kappa = 0.05 ~{\rm cm}^2~{\rm g}^{-1}$ is the opacity, normalized to a value appropriate for Fe-poor matter, and $\langle v_{\rm w} \rangle$ is the mass-weighted average velocity of the ejecta.  The peak luminosity of the transient approximately equals the rate of thermal heating of the ejecta at the peak time, $L_{\rm pk} \approx \dot{E}(t_{\rm pk})$. If radioactive decay of $^{56}$Ni provides the dominant heating source, then 
the optical transients are typically dim,
\begin{equation}
L_{\rm pk} \approx 3 \times 10^{40}~{\rm erg~s}^{-1}~ \frac{M_{\rm w}(^{56}{\rm Ni})}{10^{-3} M_\odot} \exp \left[ 0.8 \left( 1 - \frac{t_{\rm pk}}{ 7{\rm day}} \right) \right]~,
\end{equation}
given the modest $^{56}$Ni yields of our disk wind solutions, $M_{\rm w}(^{56}{\rm Ni}) \sim 10^{-4}-3\times 10^{-3} M_{\odot}$ (Table \ref{tab:OutflowProperties}).  The amount of nickel in the ejecta could in principle be increased due to outflows from the very inner portions of the accretion disk near the central compact object.  Here the midplane is composed of alpha particles and free nucleons, but the temperature is high enough that heavy elements are synthesized above the disk midplane, i.e. within the outflow itself (\citealt{MacFadyen&Woosley99}). 

The luminosity of the transient could also be enhanced by additional energy deposited within the wind ejecta by shocks.  High velocity winds which are launched off the disk at late times ($\gg t_{\rm visc,0}$) could collide with the bulk of ejecta shell launched earlier, thermalizing the kinetic energy of the late-time winds.  The kinetic power released from the disk winds at late times and small radii $\sim r_*$ based on our analytic estimates in $\S \ref{sec:latetime}$ may be crudely estimated as
\begin{align}
\dot{E}_{\rm w}(r_*,t) \sim \frac{GM M_{\rm d}}{r_* t_{\rm visc, 0}} \left(\frac{r_*}{R_{\rm d,0}}\right)^p \left(\frac{t}{t_{\rm visc, 0}}\right)^{-(2p+4)/3}~.
\end{align}
For characteristic parameters ($p=0.5$), this yields peak transient luminosities of 
\begin{align}
L_{\rm pk} &\sim 2 \times 10^{43}~{\rm erg~s}^{-1} \left(\frac{M}{1.4 M_\odot}\right)^{2/3} \left(\frac{M_{\rm d}}{0.6 M_\odot}\right) \left(\frac{\alpha}{0.1}\right)^{-2/3} \\ \nonumber 
\times &   \left(\frac{\theta}{0.4}\right)^{-4/3}  \left(\frac{R_{\rm d,0}}{10^9~{\rm cm}}\right)^{1/2} \left(\frac{r_*}{5 \times 10^6~{\rm cm}}\right)^{-1/2} \left(\frac{t_{\rm pk}}{7 ~{\rm day}}\right)^{-5/3} ~,
\end{align}
comparable to those of normal SNe.  Such a scenario might give rise to more luminous, rapidly-evolving transients, such as SN 2002bj (\citealt{Poznanski+10}; \citealt{Drout+14}; \citealt{Shivvers+16}). 

\begin{figure*}
\centering
\begin{subfigure}[]{ \epsfig{file=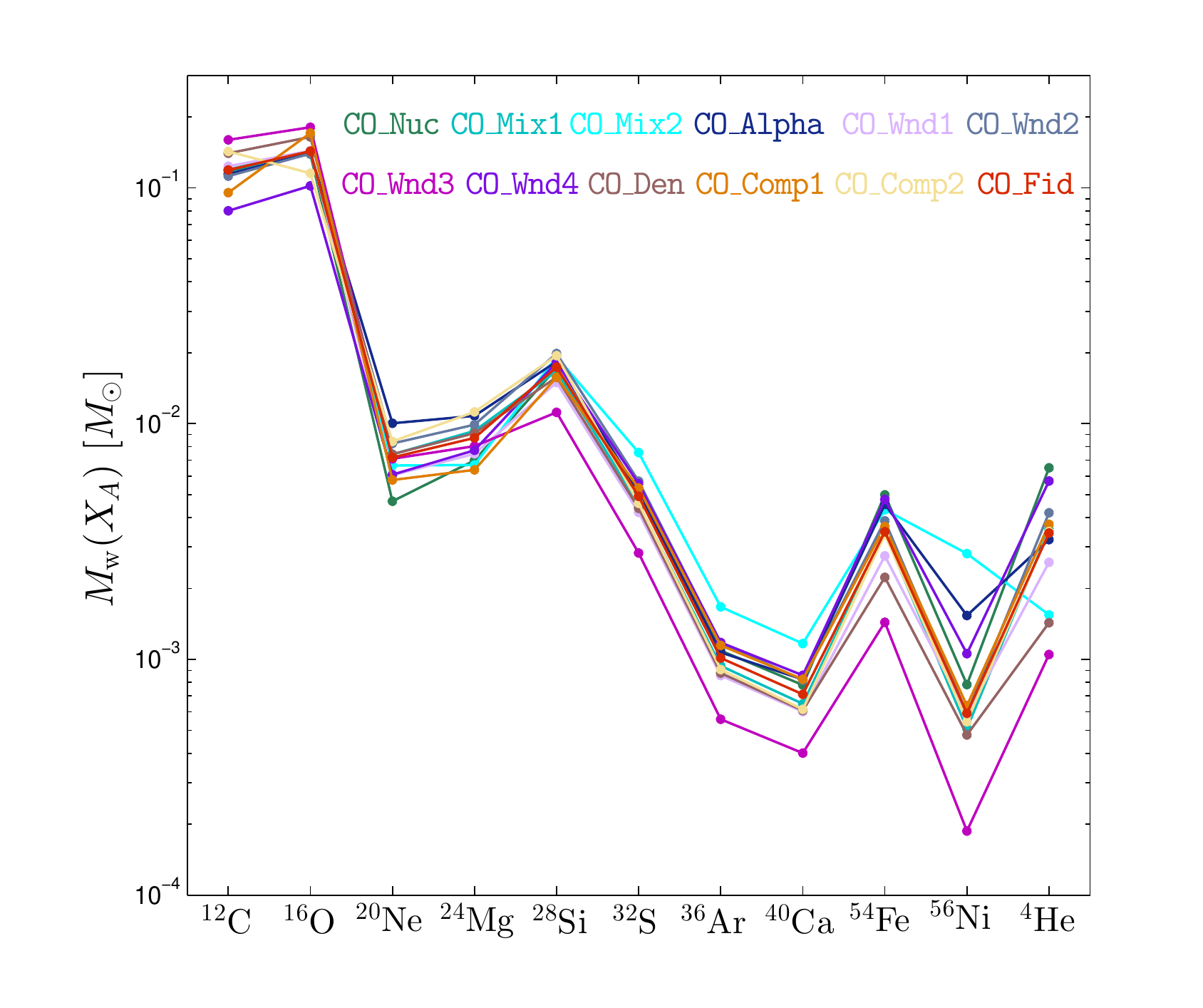,angle=0,width=0.45\textwidth} }\end{subfigure}
~
\begin{subfigure}[]{ \epsfig{file=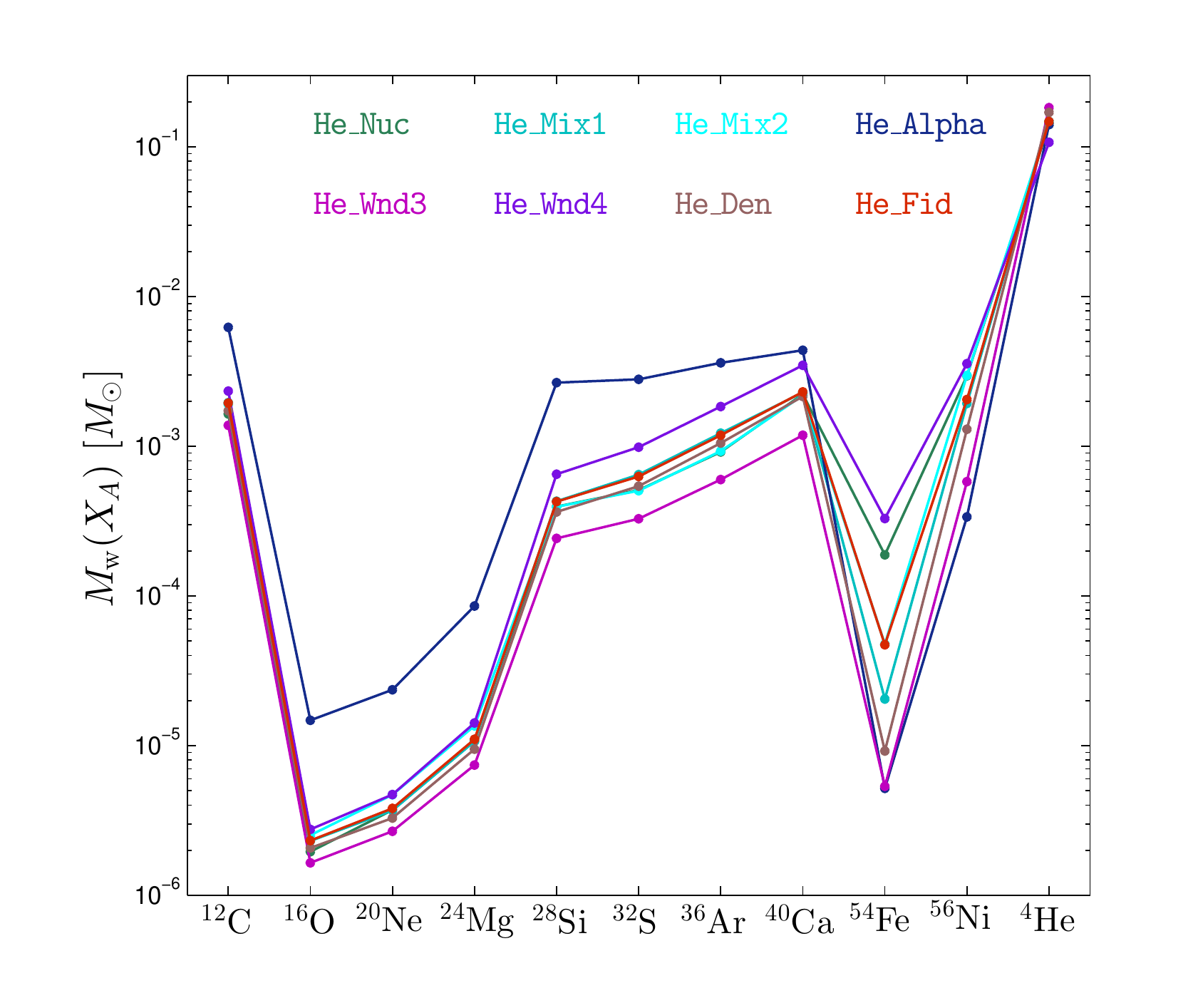,angle=0,width=0.45\textwidth} }\end{subfigure}
\caption{Total ejecta mass of each isotope at the end of the simulation for the C/O WD models {\bf (a)}, and He WD models {\bf (b)}. Tabulated data is also presented in Table \ref{tab:OutflowProperties}.
} \label{fig:Outflow_X_Runs}
\end{figure*}

The total nucleosynthetic yields of our models are summarized in Fig. \ref{fig:Outflow_X_Runs} and Table \ref{tab:OutflowProperties}.  For the C/O WD models, the nucleosynthesis is relatively robust for lighter isotopes, varying by factors of a few between models. Heavier elements, in particular $^{56}$Ni, show a significant scatter of nearly an order of magnitude between models.\footnote{Note that in these models helium is only synthesized by photodisintegrations at small radii, which are not entirely resolved in the boundaries of our numerical grid, and therefore the helium abundances are only lower bounds.}

Although $^{40}$Ca is among the most abundant isotopes produced in our He WD models, its total mass within the ejecta of $\lesssim 10^{-2}M_\odot$ appears to be insufficient to explain the inferred calcium abundances of the Ca-rich transients of $\gtrsim 0.1 M_{\odot}$ \citep[e.g. Table 4 of][]{Perets+2010}.  If these abundance measurements are robust, then WD-NS mergers as envisioned in this paper could represent at most only a subclass of these transients.  

Explosive burning in the accretion disk if a detonation wave develops (\citetalias{Fernandez&Metzger2013}) could also produce significantly larger calcium yields. Indeed, the hybrid C/O/He WD models show strong, dynamical, nuclear burning indicative of their possible susceptibility to explosions.  Their composition is similar to that found to give large Ca abundances following dynamical burning (e.g., \citealt{Perets+2010}; \citealt{Waldman+11}).   Similar dynamical burning and detonation of a disk-like configuration may occur during the core collapse of rapidly rotating stars (\citealt{Kushnir15}), also as a result of the low temperature threshold of the $^{16}{\rm O}(\alpha,\gamma)^{20}{\rm Ne}$ reaction.  Nuclear burning in this context may be particularly relevant to collapsar accretion disks and the source of $^{56}$Ni in gamma-ray burst SNe (\citetalias{Metzger2012}). 

We conclude with a discussion of the fate of the accreting NS in a WD-NS merger.  For our fiducial model, only a small fraction of the disrupted WD mass $\lesssim (r_*/R_{\rm d})^p$ is accreted onto the NS, with the remainder unbound from the system by outflows. For our fiducial C/O WD model, only $2.7 \times 10^{-2} M_\odot$ crosses the numerical boundary of our grid at the simulation end time, which corresponds to a conservative upper bound of $\lesssim 4.8 \times 10^{-2} M_\odot$ reaching the NS surface at $t=\infty$. A more realistic value of $2 \times 10^{-2} M_\odot$ is obtained if we extrapolate the wind mass loss to small radii, so that only a portion of the matter crossing the inner boundary of our numerical grid reaches the NS surface.

Given the large maximum NS mass inferred from recent observations of $\sim 2M_\odot$ NSs \citep{Demorest+2010,Antoniadis+2013}, it is improbable that a less massive NS will accrete enough matter to collapse to a BH. In fact, even if winds are inefficient at cooling the disk and $p$ is small (in contradiction with global MHD simulations), nearly the entire mass of the disrupted WD must be accreted to induce a collapse.  This contrasts with previous studies that neglect disk winds and nuclear burning \citep{Paschalidis+11}, which predict a collapse once the envelope sheds its angular momentum and cools.

Assuming that the final merger outcome is an isolated NS, a natural question is whether the small amount of mass accreted onto the NS surface can spin it up, forming a recycled millisecond pulsar. From our fiducial C/O WD model, we estimate that approximately $\approx 6 \times 10^{47}~{\rm g~cm}^2~{\rm s}^{-1}$ of angular momentum is accreted with the inflowing mass, which, for characteristic NS moments of inertia $\sim 10^{45}~{\rm g~cm}^{-2}$ \citep{Lattimer&Schutz2005} is equivalent to spinning up the NS from rest to a rotation period of $P \sim 10~{\rm ms}$. A more detailed analysis of the accretion process in the final region up to the NS surface is required to more accurately quantify this, but at the level of uncertainty of our current model, WD-NS mergers appear to provide another channel for producing isolated recycled millisecond pulsars (e.g.~\citealt{Lorimer+04}).

\section{Conclusions}
\label{sec:Conclusions}
We have presented a vertically averaged time-dependent model of the accretion disks from WD-NS mergers that incorporates nuclear burning. Such disks are expected as an outcome of unstable mass transfer between a WD and a binary NS companion, which may set in once GW emission drives the binary into Roche lobe contact.  
Note that besides characteristic masses, we have not assumed any parameters specific to the NS. As such, our model applies in its entirety to mergers of a WD with a stellar mass BH companion.

The extremely high density of the accretion flow renders it radiatively inefficient, necessitating an alternative means of cooling (other than photon radiation) to offset the nuclear and gravitational (viscous) heating. Following \cite{Blandford&Begelman1999} we have assumed that disk outflows provide this mechanism, and locally regulate the disk's enthalpy.  The properties of disk outflows predicted by our model (in particular, the mass loss coefficient, $p$) qualitatively agree with the results of global hydrodynamical and MHD disk simulations.

Nuclear burning plays a non-trivial role in both the dynamics and the nucleosynthesis in the accretion disk, as first described by \citetalias{Metzger2012}.  The radial composition profile resembles the `onion-skin' structure of evolved stars, where the initial WD matter is successively synthesized into heavier elements at sequentially smaller radii.  The temperature and density of the disk midplane at at any radius $\lesssim R_{\rm d}$ rise until the time of peak accretion $\sim t_{\rm visc}$, and subsequently decrease.  This shifts the radial composition profiles in C/O models to larger/smaller radii, respectively, and effectively inhibits nuclear burning for He models (which is limited by the triple-$\alpha$ barrier) at early/late times.

Unbound outflows from the disk carry away the majority of the initial WD mass at velocities of $\langle v_{\rm w} \rangle \sim 10^9~{\rm cm~s}^{-1}$.  Most of the wind ejecta is unburned, with a composition matching that of the initial WD.  However, the ejecta also contains a significant fraction of freshly synthesized intermediate-mass and heavy isotopes, including  $\sim 10^{-3} M_\odot$ of $^{56}$Ni. These outflows may give rise to a short-lived $\sim$week long optical transient similar to SNe, as well as a long-term radio relic due to the interaction of the fast ejecta with the interstellar medium.  We additionally find that accretion onto the NS surface is relatively limited ($\sim 10^{-2} M_\odot$); it is thus unlikely that the NS will collapse to a BH, but it might accrete sufficient angular momentum to be spun up to periods of $P\sim10~{\rm ms}$. 

The qualitative features of our numerical models are summarized as follows:
\begin{enumerate}
\item The radial composition profiles of C/O WD models preserve a fixed morphology, which evolves self-similarly with time. This transcends any specific model parameter assumptions and applies globally to all our C/O WD simulations. In this sense, one can approximate the flow as a steady-state model at any given time, with only the outer mass feeding rate $\dot{M}_{\rm in}(R_{\rm d})$ secularly changing between epochs.

\item For C/O models, nuclear burning only moderately impacts the disk dynamics at the carbon burning front (and to a lesser extent at the oxygen burning front and the photodisintegration region). He WD models are affected more significantly by nuclear burning, especially for small $\alpha$ or large $\eta_{\rm w}$, in which case nuclear heating becomes the dominant energy source in a large portion of the disk.

\item The outflow rate in isotopes other than the initial WD composition peaks on short timescales of $\sim t_{\rm visc}$, and is therefore well captured by our simulations. Extrapolations of the ejecta composition to late times yields in most cases nearly identical results.

\item The results are robust to the initial density distribution, the `chemical' mixing efficiency, and the (regulated) disk Bernoulli parameter. The C/O models are also relatively unaffected by changes to the Shakura-Sunyaev alpha-viscosity parameter, except by scaling the evolution time of the disk ($t_{\rm visc} \propto \alpha^{-1}$). He WD models on the other hand are sensitive to the value $\alpha$. Both C/O and He WD models depend strongly on the mass inflow exponent, $p$, which is set primarily by the wind efficiency parameter $\eta_{\rm w}$.
\item `Hybrid' C/O/He WD models with even modest helium mass fractions exhibit strong, dynamical, nuclear burning, indicative of their possible explosive nature. None of our C/O or He WD models showed signs of dynamic burning.
\end{enumerate}

The one-dimensional model presented here is only an approximate starting point to accurately modeling the aftermath of WD-NS/BH mergers.  It is nevertheless justified given the limited number of previous studies of these systems, and the rich behavior even this simple model already reveals.  Future work, including multi-dimensional hydrodynamic models, is needed to explore outstanding issues such as the role of dynamical burning and relative importance of convection in transporting energy outwards in the accretion flow \citep{Narayan&Yi1994,Narayan+2000,Quataert&Gruzinov2000}.  Despite its limitations, our approach has allowed us to extensively explore the parameter space of WD-NS merger accretion disks (which would be computationally prohibitive with multi-dimensional simulations), and to develop analytic estimates to aid future studies. In particular, we plan to pursue in future work more detailed models for the optical and radio transients of the disk outflows calculated here.

\section*{Acknowledgments}

The authors gratefully acknowledge support from the NSF grant AST-1410950, NASA grants NNX15AR47G and NNX16AB30G, and the Alfred P.~Sloan Foundation. 

\bibliography{WD_NSBH_Bibliography}

\appendix
\section{Initial Conditions} \label{subsec:Appendix_InitialConditions}
This section provides additional details on the initial radial profile of the WD accretion disk (\S \ref{sec:WD_Disruption}). Assuming a radial surface density profile as parameterized in equation (\ref{eq:Density_initial}), the normalization factor $\mathcal{N}$ is found by requiring that the disk mass equal that of the disrupted WD, 
\begin{equation}
\int 2 \uppi r \Sigma \,dr = M_{\rm WD} ~.
\end{equation}
This yields
\begin{align}
\mathcal{N}(m,n) \equiv \left(\frac{m+2}{n-2}\right)^{m+2} 
\frac{\Gamma(m+n)}{\Gamma(m+2) \Gamma(n-2)} ~,
\end{align}
where $\Gamma(x)$ is the gamma function.

The characteristic radius of the disk $R_{\rm d}$ is determined by requiring that the total angular momentum of the torus,
\begin{equation}
\int 2 \uppi r \Sigma \times \Omega r^2 \,dr = J_\mathrm{tot} ~,
\end{equation}
 equal that of the binary at the time of disruption, 
\begin{equation}
J_\mathrm{tot} = M_{\rm WD} \sqrt{G M_{\rm NS} R_{\rm c}},
\end{equation}
where Keplerian rotation, $\Omega=\Omega_{\rm k}$, is assumed.  The proportionality constant which relates $R_{\rm d}$ to the circularization radius, $R_{\rm c}$ (equation \ref{eq:R_circ}), is given by
\begin{equation} \label{eq:Rmn}
\mathcal{R}(m,n) \equiv \frac{R_{\rm d}}{R_{\rm c}} = \frac{m+2}{n-2} \left[ \frac{\Gamma(m+2)\Gamma(n-2)}{\Gamma(m+5/2)\Gamma(n-5/2)} \right]^2 ~.
\end{equation}

\begin{figure} \label{fig:Rmn_contours}
\centering
\epsfig{file=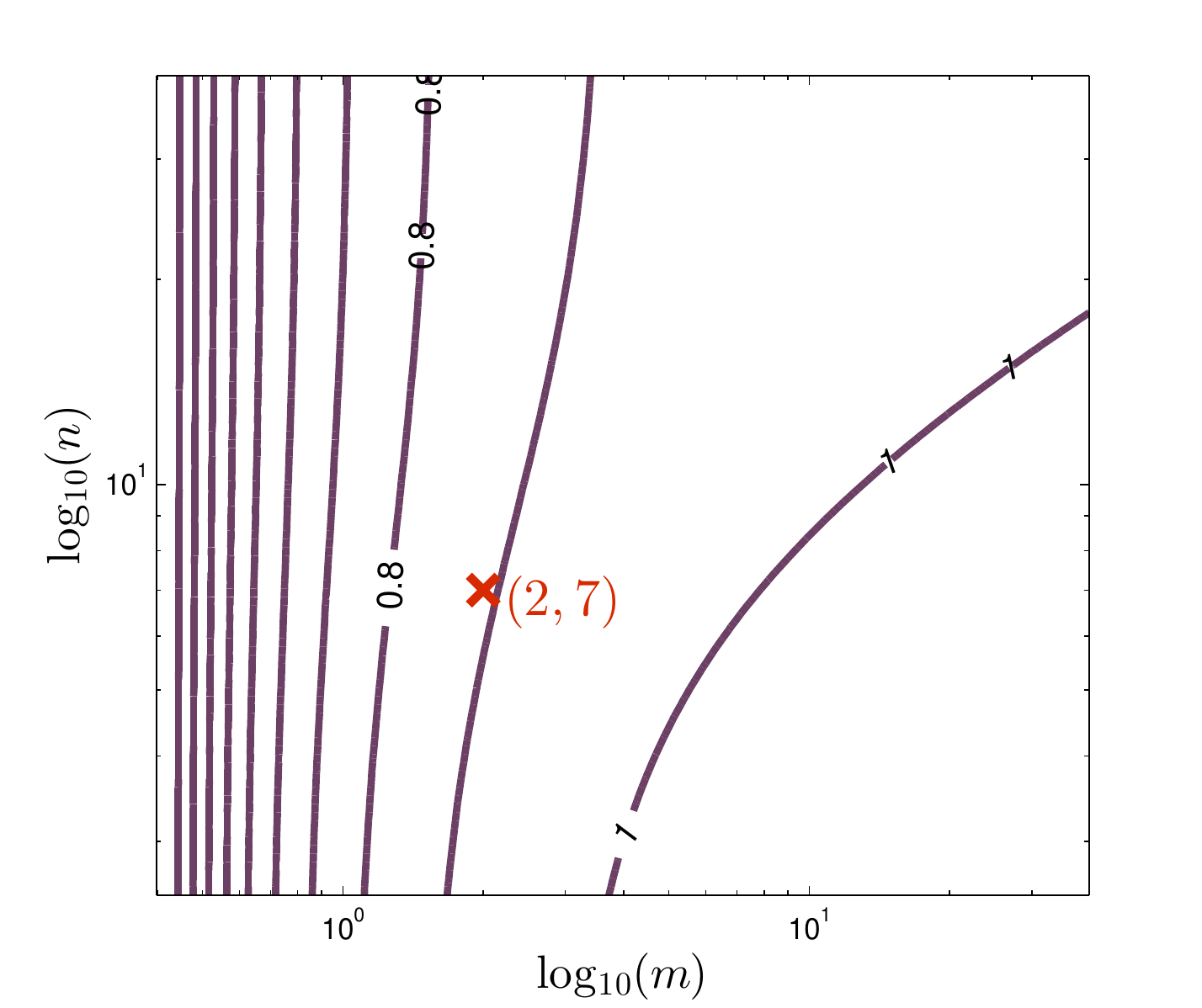,angle=0,width=0.5\textwidth}
\caption{Contours of constant $\mathcal{R}(m,n) = R_{\rm d}/R_{\rm c}$ (equation~\ref{eq:Rmn}) as a function of the power-law indices $m$ and $n$ entering the initial density profile $\Sigma_0(r)$ (equation \ref{eq:Density_initial}).  The curves are spaced equally in steps of $\Delta \mathcal{R} = \pm 0.1$. The fiducial parameters used in our numerical analysis $(m,n)=(2,7)$ are marked as a red cross.}
\end{figure}

As shown in Fig. \ref{fig:Rmn_contours}, the value of this constant is typically $\mathcal{R}(m,n) \lesssim 1$, indicating that the disk radius $R_{\rm d}$ is generally smaller than the circularization radius.

Finally, the function $\mathcal{T}(m,n)$, which determines the initial disk aspect ratio through equation (\ref{eq:theta_initial}), is found by equating the energy of the binary at disruption,
\begin{equation}
E_{\rm tot} = -\frac{G M_{\rm WD} M_{\rm NS} }{ 2 (1+q) R_{\rm c} } ~,
\end{equation}
with the combined gravitational, kinetic, and internal energy (which is proportional to $\theta^2$) of the disk, i.e. 
\begin{equation}
\left[ 1 - \frac{2 \theta^2 }{(\gamma - 1)} \right] \int 2 \uppi r \Sigma \times \left( - \frac{1}{2} \Omega_{\rm k}^2 r^2 \right) \,dr = E_\mathrm{tot} ~.
\end{equation}
For our initial density prescription, this yields
\begin{align}
\mathcal{T}(m,n) \equiv \frac{n-2}{m+1} \left[ \frac{\Gamma(m+2)\Gamma(n-2)}{\Gamma(m+5/2)\Gamma(n-5/2)} \right]^{-2} ~.
\end{align}

\section{Derivation of `Precursor' Outflow} \label{subsec:Appendix_InitialOutflow}
Here we estimate the fraction of the initial disk mass which is lost to `precursor' winds at very early times within the framework of our outflow model.  As described in \S \ref{subsec:MassInflowIndex}, these winds occur if the initial aspect ratio of the disk exceeds its steady-state value, $\theta_{\rm ss}$ (equation \ref{eq:theta_ss}).  On a short timescales $\sim t_{\rm w} < t_{\rm dyn} \ll t_{\rm visc}$, the disk aspect ratio is regulated by this excess energy loss until $\theta \simeq \theta_{\rm ss}$.

Since the mass loss timescale is shorter than other timescales in the problem, we assume that wind cooling dominates during the initial transient phase, such that
\begin{equation}
\frac{\dot{\theta}}{\theta} = \frac{1}{2} \frac{\dot{q}_\mathrm{w} / \Sigma}{u} = - \frac{(\gamma-1) \left[\eta_\mathrm{w} - \mathrm{Be^\prime_{d}}(\theta)\right]}{2\Sigma} \theta^{-2} \dot{\Sigma}_\mathrm{w} ~,
\end{equation}
where we have used the wind cooling prescription (equation~\ref{eq:q_dot_w}) and the fact that ${\dot{\theta}}/{\theta} = {\dot{u}}/2{u}$ for a gamma-law EOS (for which $\theta \propto \sqrt{u}$).  The short transient timescale also implies that the disk density changes almost entirely due to wind mass losses, such that $d\Sigma_{\rm w} \approx -d\Sigma$, and therefore
\begin{equation}
\frac{\gamma-1}{\gamma} \Sigma^{-1} d\Sigma = \frac{\theta}{\eta_{\rm w} + {1}/{2} - \theta^2 {\gamma}/({\gamma-1})} d\theta ~.
\end{equation}
Here we have explicitly used equation (\ref{eq:Bernoulli_theta}) for the Bernoulli parameter.  Integrating from the initial to final density/aspect-ratio, we obtain
\begin{equation}
\frac{\Sigma_i}{\Sigma_f} = \left( \frac{\chi - \theta_i^2}{\chi - \theta_f^2} \right)^{1/\gamma} ~,
\end{equation}
where we have defined $\chi \equiv (\eta+1/2) \gamma(\gamma-1)^{-1}$.

Finally, we obtain the amount of wind launched off the disk during the transient by radially integrating the disk surface density. This yields
\begin{align} \label{eq:M_precursor1}
M_{\rm w}^{\rm (precursor)} &= - \int 2 \uppi r \left( \Sigma_i - \Sigma_f \right) \, dr \\ \nonumber
&= M_{\rm d} \times \left[ 1 - \left( \frac{\chi - \theta_{\rm initial}^2}{\chi - \theta_{\rm ss}^2} \right)^{1/\gamma} \right] ~.
\end{align}
Here we have identified the initial mass as $M_{\rm d}$, and the initial (final) aspect ratios as $\theta_{\rm initial}$ ($\theta_{\rm ss}$) respectively, in accordance with previous notation. Expanding equation (\ref{eq:M_precursor1}) in powers of $\Delta \theta / \theta_{\rm ss} \ll 1$, where $\Delta \theta \equiv \theta_{\rm initial}-\theta_{\rm ss}$, and plugging in our definition for $\chi$ as well as the explicit solution for $\theta_{\rm ss}$ (equation \ref{eq:theta_ss}), we obtain our final result
\begin{equation}
M_{\rm w}^{\rm (precursor)} \approx \frac{1+2{\rm Be^\prime_{crit}}}{\gamma \left(\eta_{\rm w} - {\rm Be^\prime_{crit}} \right)} \left(\frac{\Delta \theta}{\theta_{\rm ss}}\right) \times M_{\rm d} ~.
\end{equation}

\section{Solution for Mass Inflow Exponent} \label{subsec:Appendix_pExponent}
Here we derive an explicit analytic expression for the steady-state mass inflow exponent, $p$, as a function of the model parameters. Along the way we obtain a few additional results of interest.

Using the definition of $p$ (equation~\ref{eq:p_index_definition}) and equation~(\ref{eq:v_r}) for the radial accretion velocity, we find by solving the continuity equation~(\ref{eq:continuity}) with $\partial_t = 0$, a steady-state wind mass loss rate of
\begin{equation} \label{eq:Sigma_dot_w_ss}
\dot{\Sigma}_{\rm w_{ss}} = 3 \alpha \theta_{\rm ss}^2 p \left( p + \frac{1}{2} \right) \Sigma \Omega_{\rm k} ~.
\end{equation}
The steady-state disk aspect ratio, $\theta_{\rm ss}$, can be substituted into this expression using equation (\ref{eq:theta_ss}). Note that in contrast to previous expressions for the wind mass loss rate (such as equation \ref{eq:Sigma_dot_w}), this result is independent of our adopted wind prescription, being entirely a consequence of mass conservation.

Turn now to energetic considerations. It is straightforward to show that the ratio of advective cooling relative to the viscous heating rate in steady state is a constant value, 
\begin{equation} \label{eq:Appendix_q_dot_adv}
\left\vert \frac{\dot{q}_\mathrm{adv}}{\dot{q}_\mathrm{visc}} \right\vert_\mathrm{ss} =
\frac{4}{3} \left(p+\frac{1}{2} \right) \left( \frac{1}{\gamma-1} + \frac{1}{2} - p\right) \theta^2_\mathrm{ss} ~.
\end{equation}
The specific advective cooling rate in the steady-state regime is defined as $\dot{q}_\mathrm{adv} / \Sigma = v_r \partial_r u + c_{\rm s}^2 v_r \partial_r \ln \Sigma$, and this expression is derived assuming a gamma-law EOS (equation \ref{eq:gamma-law_EOS}).

Using for the first time the specifics of our wind parameterization, equation (\ref{eq:q_dot_w}), the ratio of wind cooling to viscous heating is given by
\begin{align} \label{eq:Appendix_q_dot_w_ss}
\left\vert \frac{\dot{q}_\mathrm{w}}{\dot{q}_\mathrm{visc}} \right\vert_\mathrm{ss} &= \frac{4}{9} \alpha^{-1} \theta_{\rm ss}^{-2} \frac{\dot{\Sigma}_{\rm w_{ss}}}{\Sigma \Omega_{\rm k}} \left(\eta_\mathrm{w} - \mathrm{Be^\prime_{crit}} \right) \\ \nonumber
&= \frac{4}{3} p \left( p + \frac{1}{2} \right) \left(\eta_\mathrm{w} - \mathrm{Be^\prime_{crit}} \right),
\end{align}
where in the second equality we have substituted $\dot{\Sigma}_{\rm w_{ss}}$ from equation (\ref{eq:Sigma_dot_w_ss}).  The only inherent assumption in the wind parameterization of equation (\ref{eq:q_dot_w}) which we have used in deriving this result, is that the specific energy of the wind scales with the escape velocity $v_{\rm k}$, and that the disk is regulated to a fixed Bernoulli parameter $\mathrm{Be^\prime_{crit}}$. Equation (\ref{eq:Appendix_q_dot_w_ss}) does not depend on the less certain form of $\dot{\Sigma}_\mathrm{w}$ given by equation (\ref{eq:Sigma_dot_w}).

This last result allows us to solve for the mass inflow exponent $p=p(\eta_{\rm w}, {\rm Be^\prime_{crit}}, \gamma)$, by requiring energy conservation in steady-state, i.e.
\begin{equation}
\left\vert \frac{\dot{q}_\mathrm{w}}{\dot{q}_\mathrm{visc}} \right\vert_{\rm ss} + \left\vert \frac{\dot{q}_\mathrm{adv}}{\dot{q}_\mathrm{visc}} \right\vert_{\rm ss} = 1 ,
\end{equation}
as implied by equation (\ref{eq:internal_energy}) for $\partial_t=0$ (and neglecting nuclear heating, $\dot{q}_{\rm nuc}$).
Substituting equations (\ref{eq:Appendix_q_dot_adv}) and (\ref{eq:Appendix_q_dot_w_ss}) into the last expression and using $\theta_{\rm ss}$ from equation (\ref{eq:theta_ss}), we rearrange to find 
\begin{align} \label{eq:Appendix_p_index}
p &= p(\eta_{\rm w}, {\rm Be^\prime_{crit}}, \gamma) = \\ \nonumber
&= \frac{1}{2} \bigg[ 1 - 2\mathrm{Be^\prime_{crit}} + \mathrm{Be^\prime_{crit}} \gamma - \eta_\mathrm{w} \gamma \\ \nonumber
&~~~~~~~+ \Big(12 \mathrm{Be^\prime_{crit}} \gamma - 5 \gamma^2 - 18 \mathrm{Be^\prime_{crit}} \gamma^2 + 10 \eta_\mathrm{w} \gamma^2 \\ \nonumber
&~~~~~~~~~~~~+ 9 \mathrm{Be^\prime_{crit}}^2 \gamma^2 + \eta_\mathrm{w}^2 \gamma^2 - 6 \mathrm{Be^\prime_{crit}} \eta_\mathrm{w} \gamma^2 + 6 \gamma \Big)^{1/2} \bigg] \\ \nonumber
&~~~\Bigg/ \bigg[ 2 \mathrm{Be^\prime_{crit}} - \gamma - 4 \mathrm{Be^\prime_{crit}} \gamma + 2 \eta_\mathrm{w} \gamma + 1 \bigg]
\end{align}

\section{Tabulated Outflow Properties}
\begin{table*}
\begin{tabular}{c  c  c  c  c  c  c  c  c  c  c  c}
\hline
Model & $^{4}$He & $^{12}$C & $^{16}$O & $^{20}$Ne & $^{24}$Mg & $^{28}$Si & $^{32}$S & $^{36}$Ar & $^{40}$Ca & $^{52}$Fe & $^{56}$Ni \\ 
 $(M_\odot)$ & $\times 10^{-3}$ & $\times 10^{-1}$ & $\times 10^{-1}$ & $\times 10^{-3}$ & $\times 10^{-3}$ & $\times 10^{-2}$ & $\times 10^{-3}$ & $\times 10^{-3}$ & $\times 10^{-4}$ & $\times 10^{-3}$ & $\times 10^{-4}$ \\ \hline

$\mathtt{CO\_Fid}$ & $3.4$ & $1.19$ & $1.43$ & $7.2$ & $8.7$ & $1.7$ & $4.9$ & $1.0$ & $7.1$ & $3.5$ & $5.9$ \\ 

$\mathtt{CO\_Nuc}$ & $6.5$ & $1.18$ & $1.39$ & $4.7$ & $7.0$ & $1.7$ & $5.1$ & $1.1$ & $7.8$ & $5.0$ & $7.8$ \\ 

$\mathtt{CO\_Mix1}$ & $3.6$ & $1.20$ & $1.43$ & $7.4$ & $9.3$ & $1.7$ & $4.6$ & $9.4$ & $6.5$ & $3.4$ & $5.1$ \\ 

$\mathtt{CO\_Mix2}$ & $1.5$ & $1.15$ & $1.42$ & $6.6$ & $6.7$ & $1.9$ & $7.5$ & $16.7$ & $11.7$ & $4.3$ & $28.1$ \\ 

$\mathtt{CO\_Alpha}$ & $3.2$ & $1.14$ & $1.41$ & $10.0$ & $10.8$ & $1.8$ & $5.1$ & $10.7$ & $8.2$ & $4.6$ & $15.4$ \\

$\mathtt{CO\_Wnd1}$ & $2.6$ & $1.24$ & $1.44$ & $6.0$ & $7.4$ & $1.5$ & $4.2$ & $0.86$ & $6.0$ & $2.8$ & $5.8$ \\

$\mathtt{CO\_Wnd2}$ & $4.2$ & $1.12$ & $1.40$ & $8.2$ & $9.9$ & $2.0$ & $5.7$ & $1.2$ & $8.2$ & $3.9$ & $5.9$ \\

$\mathtt{CO\_Wnd3}$ & $1.1$ & $1.60$ & $1.81$ & $7.1$ & $8.0$ & $1.1$ & $2.8$ & $0.56$ & $4.0$ & $1.4$ & $1.9$ \\

$\mathtt{CO\_Wnd4}$ & $5.7$ & $0.80$ & $1.02$ & $6.1$ & $7.7$ & $1.8$ & $5.6$ & $1.2$ & $8.6$ & $4.8$ & $10.6$ \\

$\mathtt{CO\_Den}$ & $1.4$ & $1.40$ & $1.65$ & $7.4$ & $9.1$ & $1.6$ & $4.4$ & $0.88$ & $6.1$ & $2.2$ & $4.8$ \\

$\mathtt{CO\_Comp1}$ & $3.7$ & $0.95$ & $1.71$ & $5.8$ & $6.4$ & $1.6$ & $5.3$ & $1.1$ & $8.2$ & $3.7$ & $6.4$ \\

$\mathtt{CO\_Comp2}$ & $3.5$ & $1.43$ & $1.15$ & $8.4$ & $11.2$ & $1.9$ & $4.6$ & $0.91$ & $6.1$ & $3.4$ & $5.4$ \\

\hline
Model & $^{4}$He & $^{12}$C & $^{16}$O & $^{20}$Ne & $^{24}$Mg & $^{28}$Si & $^{32}$S & $^{36}$Ar & $^{40}$Ca & $^{52}$Fe & $^{56}$Ni \\ 
 $(M_\odot)$ & $\times 10^{-1}$ & $\times 10^{-3}$ & $\times 10^{-6}$ & $\times 10^{-6}$ & $\times 10^{-6}$ & $\times 10^{-4}$ & $\times 10^{-4}$ & $\times 10^{-4}$ & $\times 10^{-4}$ & $\times 10^{-4}$ & $\times 10^{-3}$ \\ \hline
 
$\mathtt{He\_Fid}$ & $1.48$ & $1.9$ & $2.3$ & $3.8$ & $11.0$ & $4.3$ & $6.3$ & $11.8$ & $23.0$ & $0.47$ & $2.0$ \\ 

$\mathtt{He\_Nuc}$ & $1.41$ & $1.6$ & $2.0$ & $3.7$ & $10.7$ & $3.9$ & $5.1$ & $9.2$ & $22.2$ & $1.9$ & $3.0$ \\ 

$\mathtt{He\_Mix1}$ & $1.48$ & $2.0$ & $2.3$ & $3.7$ & $10.7$ & $4.3$ & $6.5$ & $12.2$ & $22.8$ & $0.20$ & $1.9$ \\ 

$\mathtt{He\_Mix2}$ & $1.47$ & $1.9$ & $2.5$ & $4.7$ & $13.6$ & $3.9$ & $5.0$ & $9.3$ & $21.9$ & $0.48$ & $2.0$ \\ 

$\mathtt{He\_Alphs}$ & $1.41$ & $6.2$ & $14.8$ & $23.6$ & $85.5$ & $26.6$ & $28.0$ & $36.1$ & $43.7$ & $0.052$ & $0.34$ \\ 

$\mathtt{He\_Wnd3}$ & $1.83$ & $1.4$ & $1.6$ & $2.7$ & $7.4$ & $2.4$ & $3.3$ & $6.0$ & $11.9$ & $0.054$ & $0.58$ \\ 

$\mathtt{He\_Wnd4}$ & $1.07$ & $2.3$ & $2.8$ & $4.7$ & $14.1$ & $6.5$ & $9.9$ & $18.4$ & $34.7$ & $3.3$ & $3.6$ \\ 

$\mathtt{He\_Den}$ & $1.70$ & $1.7$ & $2.1$ & $3.3$ & $9.5$ & $3.6$ & $5.4$ & $10.5$ & $21.5$ & $0.092$ & $1.3$ \\ 

$\mathtt{He\_Mass}$ & $1.85$ & $7.3$ & $8.9$ & $17.6$ & $52.6$ & $20.0$ & $25.6$ & $38.4$ & $62.2$ & $4.1$ & $2.9$ \\ 

\hline
Model & $^{4}$He & $^{12}$C & $^{16}$O & $^{20}$Ne & $^{24}$Mg & $^{28}$Si & $^{32}$S & $^{36}$Ar & $^{40}$Ca & $^{52}$Fe & $^{56}$Ni \\ 
 $(M_\odot)$ & $\times 10^{-3}$ & $\times 10^{-1}$ & $\times 10^{-1}$ & $\times 10^{-3}$ & $\times 10^{-3}$ & $\times 10^{-2}$ & $\times 10^{-3}$ & $\times 10^{-3}$ & $\times 10^{-4}$ & $\times 10^{-3}$ & $\times 10^{-4}$ \\ \hline

$\mathtt{CO\_He1}$ & $34.8$ & $1.26$ & $0.63$ & $35.7$ & $45.9$ & $3.2$ & $1.7$ & $0.22$ & $1.3$ & $0.99$ & $1.0$ \\ 

$\mathtt{CO\_He1}$ & $9.5$ & $1.24$ & $1.17$ & $31.0$ & $13.4$ & $1.7$ & $3.8$ & $0.74$ & $5.0$ & $2.6$ & $4.2$ \\ 

\hline
\end{tabular}
\caption{Ejected mass in various elements at the simulation end time, $t_{\rm end}$.
} \label{tab:OutflowProperties}
\end{table*}

\end{document}